\documentclass[journal]{IEEEtran}
\usepackage{tikz}
\usepackage{amsmath}
\usepackage{tabularx}
\usepackage{tikz}
\usepackage{amsmath}
\usepackage{wasysym}
\usepackage{algorithmic}
\usepackage{graphicx}
\usepackage{textcomp}
\usepackage{xcolor}
\usepackage{amsmath, bm}
\usepackage{hyperref}
\usepackage{tikz}
\usetikzlibrary{positioning, shapes.geometric}
\usepackage{amsmath}
\usepackage{tikz}
\usetikzlibrary{shapes, arrows.meta, positioning, calc}
\usepackage{tikz}
\usepackage{pgfplots}
\usepackage{textcomp}
\usepackage{stfloats}
\usepackage{url}
\usepackage{tcolorbox}
\usepackage{verbatim}
\usepackage{graphicx}
\usepackage{amsmath,amssymb,amsfonts}
\usepackage{algorithmic}
\usepackage{graphicx}
\usepackage{textcomp}
\usepackage{xcolor}
\usepackage{amsmath, bm}
\usepackage{soul}

\usepackage{mathtools}
\usepackage{relsize}

\usepackage{booktabs}
\usepackage[para,online,flushleft]{threeparttable}
\usepackage{amsmath}
\usepackage{subcaption}
\usepackage{graphicx}
\usepackage{mathrsfs}  
\usepackage{pifont}
\usepackage{xcolor}
\usepackage[linesnumbered,ruled,vlined]{algorithm2e}

\SetCommentSty{mycommfont}

\SetKwInput{KwInput}{Input}                
\SetKwInput{KwOutput}{Output}              
\usepackage{algorithmic}
\usepackage{graphicx}
\usepackage{amsmath}
\usepackage{textcomp}
\usepackage{xcolor}
\def\BibTeX{{\rm B\kern-.05em{\sc i\kern-.025em b}\kern-.08em
    T\kern-.1667em\lower.7ex\hbox{E}\kern-.125emX}}

\usepackage{adjustbox}
\usepackage{multirow}
\usepackage{amssymb,mathtools}
\usepackage{array}
\newcolumntype{P}[1]{>{\centering\arraybackslash}p{#1}}
\newcolumntype{M}[1]{>{\centering\arraybackslash}m{#1}}

\usepackage{tikz}

\usepackage{xcolor}
\usepackage[linesnumbered,ruled,vlined]{algorithm2e}

\usepackage[symbol]{footmisc}

\usepackage{adjustbox}
\usepackage{booktabs}
\usepackage{todonotes}
\usepackage{multirow}
\usepackage{threeparttable}
\usepackage{tcolorbox}

\newtcolorbox{notebox}[1]{
    colback=gray!5!white,
    colframe=gray!50!black,
    boxrule=1pt,
    arc=4pt,
    left=6pt,
    right=6pt,
    top=6pt,
    bottom=6pt,
    fontupper=\itshape,
    title={#1}
}
\SetKwInput{KwInput}{Input}                
\SetKwInput{KwOutput}{Output}              

\begin{document}

\date{}

\title{Baiting AI: Deceptive Adversary Against AI-Protected Industrial Infrastructures}

\author{Aryan~Pasikhani, 
        Prosanta~Gope,~\IEEEmembership{Senior~Member,~IEEE}, Yang Yang,~\IEEEmembership{Student~Member,~IEEE} \\ Shagufta Mehnaz, Biplab Sikdar, \IEEEmembership{IEEE Fellow} 

\IEEEcompsocitemizethanks{

\IEEEcompsocthanksitem A. Pasikhani and  P. Gope are with the Department of Computer Science, University of Sheffield, Regent Court, Sheffield S1 4DP, United Kingdom.
(E-mail: a.mohammadipasikhani@sheffield.ac.uk p.gope@sheffield.ac.uk)

\IEEEcompsocthanksitem Yang Yang and Biplab~Sikdar are with the Department of Electrical and Computer Engineering, National University of Singapore, Singapore. 
(E-mail: y.yang@u.nus.edu, bsikdar@nus.edu.sg)

\IEEEcompsocthanksitem \textbf{Corresponding author:} Dr. Aryan Pasikhani
}
}

\markboth{IEEE Transactions on Dependable and Secure Computing\ }%
{Shell \MakeLowercase{\textit{Pasikhani et al.}}: A Sample Article Using IEEEtran.cls for IEEE Journals}

\maketitle

\begin{abstract}
This paper explores a new cyber-attack vector targeting Industrial Control Systems (ICS), particularly focusing on water treatment facilities. Developing a new multi-agent Deep Reinforcement Learning (DRL) approach, adversaries craft stealthy, strategically timed, wear-out attacks designed to subtly degrade product quality and reduce the lifespan of field actuators. This sophisticated method leverages DRL methodology not only to execute precise and detrimental impacts on targeted infrastructure but also to evade detection by contemporary AI-driven defence systems. By developing and implementing tailored policies, the attackers ensure their hostile actions blend seamlessly with normal operational patterns, circumventing integrated security measures. Our research reveals the robustness of this attack strategy, shedding light on the potential for DRL models to be manipulated for adversarial purposes. Our research has been validated through testing and analysis in an industry-level setup. For reproducibility and further study, all related materials, including datasets and documentation, are publicly accessible.

\end{abstract}

\begin{IEEEkeywords}

Stealthy Attack, AI-assisted Adversary, Industrial Control Systems, Deep Reinforcement Learning
\end{IEEEkeywords}

\section{Introduction} 

Industrial Control Systems (ICS) have become indispensable for managing critical infrastructures across various sectors, from chemical manufacturing and biotechnology to water and wastewater treatment facilities \cite{gong2024possibilities, degrave2022magnetic}. Unlike traditional Information Technology (IT) systems designed primarily for data management, ICSs control physical processes and are vital for Operational Technology (OT). Integrating Artificial Intelligence (AI) into these systems has been transformative, offering enhanced efficiency and predictive capabilities.

In early 2023, cyberattacks on critical infrastructure surged, with 60 incidents reported—more than double the previous year's total \cite{MOREHOUSE}.
Recent cyber incidents \cite{WildPressure, mukherjee2023evading} have further highlighted the limitations of traditional security measures, prompting the rise of Machine Learning (ML)-based defences \cite{umer2022machine}. These systems effectively process ICS, enhancing security capabilities. The trend towards more advanced system monitoring and efficient event collection supports the transformation from static, artefact-based security measures (e.g., statistical intrusion detection and prevention systems) to dynamic, ML-enhanced defences.
In this regard, researchers have developed security measures using Deep Neural Network (DNN)-based and Deep Reinforcement Learning (DRL) technologies \cite{umer2022machine} to secure ICS against various threats. Although integrating AI into ICS enhances operational efficiency, process optimisation, system monitoring, and automated decision-making, this evolution also introduces new vulnerabilities.
In this context, recently, there has been a growing research interest in adversarial attacks against DNN and DRL models, degrading the prediction accuracy \cite{9536399, sun2020stealthy} of these models. Some methods \cite{kos2017delving} involve injecting perturbations at intervals, guided by the value function. \cite{lin2017tactics} introduced two attack mechanisms. In the \emph{Strategically-Timed Attack}, the attacker injects adversarial perturbations into the victim's state when the relative action preference function exceeds a pre-calculated heuristic threshold. Meanwhile, the \emph{Enchanting Attack} involves using a prediction model to anticipate the agent's future states based on a sequence of actions, followed by crafting an action sequence to steer the victim towards a designated harmful state. \cite{sun2020stealthy} developed the \emph{Critical Point Attack} and the \emph{Antagonist Attack}. In the Critical Point Attack, the attacker uses a domain-specific prediction model to forecast the agent’s future states and assesses the potential damage of all possible strategies using a Damage Awareness Metric. The Antagonist Attack employs a DRL model that is optimised with the victim agent’s reward function, guiding the attacker in launching effective assaults with fewer steps.

Adversarial attacks on DNN and DRL models reveal vulnerabilities such as data poisoning, model hijacking, and adversarial sample creation \cite{9536399, sun2020stealthy}. However, there is a remarkable gap in research studying the impact of DRL-driven adversaries that evade AI-driven defence systems and adversely affect physical system processes. Employing a self-adaptive AI approach, the adversary may strategically introduce minor disturbances, learning from the system’s responses to gradually shift the `normal' operational baseline. This `low and slow' strategy allows them to covertly undermine both the integrity and availability of ICS, leading to significant yet covert compromises in system performance. 
Formally, we can define a control system by a state space \( S \), where each state \( s_t \in S \) describes the system at a given time \( t \). Adversarial actions \( a \in A \) selected by a DRL policy \( \pi \) aim to transition the system from state \( s_t \) to state \( s_{t+1} \) in a manner that avoids detection:
\[
s_{t+1} = T(s_t, a), \quad a \sim \pi(s_t)
\]
where \( T: S \times A \rightarrow S \) is the transition function of the system. The DRL agent optimizes \( \pi \) to maximise a cumulative reward function that, perversely, rewards states that undermine system integrity yet evade detection by existing Intrusion Detection Systems (IDS).
These IDS, typically trained to detect anomalies based on a defined normative behaviour profile, fail to recognise subtle deviations introduced by DRL-optimised actions. Mathematically, if \( d(s, \hat{s}) \) represents a distance metric measuring deviation from expected state \( \hat{s} \), then for IDS with detection threshold \( \tau \), an undetected state transition by adversarial action satisfies:
\[
d(s_{t+1}, \hat{s}_{t+1}) < \tau
\]
The ability of DRL-based adversaries to operate beneath this detection threshold underscores the urgent need for innovative research focused on advanced defensive strategies that can detect and mitigate the effects of these subtle but dangerous manipulations.

\subsection{Motivations and Contributions}
\label{sec:Motivations}
After thoroughly investigating the existing literature on AI-driven attacks, we find a significant gap that becomes apparent in the absence of studies on DRL-assisted adversaries targeting physical processes. This gap poses a considerable challenge in the realm of cybersecurity, as it obstructs the evaluation of defence mechanisms. In this context, there is a growing concern about the potential of AI-assisted attacks against OT (Operational Technology) networks \cite{Amri_2023}. The impact of such stealthy attacks on industrial operations, such as the precise mixing of liquids, can be profound \cite{makrakis2021industrial, 9581186, alsabbagh2023security}. These DRL-based adversaries can subtly alter control systems to evade detection by AI-driven Intrusion Detection Systems (IDS) (as shown in Section \ref{sec:Experiments Results}). They can evade security measures by mimicking normal operations in control systems and avoiding obvious actions. 
They can disrupt safety and efficiency without triggering alarms. This highlights the urgent need to explore and develop robust defences against such covert attacks. In this paper, we focus on studying a \emph{novel} attack strategy and its negative impacts on the target system.

Our research primarily adopts the black-box threat model, limiting adversaries' insight into the detection mechanisms. However, to ensure a comprehensive assessment, we also evaluate the efficacy of our approach under grey-box and white-box models. We validate our findings with a real industrial system testbed, including a Siemens SIMATIC S7-1200 PLC, Siemens SIMATIC IPC127E, Cisco C9300-24U-A switch, Cisco ASA 5555-X Firewall/IPS, Cisco C8200-1N-4T Edge router, and Nvidia Jetson AGX Orin (further details provided in Section \ref{sec:Design and implementation}). To ensure high reliability and real-time performance, we employ the PROFINET protocol \cite{8627173}, reflecting industry preference for flexibility, speed, and efficiency in data exchange and communication between field devices and controllers. A series of experiments are conducted to assess the negative impact of such adversaries and the effectiveness of different DNN-based countermeasures. 
When tested against DNN-based IDS models (DenseNet \cite{huang2017densely}, CNN \cite{krizhevsky2012imagenet}, ResNet \cite{he2016deep}, LSTM \cite{hochreiter1997long}, and Transformer \cite{vaswani2017attention}) on a mixing liquid control system test-bed, DRL-assisted adversaries were able to create and execute stealthy attack variants, reducing intrusion detection sensitivity (a.k.a recall) by up to approximately 99.95\%. Specifically, in black-box settings, recall reductions ranged from 17.34\% to 27.61\%, in Grey-box settings from 28.96\% to 69.24\%, and in White-box settings up to 99.95\% across different models, demonstrating significant vulnerabilities under varying levels of adversary knowledge (depicted in Table \ref{tab:benchmarking}). 
This study targets explicitly DNN-based detection systems, with other security systems, such as side-channel \cite{9347104} or invariant-based \cite{adepu2020control, feng2019systematic, yang2020plc, 9973658} approaches, being outside the scope of this article.

Our work primarily demonstrates a stealthy attack strategy while exploring how the same adversarial policies can serve as a \textit{training curriculum} to strengthen defence systems. By repurposing the attack model into a Min-Max framework, these adversarial strategies enhance the robustness of DNN-based detectors. Details on \emph{adversarial training} are provided in \textbf{Appendix~\ref{sec:appendix-adv-training}}.

\textbf{Our Contributions:} This article introduces a novel algorithm that has been designed to perform strategically timed attacks by low \& slow adversarial strategy to disrupt control systems while concurrently avoiding detection by AI-driven defence systems. Our contributions are outlined as follows:

    \ding{111}  Designed a new adversarial algorithm for finding a strategically timed, low \& slow attack that can disrupt control systems while avoiding detection by AI-driven defence systems. 

    \ding{111}  Rigorously assessed the impact and effectiveness of our proposed adversarial algorithm against state-of-the-art AI-driven countermeasures and the Cisco ASA 5555-X Firewall/Intrusion Prevention System.

     \ding{111} Developed an Industrial Control System (ICS) testbed equipped with a Siemens SIMATIC S7-1200 PLC, Siemens SIMATIC IPC127E, Cisco C9300-24U-A switch, Cisco ASA 5555-X Firewall/IPS, Cisco C8200-1N-4T Edge router, and Nvidia Jetson AGX Orin etc. This testbed is meticulously engineered to mirror real-world operational environments\footnote[2]{\textbf{All related materials, including datasets and codes, are available for the research community here: \href{https://tinyurl.com/ycncu57k}{https://tinyurl.com/ycncu57k}}}.
     
 \paragraph*{Positioning and Novelty.
Prior DRL-enabled attack studies \cite{shereen2023reinforcement, maiti2023targeted, mohamed2023use, chandratre2023stealthy} largely operate in simulation and target cyber variables rather than \emph{field-network timing} in real ICS plants. \emph{To the best of our knowledge}, this work is the first to (i) validate a \emph{multi-agent} (Scheduler+Disturber) DRL adversary on an industry-level ICS testbed with Siemens S7 PLCs and PROFINET, (ii) optimise \emph{timing-driven low\&slow wear-out} to degrade actuator lifetime and product quality while \emph{evading} DNN-based IDS, and (iii) \emph{reuse} the learned adversary in a \emph{min--max} loop to harden detectors.}

\section{Background and Related Works}
\label{sec:Background and Related Work}

In this section, we discuss the fundamental aspects of Industrial Control Systems (ICS) and their vulnerability to emerging cyber threats. We provide an overview of AI-based attacks, with a particular emphasis on DRL, and review the existing literature on vulnerabilities and attacks in ICS and AI. To the best of our knowledge, there is no literature on AI-assisted attacks against ICS. While AI-assisted attacks on Cyber-Physical Systems (CPS) and Power Grids have been studied \cite{khazraei2022learning, chandratre2023stealthy, jiang2024vulnerability, mohamed2023use, shereen2023reinforcement, maiti2023targeted}, many of these studies only concentrate on attacks targeting electronic grids, including False Data Injection, and Load Frequency Control. However, ICS presents unique challenges due to its critical role in industrial processes, real-time requirements, and the severe consequences of breaches. Unlike CPS, where attacks often target cyber components, ICS attacks must simultaneously evade detection and directly influence physical processes. This paper addresses this critical gap by focusing on stealthy AI-assisted attacks targeting ICS vulnerabilities.

\begin{figure}[t!] 
\center{\includegraphics[width=60mm]{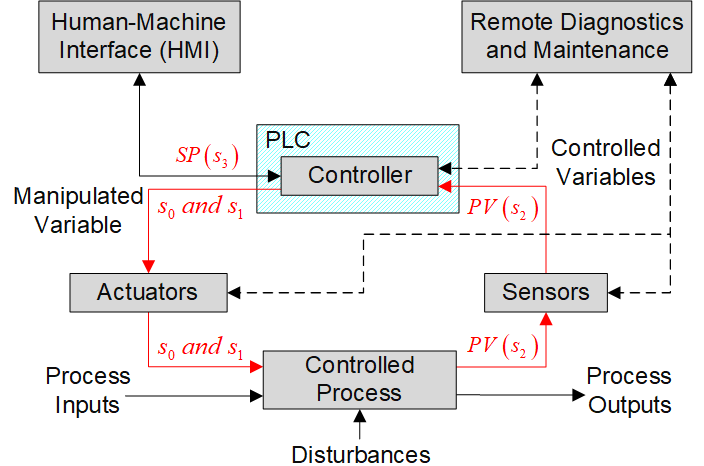}} 

\caption{Overview of ICS Operation (NIST SP 800-82)}
\label{fig:Overview of ICS Operation} 
\vspace{-4mm} 

\end{figure}

\textbf{Security of Industrial Control System:} 
At the core of any process control system is the controller, which compares input values against a set condition or reference and decides on the necessary actions. It then dispatches corrective error signals to adjust the manipulated variable, which represents dynamic aspects such as temperature, pressure, flow rate, level, rotational speed, or position. 
Moreover, in control systems, different components operate on varying timing cycles—for example, field devices often have sub-1 millisecond response times, while controllers typically update between 1 millisecond and 50 milliseconds \cite{yoo2019control}.
Any physical parameter that can change spontaneously or in response to external influences is considered a dynamic variable. This variable, termed the Process Variable (PV), is intended to be maintained at a predetermined value, known as the Set Point (SP). Whether fixed or adjustable, the SP dictates the target state for the PV, requiring continuous monitoring and corrective actions from the control system to ensure alignment with the desired SP. Process control plays a crucial role in ensuring control systems' safe and efficient operation, as depicted in Fig. \ref{fig:Overview of ICS Operation}. Human-machine interfaces (HMI), remote diagnostics, and maintenance processes support the primary control mechanisms.  Interrupting these systems can lead to disastrous consequences, leading to financial losses or even costing human lives, making vigilant monitoring and corrective actions crucial. 

Historically, the security of ICS has been primarily safeguarded by their isolation from external networks; the concept is often referred to as being `air-gapped' \cite{ylmaz2018cyber}. However, with the advent of Industry 4.0, ICSs have become increasingly interconnected, thereby expanding their attack surface \cite{robles2019plc}. As a result, traditional statistical-based security systems are no longer sufficient for protecting ICSs against sophisticated attacks \cite{robles2019plc}. In response, researchers \cite{umer2022machine} have proposed AI-driven intrusion detection systems designed to safeguard critical infrastructures from such threats. However, these AI models themselves are vulnerable to various types of attacks, indicating a need for continuous advancement in security technologies.
Recent research \cite{wardak2016plc, abbasi2016ghost, makrakis2021industrial, 9581186, alsabbagh2023security} has identified critical vulnerabilities in network communications between engineering stations and programmable logic controllers (PLCs). In this regard, \cite{beresford2011exploiting} revealed how modifying ISO-TSAP packets could manipulate industrial system behaviours, while \cite{lim2017attack} showed potential catastrophic outcomes from altered command packets in nuclear facilities. Furthermore, \cite{ylmaz2018cyber} outlined scenarios where Denial of Service (DoS) attacks involved spoofing IP addresses to compromise the communication between S7 PLCs and Total Integrated Automation (TIA) Portal software. Similarly, \cite{sayegh2013internal} executed IP packet flooding on PLC ports, targeting the interconnectivity between PLCs and Human Machine Interfaces (HMIs) with various techniques. 
Furthermore, \cite{niedermaier2018you} indicated that attackers could employ a variety of packet types, such as SYN, S7Comm, UDP, and TCP/IP packets, to exploit PLC vulnerabilities effectively.

\textbf{AI-driven Attacks against Power Grid and CPS:}
Recent research in cyber-physical system security has examined AI for offensive and defensive strategies in power grids. Mohamed and Kundur  \cite{mohamed2023use} use the Deep Deterministic Policy Gradient (DDPG) algorithm for generating False Data Injection (FDI) and load-switching attacks in Load Frequency Control (LFC) systems. Their grey-box model allows the RL agent to destabilise the system by observing and manipulating key variables.  Similarly, Maiti et al. \cite{maiti2023targeted} employ DDPG  to generate attacks that maximise disruption in smart grids by exploiting the most vulnerable time instances. Shereen et al. \cite{shereen2023reinforcement} employ the Proximal Policy Optimisation (PPO) algorithm to generate stealthy attacks against Automatic Generation Control (AGC) systems, assuming the adversary can infiltrate and manipulate control signals within a network using protocols lacking encryption, such as Modbus and DNP3. 

On the other hand, Khazraei et al. \cite{khazraei2022learning} use deep learning for vulnerability analysis in CPS, creating grey-box attack generators with feed-forward and recurrent neural networks to design stealthy sensor attacks that degrade system performance without detection. They conduct their experiments in simulation environments. 
Aniruddh Chandratre et al. \cite{chandratre2023stealthy} introduces a framework for simulating falsification attacks on CPS, namely inverted pendulum and unmanned aerial vehicles, that push the system into unsafe or undesired states by violating the system's Signal Temporal Logic specifications. The framework uses PSY-TaLiRo and Bayesian optimisation to identify effective attack vectors \footnote[2]{\textbf{Due to page limit, further related work is discussed in Appendix G.
}}. 
   All of these studies \cite{shereen2023reinforcement, maiti2023targeted, mohamed2023use, chandratre2023stealthy} conduct their experiments in simulation environments (e.g. \cite{shereen2023reinforcement} using MATPOWER, \cite{mohamed2023use, maiti2023targeted} utilising MATLAB and Simulink), which, while useful for initial exploration, may not fully capture the complexities of real-world operations. Moreover, none of the existing literature addresses the complexities and dynamics of real-world industrial control processes within ICS environments. Our work distinguishes itself by targeting the field network process (with a focus on controlling industrial processes under strict real-time and safety requirements), using an innovative multi-agent DRL approach to conduct stealthy, strategically timed disruptions. This approach undermines existing AI-driven defences and emphasises the complexities of low \& slow attacks, thereby extending the scope of RL-based adversarial models beyond the power sector and into broader industrial applications.

\textbf{Attacks against AI:}
The vulnerability of Deep Neural Networks (DNNs) to adversarial attacks has been a prominent area of research in recent years \cite{ozdag2018adversarial}. It has been established that DNNs can be deceived into making incorrect predictions with high confidence through the introduction of small, human-imperceptible perturbations \cite{goodfellow2014explaining}. Adversarial attacks are generally classified into two types: white-box attacks, where the attacker has complete knowledge of the DNN model, and black-box attacks, where the attacker does not know the internal workings of the model.
In the white-box context, numerous attack methods have been developed. These include the Limited-memory Broyden-Fletcher-Goldfarb-Shanno (L-BFGS) algorithm \cite{szegedy2013intriguing}, Fast Gradient Sign Method (FGSM) \cite{goodfellow2014explaining}, Basic Iterative Method (BIM) \cite{kurakin2018adversarial}, Projected Gradient Descent (PGD) \cite{madry2017towards}, Distributionally Adversarial Attack (DAA) \cite{zheng2019distributionally}, Carlini and Wagner attacks \cite{carlini2017towards}, and Jacobian-based Saliency Map Attack (JSMA) \cite{papernot2016limitations}, among others. These methods can also be modified for use in black-box scenarios due to the transferability of adversarial examples \cite{papernot2016transferability}.
Specifically designed for black-box environments are techniques such as the Zeroth Order Optimisation (ZOO) Attack \cite{chen2017zoo}, AutoZOOM Attack \cite{tu2019autozoom}, Decision Based Attack \cite{brendel2017decision}, Universal Adversarial Perturbations (UAPs) \cite{co2019procedural}, Derivative-Free Attack \cite{bu2021taking}, and Simultaneous Perturbation Stochastic Approximation (SAI-FGSM) \cite{liu2022efficient}. These methods demonstrate the adaptability of adversarial techniques across different knowledge bases of the target models.

\textbf{Deep Reinforcement Learning:} 
Reinforcement Learning (RL) is concerned with decision-making in evolving environments. RL agents engage in a learning process through interaction with their environment, taking actions, and receiving feedback in the form of numerical rewards.
Formally, at each time step \(t\), an agent observes a state \(s_t\) from the state space \(S\), takes an action \(a_t\) from the action space \(A\), and receives a reward \(r_t\). 
The primary objective of an RL agent is to develop a policy \(\pi\), which is a function \(\pi: S \rightarrow A\) that maps states to actions, aiming to maximise the expected cumulative rewards over time.
Deep Reinforcement Learning (DRL) extends RL's scope by leveraging Deep Neural Networks (DNNs) to refine the policy construction process, either through Q-function or Value-function approximations or direct policy learning from experiences.
At the core of a DRL system is the policy mechanism, which directs agents to perform optimal actions based on the environmental state, significantly enhanced by advanced algorithms. These innovative methodologies have been instrumental in generating DRL policies, enabling them to demonstrate remarkable achievements across various tasks requiring AI. Unlike traditional machine learning tasks that are categorised into supervised, unsupervised, or semi-supervised learning, involving data represented as a tuple \( \langle X, Y \rangle \), where \( X \) is the inputs (independent features) and \( Y \) is the label (a.k.a ground truth), DRL tasks feature a more complex structure. They comprise at least four elements: \( \langle S, A, \pi, R \rangle \)—state set, action set, policy network, and reward set, respectively.  Although this formulation enables DRL to learn from environmental interactions, offering broader applications, it also introduces greater vulnerability to attacks, as evidenced by several studies \cite{9536399, sun2020stealthy}.

\textbf{Attacks against Deep Reinforcement Learning:} Adversarial attacks on DRL agents differ significantly from those on classification systems, primarily because DRL agents interact with their environment by executing a sequence of actions, thus affecting and receiving feedback from the environment \cite{9536399}. This dynamic allows adversaries multiple points of attack during an episode, with each step providing opportunities to introduce perturbations. The objectives of these adversaries also vary, from reducing the agent’s overall rewards to navigating them into dangerous states, contrasting with attacks on classification systems that aim to degrade accuracy. Despite the importance of both short-term and long-term attack strategies, their integration has rarely been explored. Short-term attacks produce immediate effects by exploiting specific actions, while long-term strategies aim for a sustained impact, targeting the foundational behaviour of DRL agents \cite{sun2020stealthy}. The challenge of integrating short and long-term strategies lies in identifying critical points (times) that can significantly affect future actions, given their indirect connections. 

\begin{figure}[htbp] 
\center{\includegraphics[width=85mm]{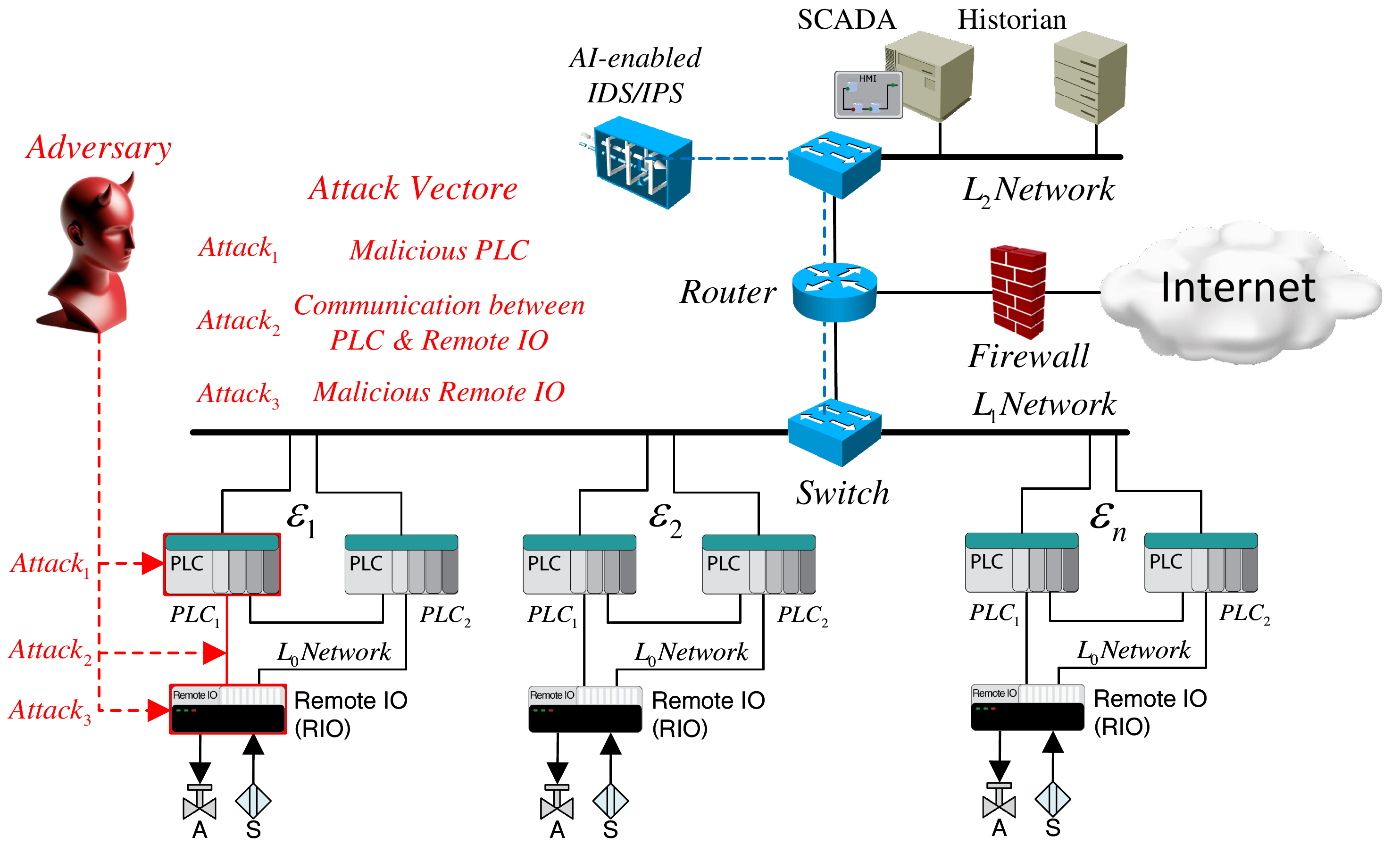}} 

\caption{AI-enabled Adversary in a Field Network (adversary's attack vectors, includes malicious PLC, compromised Remote I/O, or the interception and manipulation of communications between PLC and Remote I/O at $L_{0}$ network).}
\label{fig:Adversary in Field Network} 
\end{figure}

\subsection{Scope, Generalizability, and Limitations}
\label{sec:scope}
\textbf{Where low\&slow transfers.} The timing-driven disturbance transfers most naturally to \emph{batch/process} ICS (blending, dosing, filtration), where buffering and stochastic variability can mask small cumulative delays.

\textbf{Harder domains.} In tightly time-constrained motion and safety loops, even millisecond delays may violate invariants or be conspicuous.

\textbf{Enablers for transfer.} We emphasise \emph{process-agnostic timing features} (order statistics of inter-arrival and actuation latencies), adopt \emph{domain randomisation} over plant gains, sensor noise, sampling cadence, and feature sets during training, and calibrate actions to PLC \emph{scan-cycle budgets}. 

\textbf{Quantifying robustness.} Our Out-of-Context evaluation (Section~\ref{sec:Evaluation Framework}) alters dynamics, noise, cadence, features, and detector windowing to induce drift and measure transferability.

\section{Attack Methodology}
\label{sec: Attack Method}

\subsection{Threat Model}
\label{sec: Threat Model}

In our threat model, we introduce two distinct Advantage Actor-Critic (A2C) agents, the scheduler and the disturber, which exploit DRL techniques to conduct wear-out attacks against industrial control processes, depicted in Fig. \ref{fig: Adversary Agents}. The scheduler agent is responsible for selecting optimal attack times (waiting time $t^{wait}$) and delay periods ($t^{delay}$), using real-time observations from actuators, sensors, and network communications to inform its decisions (actions). Concurrently, the disturber agent waits for $t^{wait}$ and selects which packets to delay for the periods specified by the scheduler ($t^{delay}$), effectively integrating the scheduler's actions into its decision-making process to synchronise the attack. 
We initially considered a single-agent approach (discussed in the Appendix); however, it proved challenging to balance both strategic timing and execution control within a unified policy. The complexity of jointly optimising when to launch an attack and how to execute it without detection created learning inefficiencies and reduced adaptability. By \emph{decoupling these roles}, the scheduler agent can focus exclusively on identifying optimal attack windows, while the disturber agent handles the fine-grained execution of stealthy packet delays. This modular design improves convergence during training and enhances the adversary’s ability to adapt to evolving defensive mechanisms in real-time.

These adversaries' self-adaptive capabilities enable them to dynamically adjust strategies in response to environmental changes and defensive measures, making them stealthy and robust.

To ensure the feasibility of the proposed attack, we position the adversary's actions within the Cyber Kill Chain framework, which provides a structured approach to understanding the lifecycle of cyberattacks. Specifically, the adversaries operate within the \textbf{Exploitation} and \textbf{Command \& Control (C2)} phases. During the Exploitation phase, vulnerabilities such as {\textbf{CVE-2024-2442} \cite{nistCVE20242442}, which expose ICS communication protocols, and {\textbf{CVE-2024-2882} \cite{nistCVE20242882}, which facilitates remote code execution, are leveraged to infiltrate the system and trigger stealthy wear-out attacks. In the C2 phase, the adversary establishes persistent communication channels using exploited vulnerabilities like {\textbf{CVE-2023-5885}\cite{nistCVE20235885}, enabling remote coordination and precise timing of synchronised attacks by the scheduler and disturber agents. These vulnerabilities allow the adversaries to reconnoitre the environment, establish control, and execute their objectives without detection, demonstrating the feasibility and practicality of the attack model.

\paragraph*{Vulnerability-Independent Ingress.}
The listed CVEs illustrate feasibility but are not prerequisites. In practice, an adversary can attain the same $L_{0}$ vantage via:
\begin{itemize}
  \item \textbf{Bump-in-the-wire} tap between PLC and I/O that transparently forwards PROFINET/Modbus/OPC~UA while imposing bounded, protocol-legal queueing delays (no payload edits).
  \item \textbf{Network QoS/mirroring misconfiguration} (e.g., unmanaged switch in the cell/area), enabling selective packet pacing without rewriting frames.
  \item \textbf{Remote I/O or HMI workstation foothold} (non-root), where user-space timing control of application-layer calls produces equivalent micro-delays.
  \item \textbf{Engineering-station pivot} with read-only access to project tags; timing perturbations on cyclic reads/writes remain within scan-cycle budgets.
\end{itemize}
These avenues preserve the protocol’s statefulness and align with bounded-delay constraints we enforce in all experiments.

\textbf{Adversary's Objectives:} 
The adversary has three primary objectives as given in the following:

\textbf{($\mathcal{O}_{1}$) Wear-out system components - }
The adversary manipulates the operational sequence of packet delivery in industrial control systems through calculated delays. This action is formally represented by modifying the temporal distribution of control packet transmissions, where each packet $i$, initially scheduled for $t_i$, is delayed by $\delta_i$, rescheduling its transmission to $t_i + \delta_i$. These intentional disruptions induce asynchronous and inefficient actuator operations, accelerating wear-out due to increased mechanical stress and suboptimal performance cycles. Such continuous operational anomalies significantly shorten the expected lifespan of critical system components. 

\textbf{($\mathcal{O}_{2}$) Inducing operational inaccuracies resulting in a reduction in the quality of the products -} the adversary strategically introduces deviations \(\delta\) to critical control system set points, which shifts operational parameters like temperature or pressure away from their optimal states. This action mathematically increases the process variance \(\sigma^2\), leading to increased production errors and reduced quality, quantified by the relationship \(\Delta Q \propto \sigma^2\), where \(\Delta Q\) denotes the decline in product quality. For instance, in the pharmaceutical industry, such perturbations to temperature and mixing speed set points may result in non-compliant drug potency, causing significant financial losses and endangering patient health.
   
\textbf{($\mathcal{O}_{3}$) Covert operations and evasion of detection -} the adversary employs a `low and slow' strategy, subtly introducing deviations (\(\delta\)) that accumulate over time to alter operational processes without triggering alarms from AI-powered defence systems. This method involves careful calibration of changes to stay below detection thresholds, mathematically optimising the perturbations to maintain a low profile (\(\sigma^2 < \lambda\), where \(\lambda\) is the detection limit). In industries like pharmaceuticals, where precision in the mixing of liquids is crucial, such undetected alterations can lead to significant deviations in product consistency and quality, thereby posing serious risks to both operational reliability and consumer safety.

\textbf{Adversary's Capabilities:} Understanding the adversary's level of knowledge is crucial. We delineate this within two main categories:

\textit{\textbf{Black-Box Attacks:}} The adversary has limited insight, operating without detailed internal knowledge of the defence system's configurations or logic.

\textit{\textbf{Grey-Box Attacks:}} The adversary possesses partial information about the defence system, which may include public features or certain system details that are not fully concealed (more details provided in Appendix).

\textit{\textbf{White-Box Attacks:}} The adversary possesses complete knowledge about the defence system, including features that DNN-based detectors use to detect intrusions.

Our research primarily adopts the black-box threat model, which restricts adversaries' insight into the detection mechanisms. In this context, the adversary does not know the detection system's architecture, parameters, observations, or statistical features, including time-based and historical ones. To ensure a comprehensive assessment, we also evaluate the efficacy of our approach under different adversarial assumptions, including both grey-box and white-box models.

However, we consider the adversary can influence PLC, Remote I/O, or the communication channel between them, compromising known vulnerabilities in these components \cite{wardak2016plc, abbasi2016ghost, makrakis2021industrial, 9581186, alsabbagh2023security}, as depicted in Fig. \ref{fig:Adversary in Field Network}.  Such adversaries are capable of:

    \ding{111} Interfering with PLC inputs, including setpoints ($s_{3}$) and level meters ($s_{2}$), as well as outputs such as the values for filling valves ($s_{0}$) and discharge valves ($s_{1}$).
    
    \ding{111} Delaying signals $s_{0}, s_{1},$ and $s_{2}$ to disrupt process timings and cause operational inconsistencies.
    
    \ding{111} Applying their partial knowledge of the detection agent's model to predict the system's responses and effectively construct their attack strategies.

\begin{figure}[htbp] 
\center{\includegraphics[width=85mm]{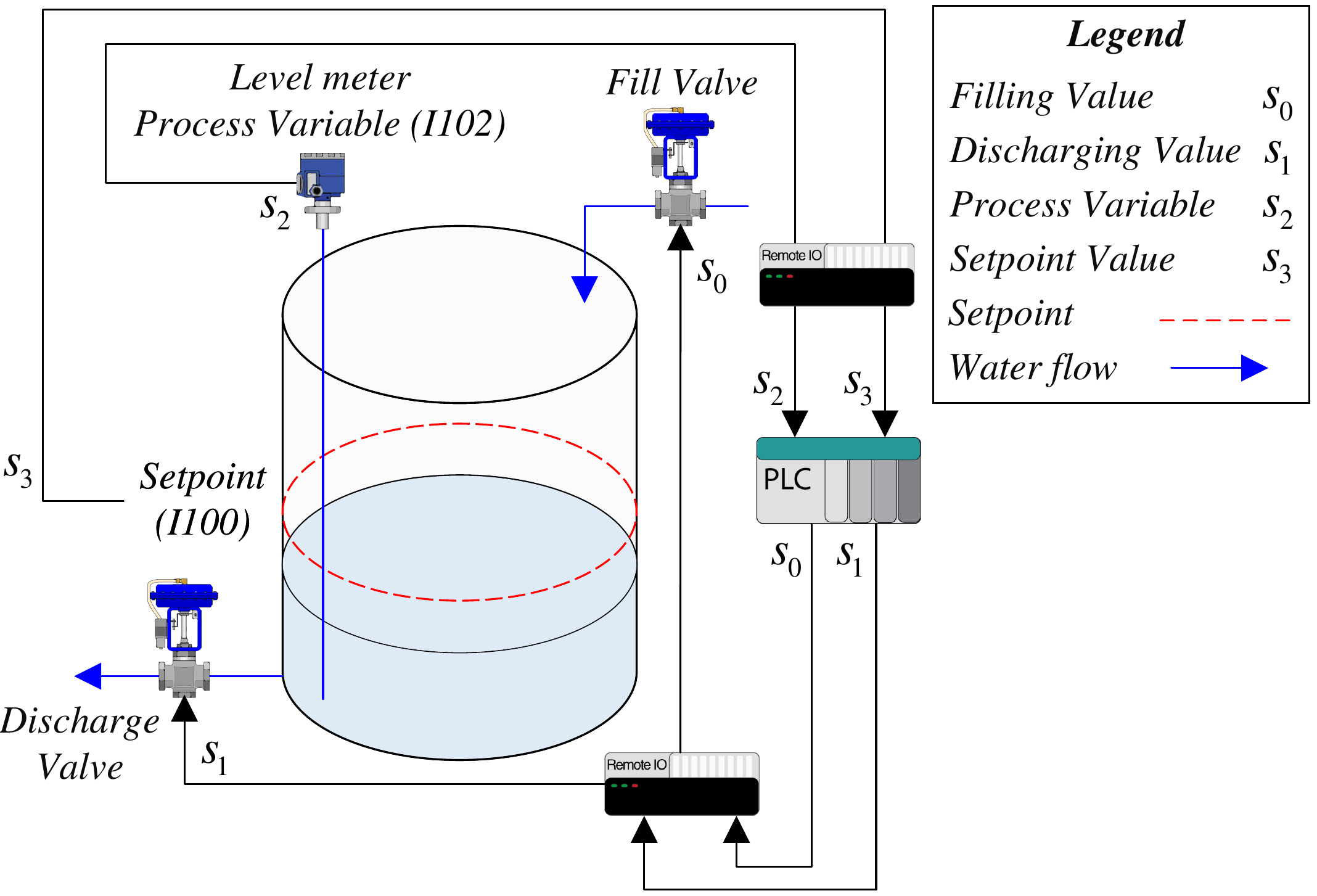}} 

\caption{Adversary Observations}
\label{fig:Adversary Observation} 
\vspace{-3mm} 
\end{figure}

\textbf{Adversary's Observation:} 
Fig. \ref{fig:Adversary Observation} demonstrates the adversary's placement within a field network, which enables them to intercept and analyse data. 
A Programmable Logic Controller (PLC) continuously reads inputs and responds by adjusting outputs in real-time to prevent malfunctions. This process involves scanning the inputs from an I/O Image Table and using programmed logic to determine the output states. Given the importance of a timely response in PLC operations, any attack that disrupts or delays this real-time capability can significantly compromise industrial control system security. In the S7 family of Siemens PLCs, such as the S7-1200, the CPU offers a variety of dedicated memory locations, each with a unique address for program access. These areas include process image inputs ($I^{PLC}$), where the state of physical inputs is copied at the start of each scan cycle, and process image outputs ($Q^{PLC}$), where outputs are derived from at the end of each scan cycle. In this context over communication over PLC with remote I/O, actuator, and sensors, the adversary can observe $I^{PLC}$ value ($s_{2}$) and $Q^{PLC}$ values ($s_{0}$ and $s_{1}$), as illustrated in Fig. \ref{fig:Overview of ICS Operation} and Fig. \ref{fig:Adversary Observation}. In this system, level meters serve as sensors to monitor the tank levels, providing inputs to a PLC. The PLC, in turn, controls actuators such as fill and discharge valves to manage outputs. It operates these valves to add chemicals to a mixing tank until the sensors indicate that the tank level has reached the setpoint (We formalise the adversary's observation space $O$ in Section \ref{sec: Formalisation of Attack}).

\begin{notebox}{Action Space Clarification:}  Here we would like to emphasise that the terms ``Low and Slow'' and ``Smash and Grab'' used in this paper represent broad categories of adversarial strategies rather than discrete or simplistic actions. Within each category, adversarial agents execute a wide spectrum of nuanced and context-driven actions, dynamically determined by real-time observations and sophisticated strategic policies. Thus, these categories encapsulate sets of actions that vary in timing, magnitude, and stealth. \end{notebox}

\begin{figure*}[t!]
\begin{subfigure}{.5\textwidth}
\centerline{\includegraphics[width=7cm]{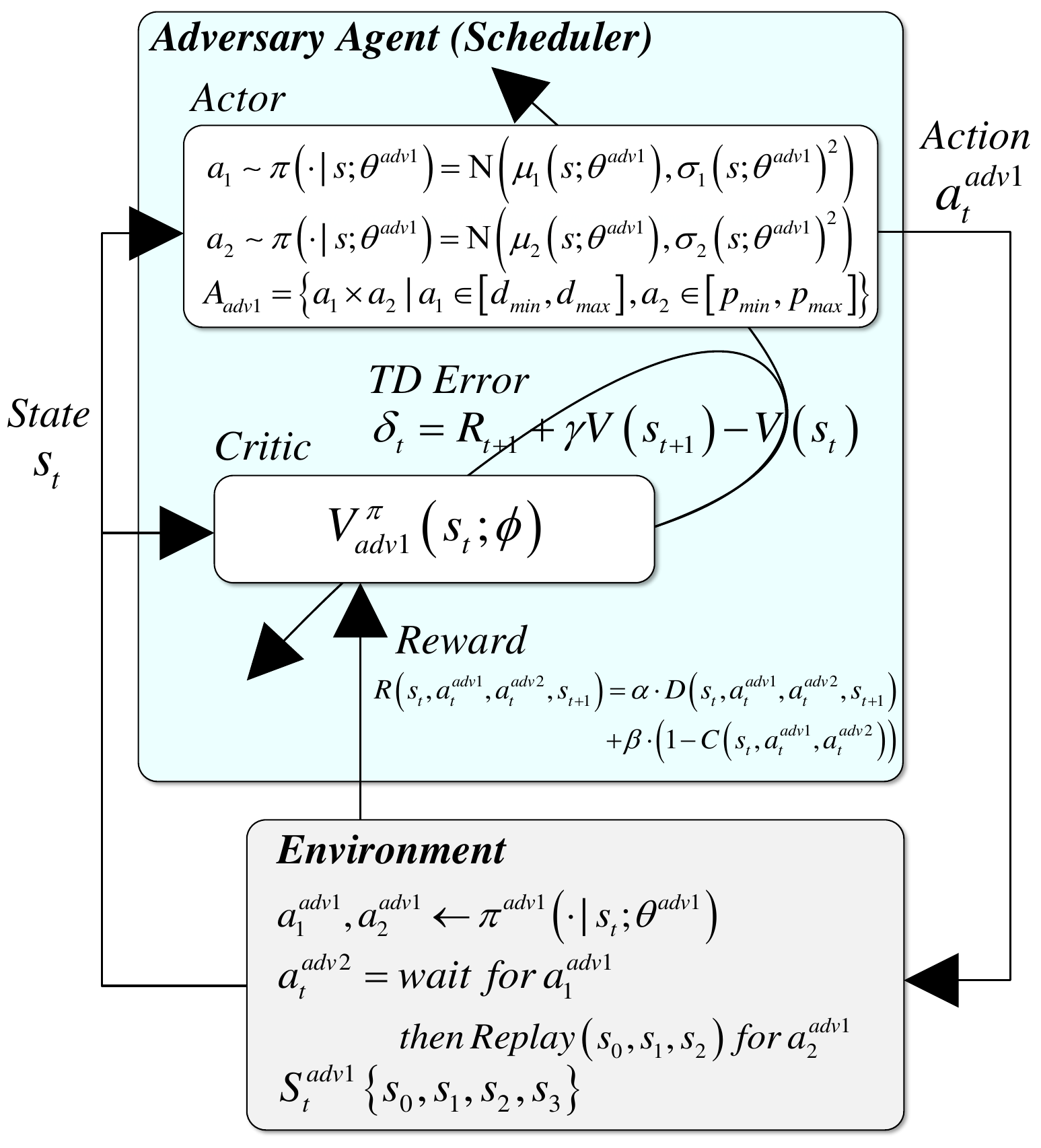}}
\caption{Adversary (Scheduler)}
\label{fig: Adversary (Scheduler)}
\end{subfigure}%
\begin{subfigure}{.5\textwidth}
\centerline{\includegraphics[width=7cm]{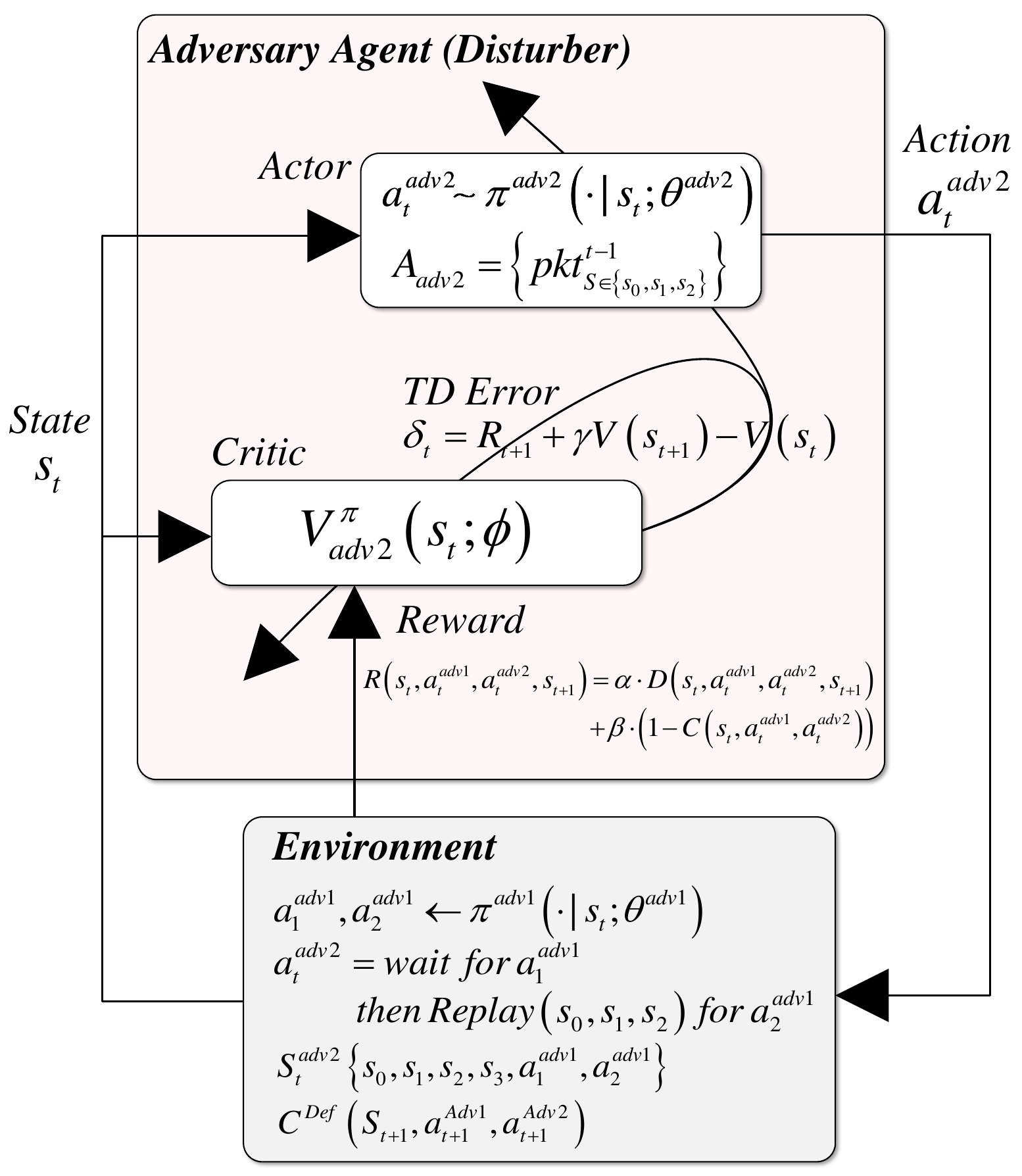}}
\caption{Adversary (Disturber)}
\label{fig: Adversary (Disturber)}
\end{subfigure}
\caption{Adversary Agents}
\label{fig: Adversary Agents}
\vspace{-3mm} 
\end{figure*}

\subsection{Formalisation of Attack}
\label{sec: Formalisation of Attack}
In this section, we consider the attack strategy using a partially Observed Markov decision process (POMDP). This model is particularly fitted to scenarios where an adversary may not have full visibility of the system states but must operate based on partial observations that can be deceptive or incomplete. In this context, the adversary employs the disturber and scheduler agents (as depicted in Fig. \ref{fig: Adversary Agents}) to identify and modify conspicuous events. This approach enables them to evade detection while still preserving their objectives.
The POMDP is defined by the tuple $(S, A, T, R, X, O)$, where:

    \ding{111}  $S$:
    The state $s_t \in S$ at time $t$  represents the vector of all relevant system parameters—including the filling variable, discharge variable, process variable, and setpoint value, as depicted in Fig. \ref{fig:Adversary Observation} (further detailed in Appendix Fig. 7).

    \ding{111}  $A$: Actions available to the adversary, primarily focusing on the delay of specific control packets. Each action $a_t \in A$ is represented as $
    a_t = \text{delay}(pkt_i, period_{t})$ where $pkt_i$ is the packet to be delayed, and $period_{t}$ is the delay duration.

    \ding{111}  $T: S \times A \times S \rightarrow [0, 1]$ defines transition probabilities, i.e., likelihood of moving between states given specific actions.
    
    \ding{111}  $R: S \times A \times S \rightarrow \mathbb{R}$ is the reward function, which in the context of adversarial strategies, could reflect the efficacy or impact of an attack (more details provided in Section \ref{sec:Crafting an Adversarial Strategy}).
    
    \ding{111}  $X$: denotes the set of observations available to the adversary.
    
    \ding{111}  $O: S \times X \rightarrow [0, 1]$ is the observation probability function, i.e., the likelihood of the adversary observing $x_t$ in state $s_t$.

We formalise the observation $O$ at time $t$ as $o_t = f(s_t, \omega(a_{t-\Delta:t-1}))$, where $f(s_t, \cdot)$ is a function that models the immediate impact of the current system state $s_t$ on the observation $o_t$. This function captures how the adversary perceives the current state of the system. $\omega(a_{t-\Delta:t-1})$ is a function that aggregates the influence of the adversary's actions over a recent time window from $t-\Delta$ to $t-1$. This function summarises how these past actions affect the current observation, considering the cumulative or delayed effects of these actions on the system.
In the black-box and grey-box threat models, the adversary's policy is primarily informed by observations rather than full states, reflecting the limited visibility into the system.  Therefore, the adversary's decision-making process relies on the observed values $o_{t}$, which are derived from the underlying state $s_{t}$ but do not represent full system visibility. Therefore, $s_t$ should be interpreted as $o_t$ in partially observable environments (black-box and grey-box threat models).

Each step within this environment involves the sequence $(s_t, x_t, a_t, s_{t+1})$, where $s_t$ is the current state, $x_t$ is the observation related to this state, $a_t$ is the action taken by the adversary, and $s_{t+1}$ is the subsequent state. The adversary aims to exploit this sequence to strategically time their actions based on observed states and anticipated system responses, maximising the disruption to operational integrity while minimising detection.
The objective of the adversary in this POMDP framework is to develop a policy $\pi$, which maps observed states to actions that maximise the expected cumulative adverse impact on the system. This can be expressed as:
\vspace{-3mm} 

\begin{equation}
\pi^* = \arg\max_{\pi} \mathbb{E}\left[\sum_{t=0}^{\infty} \gamma^t R(s_t, a_t, s_{t+1}) \mid \pi, O\right],
\label{eq: Adversary optimal policy}
\end{equation}
\vspace{-3mm} 

where $\gamma$ is the discount factor (0 $\leq \gamma < 1$), representing the diminished significance of future adverse impacts relative to immediate ones. The chosen policy $\pi^*$ is the strategy that optimally balances the effectiveness of the attack with the adversary's ability to remain undetected.
This POMDP-based approach allows for sophisticated modelling of adversarial behaviour in environments where complete state information is not available, mirroring the real-world complexities of cybersecurity in industrial settings.

\subsection{Crafting an Adversarial Strategy}
\label{sec:Crafting an Adversarial Strategy}

To develop an effective adversarial strategy using the Advantage Actor-Critic (A2C) approach, the adversary aims to maximise operational disruption while evading detection by DNN-based anomaly detectors. The A2C algorithm, which combines the benefits of policy gradient and value-based methods, is well-suited for environments where the adversary must make decisions based on incomplete or deceptive observations.
The reward function \(R\) for the adversary is carefully designed to evaluate the quantification of actions against two criteria: their contribution to increasing actuator operation and their stealthiness. The mathematical formulation of the reward function \( R \) represented as:
\vspace{-3mm} 

\begin{multline}
\label{eq: adversary reward}
    R(s_t, a_{t1}, a_{t2}, s_{t+1}) = \alpha \cdot D(s_t, a_{t1}, a_{t2}, s_{t+1}) \\
    + \beta \cdot (1 - C(s_t, a_{t1}, a_{t2})),
\end{multline}

\ding{111} \( a_{t1} \) and \( a_{t2} \) are the actions taken by the scheduler and disturber, respectively.

    \ding{111} \(D(s_t, a_{t1}, a_{t2}, s_{t+1})\) measures the operational disturbance imposed on the system, induced by the adversary's actions \( a_{t1} \) and \( a_{t2} \) from state \( s_t \) to \( s_{t+1} \). This disturbance leads to inaccuracies in the control process by causing the system to deviate from its optimal operational parameters, contributing to the degradation of control precision.

    \ding{111} \(C(s_t, a_t)\) represents the action's detectability, with higher values indicating greater likelihood of detection.
    
    \ding{111} \(\alpha\) and \(\beta\) are coefficients that balance the importance of actuator wear and stealthiness, respectively.

The adversary’s stochastic policy \( \pi_{Adv}(a|s)=Pr\{a_{t}=a|s_{t}=s\}, s\in S \) (specifies the probability of selecting action $a$ in the current state $s$), aimed at maximising the expected cumulative reward  (driving force to search for an optimal policy $\pi^{*}$), is updated through the actor-critic algorithm, where the actor suggests actions to take based on current policy, and the critic provides feedback on the action’s potential reward, using Equation \ref{eq: Adversary optimal policy}.
The A2C algorithm iteratively updates the policy by calculating the advantage function, which helps in identifying how much better an action is compared to the average action at a state, thus refining the adversary’s strategy:

\begin{equation}
A^{\pi}(s_t, a_t) = Q^{\pi}(s_t, a_t) - V^{\pi}(s_t),
\end{equation}

    \ding{111} \( Q(s_t, a_t) \) is the action-value function, estimating the expected reward of taking action \( a_t \) in the state \( s_t \).
    
    \ding{111} \( V(s_t) \) is the value function, estimating the expected reward of the state \( s_t \) under the current policy.

By optimising this framework, the adversary strategically times actions to maximise actuator exhaustion without alerting defensive systems, leveraging the capabilities of A2C to navigate complex and partially observable environments effectively.

\textbf{Scheduler Agent:} The primary role of the scheduler agent is to determine optimal attack timings and delays based on real-time data inputs. This agent operates by continuously assessing the state of the system, represented by a state vector \( s_t \) that includes observations from actuators, sensors, and network communications. The scheduler's policy \( \pi_{\text{sched}}(a^{adv1}_t | s_t) \) selects actions \( a^{adv1}_t \) which specify the waiting time \( t^{wait} \) and the delay period \( t^{delay} \), aimed at maximising the disruption effect on the system:
\[
a^{adv1}_t = \{t^{wait}, t^{delay}\}, \quad a_t \sim \pi_{\text{sched}}(s_t)
\]

\textbf{Disturber Agent:} Once the scheduler determines \( t^{wait} \) and \( t^{delay} \), the disturber agent takes over to execute the attack. This agent aims to evade detection (i.e., produce false negatives) by mimicking the execution of benign activities.
This agent waits for the duration \( t^{wait} \) before it begins to delay packets for a period \( t^{delay} \). The disturber's decision-making process is governed by its own policy \( \pi_{\text{disturb}}(a^{adv2}_t | s_t) \), synchronising its actions with those scheduled by the scheduler to maximise the impact of the attack:
\[
a^{adv2}_t = \text{delay action}, \quad a^{adv2}_t \sim \pi_{\text{disturb}}(s_t), \quad \text{after } t^{wait}
\]

The interaction between these two agents is designed to enhance the attack's effectiveness and complicate detection by standard intrusion detection systems. Both agents utilise a self-adaptive approach, allowing them to dynamically adjust their strategies in response to environmental changes and defensive measures. This adaptation is mathematically represented by the updating of their policy functions through reinforcement learning, ensuring continuous improvement and stealthiness of the attack strategy:
\[
\theta_{t+1} = \theta_t + \alpha \nabla_{\theta} \log \pi_\theta (a_t|s_t) (R_t - V(s_t))
\]
where \( \theta \) represents the parameters of the policy networks, \( \alpha \) is the learning rate, \( R_t \) is the reward received after taking action \( a_t \), and \( V(s_t) \) is the value estimate of state \( s_t \). 
(Algorithm 1 in the Appendix C, depicts our proposed adversarial architecture, which efficiently orchestrates the actions of both the scheduler and the disturber agents within a unified advantage actor-critic methodology). The following sections will provide a detailed formalisation of each agent’s operational logic and collaborative interaction within the system’s environment.

\subsubsection{\textbf{Scheduler (Strategically-Timed Attack)}}
The Scheduler within the A2C model operates using an actor-critic algorithm, where:
- The \textbf{actor} suggests actions based on the policy function \( \pi(\cdot \mid s) \), which determines the optimal actions given the current state \( s \).
- The \textbf{critic} evaluates these actions using the value function \( V^\pi(s) \), aiming to minimise the Temporal Difference (TD) Error that measures discrepancies between predicted and actual rewards (as illustrated in Fig. \ref{fig: Adversary Agents}).

The TD Error is computed and minimised as follows:
\begin{equation} \label{eq:TD Error Update}
\psi = \psi - \eta_\psi \nabla J_{V^\pi_\theta}(\psi)
\end{equation}
\vspace{-8mm} 

\begin{equation} \label{eq:Value Function Loss}
J_{V^\pi_\theta}(\psi) = \frac{1}{2} \left(\sum_{t=i}^{i+L-1} \gamma^{t-i} R_t + \gamma^L V^\pi_\theta(S_{t+1}) - V^\pi_\theta(S_i)\right)^2
\end{equation}

\textbf{Policy Optimisation:}
The actor updates its policy based on the gradient of the performance measure, computed by considering the impact of each action over the time horizon \( L \):
\begin{equation} \label{eq:Policy Update}
\theta = \theta + \eta_\theta \nabla J_{\pi_\theta}(\theta)
\end{equation}

\vspace{-8mm} 

\begin{multline}
 \label{eq:Policy Gradient}
\nabla J(\theta) = \mathbb{E}_{\tau, \theta} \left[ \sum_{i=0}^\infty \nabla \log \pi_\theta(A_i \mid S_i) \right.\\
\left. \left(\sum_{t=i}^{i+L-1} \gamma^{t-i} R_t + \gamma^L V^\pi_\theta(S_{t+1}) - V^\pi_\theta(S_i)\right) \right]
\end{multline}

\textbf{Timing Parameters:}
To execute strategically-timed attacks, the Scheduler defines two critical timing parameters:
- \( t^{\text{wait}} \): The time the Scheduler waits before executing an action, allowing for the accumulation of necessary state information or an optimal attack window.
- \( t^{\text{delay}} \): The delay imposed between actions to enhance the subtlety and effectiveness of the attack strategy, minimising detection risks.

\textbf{Function Approximators and TD Estimation:}
Function approximators, typically neural networks, are utilised for both \( \pi_\theta \) and \( V^\pi_\theta \), with shared parameters in the lower layers to form a common state representation. The parameter \( L \), usually set to 1, provides the TD(1) estimation balancing the bias-variance trade-off. This adversary model enhances the adversary's ability to adapt and respond dynamically to defensive measures in real-time within ICS environments. The Scheduler's ability to decide when to act (using \( t^{\text{wait}} \)) and when to withhold action (using \( t^{\text{delay}} \)) is crucial for maintaining operational stealth and effectiveness.

\textbf{Example Strategic Use:}
In a scenario where the RL agent observes a sequence of states \(\{s_1, \ldots, s_L\}\), rather than attacking at every timestep, the Scheduler strategically selects timesteps based on \( t^{\text{wait}} \) and \( t^{\text{delay}} \) to maximise impact and minimise detectability. This strategic timing aligns with the overall adversarial goals and adapts to the defensive responses observed in the environment.

\textbf{When to Attack?}
In the context of strategically timed attacks within industrial control systems, specifically liquid mixture tanks, the Scheduler employs a relative action preference function \( c \). This function is crucial for identifying critical moments to delay control packets rather than manipulating control actions directly. By delaying packet delivery at precise times, the Scheduler can induce subtle yet significant disruptions in the mixing process.
The relative action preference function \( c \) is defined as:
\begin{equation}
c(s_t) = \max_{a_t} \pi(s_t, a_t) - \min_{a_t} \pi(s_t, a_t)
\label{eq: relative action preference function} 
\end{equation}
Here, \( \pi \) is the policy network mapping state-action pairs to probabilities, assessing the likelihood of selecting each possible action \( a_t \) at the current state \( s_t \). A higher value of \( c(s_t) \) indicates strong disparities in action preferences, signalling moments when delaying a specific action could disproportionately impact tank operations.

The Scheduler executes packet delays when \( c(s_t) \) exceeds a predefined threshold \( \mu \), optimising the timing of these delays to coincide with operationally sensitive moments:
\begin{equation}
b_t = 1 \quad \text{if and only if} \quad c(s_t) \geq \mu
\label{eq: packet delays} 
\end{equation}
This decision protocol ensures that packet delays are strategically timed to disrupt the mixture process effectively, thereby maximising impact and maintaining stealth. The threshold \( \mu \) helps manage the frequency of these interventions, which is critical for avoiding detection and is closely tied to the parameter \( \Gamma \), limiting the total number of strategic delays.

Consider a scenario in an industrial setting where the precise timing of chemical additions is critical to product quality. If the Scheduler detects that a critical component's addition is scheduled at a peak moment (high \( c \) value), delaying this control packet can subtly alter the concentration balance. Such a delay, especially if it occurs during peak processing times, can lead to inefficiencies in blending or chemical reactions, potentially degrading the overall product quality over time without triggering immediate operational alarms. By strategically timing these packet delays, the Scheduler can effectively manipulate the operational process, gradually degrading system efficiency or product quality. This approach not only maximises the adversary's impact but also enhances the difficulty of detecting and diagnosing the problem's source, given the intervention's subtle nature.

\subsubsection{\textbf{Disturber (Selective Packet Delay)}}
The Disturber, operating within the A2C model framework, is tailored to selectively delay critical control packets to subtly undermine system stability without direct manipulation or detection. It employs an actor-critic architecture, where decisions about which packets to delay are calculated based on their estimated impact on system operations (as illustrated in Fig. \ref{fig: Adversary Agents}).

\textbf{Actor-Critic Architecture:}
The architecture is structured as follows:
- The \textbf{actor} suggests specific packets for delay by analysing the current system state \( s_t \) and predicting the impact of delaying each possible packet.
- The \textbf{critic} evaluates the potential impact of these delays using a specialised value function \( V^{adv2}(s) \), focusing on long-term operational consequences and stealth.

\textbf{Temporal Difference Error for Disturber:}
The Temporal Difference (TD) Error, crucial for adjusting the policy towards more effective disruptions, is recalculated to fit the Disturber's objectives:
\begin{equation} \label{eq:TD Error Disturber}
\delta_t = R_{t+1}^{adv2} + \gamma V^{adv2}(S_{t+1}) - V^{adv2}(S_t)
\end{equation}
Here, \( R_{t+1}^{adv2} \) is the reward received after executing a delay, considering both the impact of the delay and the importance of remaining undetected.

\textbf{Policy Optimisation for Disturber:}
The policy optimisation process adapts to prioritise packet delays that maximise disruption while minimising visibility using Equation \ref{eq:Policy Gradient} and:
\begin{equation} \label{eq:Policy Update Disturber}
\theta^{adv2} = \theta^{adv2} + \eta_{\theta^{adv2}} \nabla J_{\pi_{\theta^{adv2}}}(\theta^{adv2})
\end{equation}

\textbf{Strategic Selection and Timing of Packet Delays:}
The Disturber uses the relative action preference function \( c \) to determine the criticality of delaying specific packets at given moments (for the period $t^{delay}$, decided by the scheduler agent), using Equation \ref{eq: relative action preference function}.
High values of \( c(s_t) \) indicate moments where delaying a packet will maximally disrupt operations while maintaining operational secrecy.

\textbf{Execution Condition:}
Packet delays are executed when \( c(s_t) \) exceeds the threshold \( \mu \) in Equation \ref{eq: packet delays}.
This condition ensures that actions are taken only during the most opportune moments to effectively affect the mixture process, thus optimising the timing of attacks for maximum impact and minimal detection.
If the Disturber identifies that delaying a packet that controls the flow rate of a critical chemical component can cause significant deviations from optimal mixing ratios, it can strategically delay this control signal during critical production phases. The careful selection and timing of such delays can lead to product quality degradation over time, effectively executing a wear-out attack without immediate detection.
By employing this refined and targeted approach, the Disturber not only disrupts but also introduces a layer of complexity to the operational environment that defenders must untangle, enhancing both the effectiveness and stealth of the adversarial actions.

\begin{table*}[tb]
\scriptsize

\begin{threeparttable}
\setlength\tabcolsep{0pt} 
\caption{Performance of DNN-based detection models against Extreme Attack}
\label{tab:anomaly_detectors against extreme attack}
\begin{tabular*}{\textwidth}{@{\extracolsep{\fill}}l c ccc ccc ccc  ccc ccc }
\toprule
\multicolumn{1}{c}{\multirow{2}{*}{\parbox{1.3cm}{ \centering \textbf{Attack Strategy}}}} & \multicolumn{1}{c}{\multirow{2}{*}{\parbox{1.3cm}{ \centering \textbf{Adversary Knowledge}}}}\  
    &\multicolumn{3}{c}{\textbf{DenseNet}\tnote{\dag}}
    &\multicolumn{3}{c}{\textbf{CNN}\tnote{\dag}}
    &\multicolumn{3}{c}{\textbf{ResNet}\tnote{\dag}}
    &\multicolumn{3}{c}{\textbf{LSTM}\tnote{\dag}}
    &\multicolumn{3}{c}{\textbf{Transformer}\tnote{\dag}}
     \\
\cmidrule{3-5} 
\cmidrule{6-8} 
\cmidrule{9-11} 
\cmidrule{12-14}
\cmidrule{15-17}
\multicolumn{1}{c}{} & \multicolumn{1}{c}{}

    & Precision & Recall & F1 
    & Precision & Recall & F1 
    & Precision & Recall & F1 
    & Precision & Recall & F1 
    & Precision & Recall & F1 
    \\

\midrule
Extreme Attack & Black-box &
0.817 & 0.844 & 0.830 & 0.984 & 0.956 & 0.970 &
\textbf{\underline{0.997}} & \textbf{\underline{0.994}} & \textbf{\underline{0.995}} & 0.912 & 0.911 & 0.912 & 0.980 &   0.966 &  0.973 
\\

\bottomrule
\end{tabular*}
\scriptsize

\scriptsize

\end{threeparttable}
\label{tab:dnn-based detectors against extreme attack}%

\vspace{-10pt}
\end{table*}

\begin{table*}[tb]
\scriptsize

\begin{threeparttable}
\setlength\tabcolsep{0pt} 
\caption{Performance of Adversary against DNN-based detection models}
\label{tab:benchmarking}

\begin{tabular*}{\textwidth}{@{\extracolsep{\fill}}l c ccc ccc ccc  ccc ccc }
\toprule
\multicolumn{1}{c}{\multirow{2}{*}{\parbox{1.3cm}{ \centering \textbf{Attack Strategy}}}} & \multicolumn{1}{c}{\multirow{2}{*}{\parbox{1.3cm}{ \centering \textbf{Adversary Knowledge}}}}\  
    &\multicolumn{3}{c}{\textbf{DenseNet}\tnote{\dag}}
    &\multicolumn{3}{c}{\textbf{CNN}\tnote{\dag}}
    &\multicolumn{3}{c}{\textbf{ResNet}\tnote{\dag}}
    &\multicolumn{3}{c}{\textbf{LSTM}\tnote{\dag}}
    &\multicolumn{3}{c}{\textbf{Transformer}\tnote{\dag}}
     \\
\cmidrule{3-5} 
\cmidrule{6-8} 
\cmidrule{9-11} 
\cmidrule{12-14}
\cmidrule{15-17}
\multicolumn{1}{c}{} & \multicolumn{1}{c}{}

    & Precision & Recall & F1 
    & Precision & Recall & F1 
    & Precision & Recall & F1 
    & Precision & Recall & F1 
    & Precision & Recall & F1 
    \\

\midrule
Smash \& Grab & \multirow{2}{*}{Black-box} &
 0.5058 & 0.6989 & 0.5869 & 0.5060 & 0.6965 & 0.5861 & 
 0.5065 & \textbf{\underline{0.7012}} & 0.5881 & \textbf{\underline{0.5091}} & 0.7006 & \textbf{\underline{0.5897}} & 0.5053 &   0.7001 &  0.0099 
\\
Low \& Slow & &
0.5067 & 0.6978 & 0.5871 & 
0.5060 & 0.6923 & 0.5846 &
\textbf{\underline{0.5071}} & \textbf{\underline{0.6984}} & \textbf{\underline{0.5876}} &
0.5053 & 0.6962 & 0.5856 & 0.5055 & 0.6936 &  0.5848 
\\

\midrule

Smash \& Grab & \multirow{2}{*}{Grey-box} &
0.5102 & 0.5997 & 0.5513 & 0.5078 & 0.2941 & 0.5476 &
\textbf{\underline{0.5127}} & \textbf{\underline{0.6031}} & \textbf{\underline{0.5542}} & 0.5089 & 0.6039 & 0.5524 & 0.5074 &  0.5962 &  0.5482
\\
Low \& Slow & &
0.5377 & 0.2816 & 0.3697 & 0.5584 & 0.3131 & 0.4012 &
\textbf{\underline{0.5555}} & 0.3142 & 0.4014 & 0.5461 & 0.3688 & \textbf{\underline{0.4403}} & 0.5528 &  \textbf{\underline{0.3173}} &  0.4032 
\\

\midrule

 Smash \& Grab & \multirow{2}{*}{White-box} &
0.6338 & 0.0017 & 0.0034 & \textbf{\underline{0.8254}} & 0.0020 & 0.0039 &
0.6966 & \textbf{\underline{0.0024}} & \textbf{\underline{0.0047}} & 0.8235 & 0.0016 & 0.0032 &  0.7222&   0.0020 &  0.0039
\\

Low \& Slow &  &
\textbf{\underline{0.7500}} & 0.0006 & 0.0011 & 0.5185 & 0.0005 & 0.0011 &
0.5200 & \textbf{\underline{0.0010}} & \textbf{\underline{0.0020}} & 0.5294 & \textbf{\underline{0.0010}} & \textbf{\underline{0.0020}} & 0.6216 &   0.0009 &  0.0017 
\\

\bottomrule
\end{tabular*}
\scriptsize

\scriptsize
\begin{tablenotes}
\item[\dag] These are DNN-based detection models. The underlined values denote the highest performance metrics achieved across different attack strategies and adversary knowledge levels.

\end{tablenotes}
\end{threeparttable}

\label{tab:wear_out_results}%

\end{table*}

\section{Design and Implementation}
\label{sec:Design and implementation}

We begin our analysis by describing the implementation underlying our evaluations.
To validate our proposed threat model, we constructed an ICS testbed that facilitates the mixing of liquids using three reservoirs, depicted in Fig. \ref{fig:RI flow diagram}. 
The testbed’s system architecture, illustrated in Fig. \ref{fig: Implemetation}, includes Operational Technology (OT) and Information Technology (IT) components. Key components include a Cisco ASA 5555-X Firewall for traffic monitoring and intrusion detection/prevention, Cisco C8200-1N-4T Router and Cisco C9300-24U-A V06 Switch for robust network management, SIMATIC S7-1200 and SIMATIC IPC127E PLCs for process control, and three NVIDIA Jetson AGX Orin Developer Kits for running advanced AI-driven intrusion detection systems, utilising \emph{state-of-the-art models} such as DenseNet, CNN, ResNet, LSTM, and Transformer-based anomaly detectors. To ensure high reliability and real-time performance essential for industrial applications, we employed the PROFINET protocol \cite{8627173} across our network. This choice reflects the industry preference for PROFINET due to its flexibility, speed, and efficiency in handling extensive data exchange and real-time communication between field devices and controllers.
\emph{This setup incorporates components and technologies currently used in the industry, ensuring our findings are relevant and applicable to real-world scenarios.} Next, we provide the implementation details of our testbed, outlining the specific components and configurations used to conduct the experiments.

\begin{figure}[t!]
\centering
\includegraphics[width=70mm]{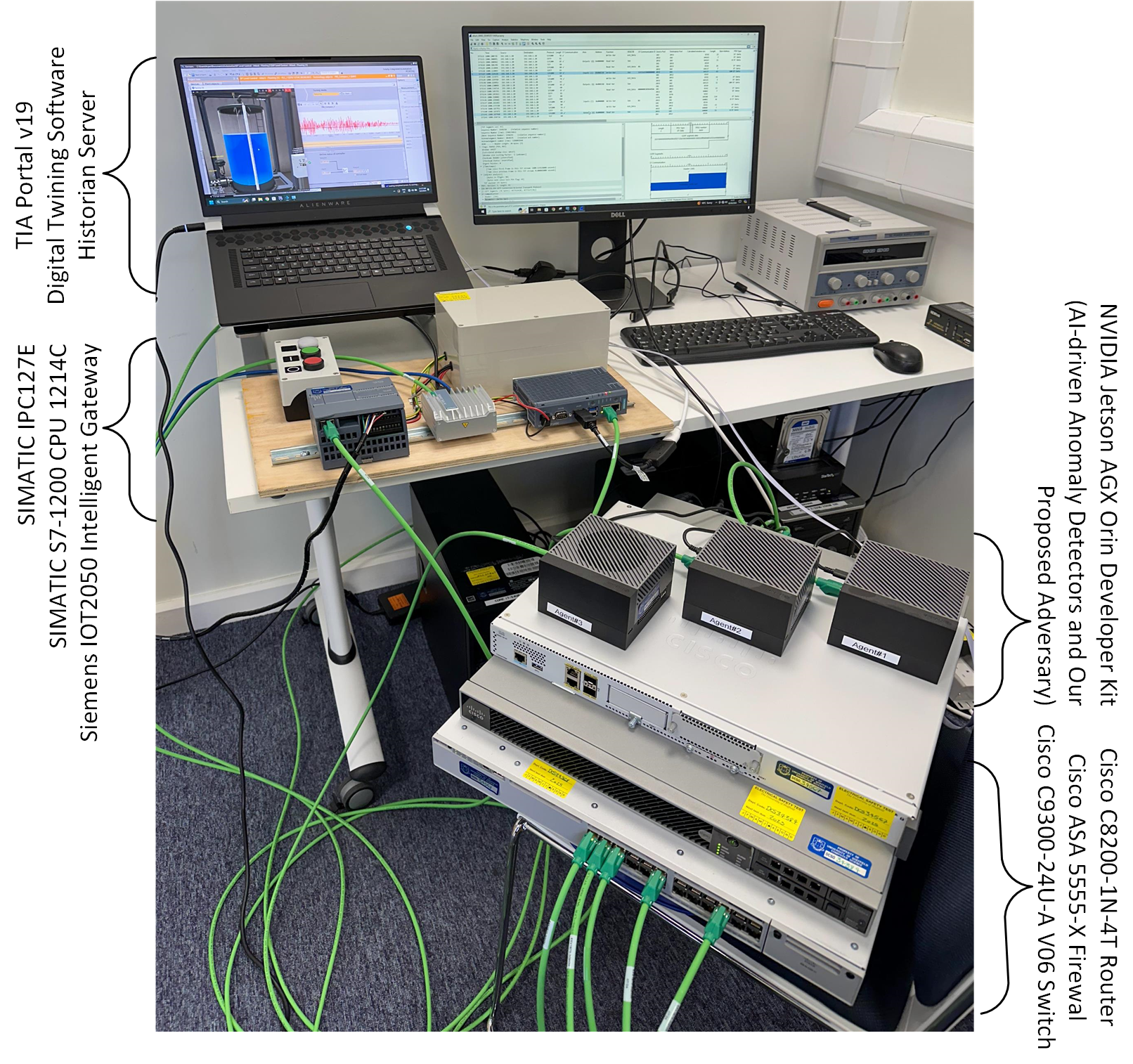}
\captionsetup{font=small}
  \caption{Our System Implementation}
  \label{fig: Implemetation}
\vspace{-5mm}
\end{figure}

\subsection{Operational Technology Architectures}
\label{subsec:implementation}
Our testbed is designed to replicate a complex industrial control process, integrating a range of digital and analogue sensors, actuators, and control elements to mimic real-world operational conditions of industrial systems, depicted in Fig. \ref{fig:RI flow diagram}. We focus on batch processing in the process model, which involves combining specific quantities of raw materials in designated sequences. Focused on the dynamics of fluid mixing processes, the architecture provides a comprehensive environment for testing and validating various control strategies and response mechanisms.
Central to the setup are two storage reservoirs (VE-1, VE-2) and one main mixing tank (VE-3), which play critical roles in the fluid management system. These reservoirs feed liquids into the main tank, where they are combined according to predefined recipes. This configuration allows for the precise adjustment of mixing ratios and the examination of fluid dynamics under controlled conditions. The liquids are moved through the system by pump PL-1, a key component ensuring the transfer and circulation of fluids throughout the testbed.
Flow rates are monitored by the sensor BF1, positioned at a critical juncture in the system to measure the throughput of the pipelines. This sensor ensures that the controlled variable `FIC BF1' remains stable, maintaining consistency in flow despite potential disturbances or variable operational demands. The setup includes several strategically placed valves (e.g. VV-1 and VV-2) equipped with pneumatic quarter-turn actuators. These valves are essential for modulating the flow, allowing precise control over the liquid movement between the reservoirs, the main tank, and further processing stations.
The process valves, particularly those with pneumatic actuators, enhance the system's flexibility and responsiveness. They manage the flow from the lower tanks (VE-1 and VE-2) into the mixing tank (VE-3) and facilitate the transfer of the finished mixture to subsequent processing stations or back into the dosing tanks, depending on the operation's current needs.

\begin{figure}[htbp] 
\center{\includegraphics[width=85mm]{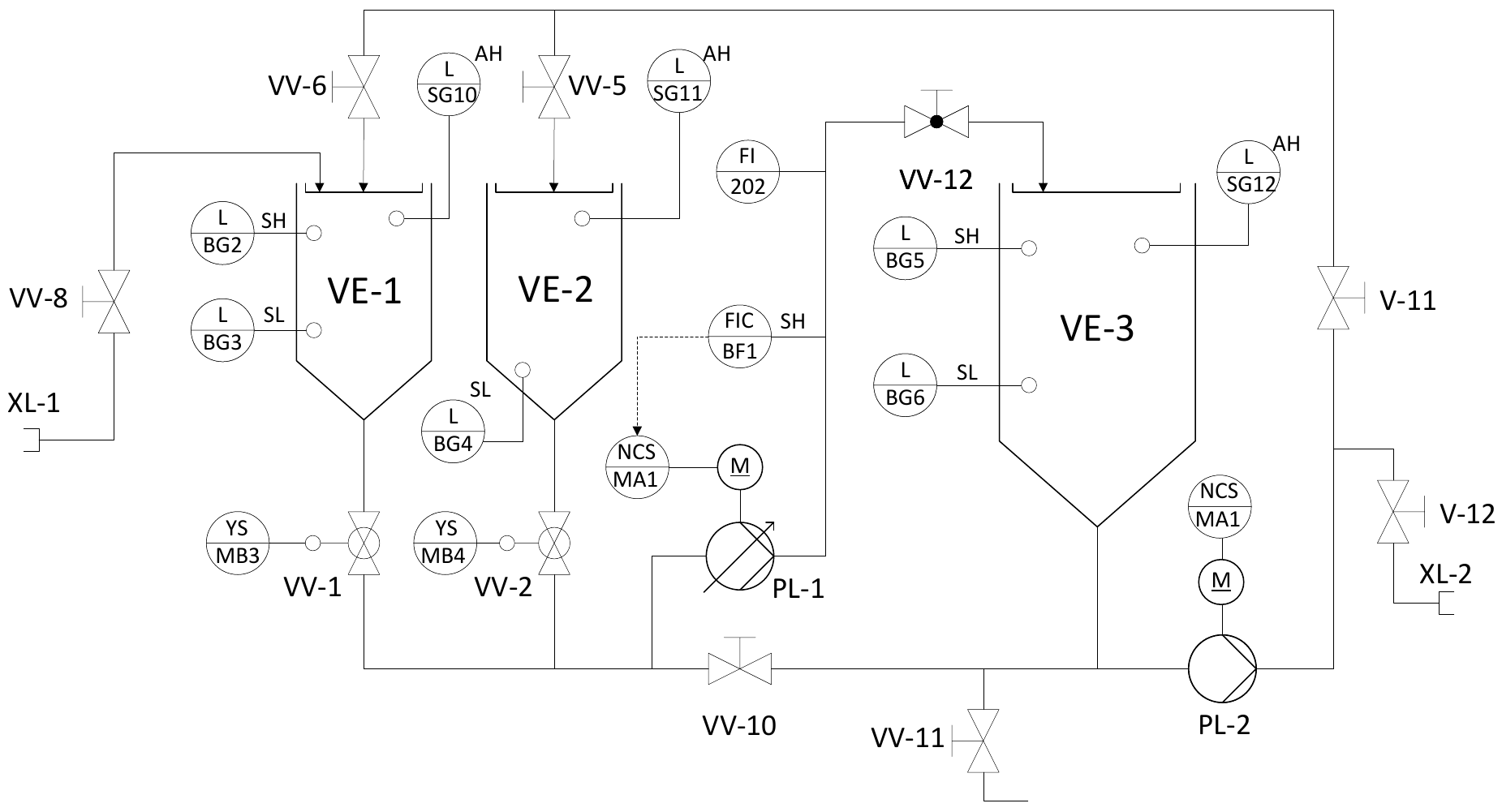}} 
\caption{Cascade mixing liquid control; two liquid reservoir tanks and one main tank.}
\label{fig:RI flow diagram} 
\end{figure}

\subsection{Evaluation Framework}
\label{sec:Evaluation Framework}
The evaluation of the proposed adversary against AI-driven anomaly detectors is structured into three distinct phases, each designed to progressively challenge the anomaly detectors and refine adversarial tactics in a controlled manner.

\textbf{Phase One: Anomaly Detector Development} -
This phase focuses on creating and training various AI-driven anomaly detectors using datasets generated in our dedicated testbed environment. These detectors, including DenseNet \cite{huang2017densely}, CNN \cite{krizhevsky2012imagenet}, ResNet \cite{he2016deep}, LSTM \cite{hochreiter1997long}, and Transformer \cite{vaswani2017attention}, employ a combination of flow-based and ratio-based features to analyse operational data from industrial control systems. The objective is to establish a robust baseline of anomaly detection capabilities to identify potential threats under normal operational conditions. Table \ref{tab:anomaly_detectors against extreme attack} illustrates the performance of these detectors against an extreme attack scenario, where the adversary randomly delays packets,

\textbf{Phase Two: Adversarial Policy Crafting} -
In the second phase, the constructed anomaly detectors are deployed within a sandboxed environment of our proposed adversarial model (discussed in Section \ref{sec:Crafting an Adversarial Strategy}). This model is tasked with developing and optimising attack strategies that aim to maximise long-term rewards while evading detection. The adversary iteratively tests and refines its policies, effectively learning to exploit potential weaknesses in the anomaly detectors without being detected.

The adversary aims to converge on an optimal set of actions that achieve the intended disruption while remaining undetected by the AI-based security systems. This approach highlights any fluctuations caused by the attacks, which might otherwise appear as flat representations in a more extended time window). 

\textbf{Phase Three:  Testing and Evaluation (Out of Context Environments)} -

The final phase involves evaluating the refined adversarial strategies by deploying them against a new set of AI-driven anomaly detectors trained on a large dataset with different sets of features. These detectors are designed with diverse architectures to challenge the adversary's tactics. Additionally, the evaluation is conducted in an \textbf{Out-of-Context scenario}, where the ICS environment differs from the one used in \textit{Phase Two}. This ensures that the adversary's strategies are robust and effective in adapting to varying system dynamics and conditions. The effectiveness of the adversarial strategies is assessed based on their ability to avoid triggering the detection mechanisms in this real control process environment.

\begin{notebox}{Takeaway:} To ensure clarity in our methodology, here we would like to highlight that the adversarial agents were initially optimised within a controlled, simulated environment, closely mirroring realistic field network dynamics and constraints. This approach enabled safe and comprehensive optimisation without triggering actual alarms or disruptions. Subsequently, the validated strategies were rigorously assessed on a physical testbed (depicted in Fig. \ref{fig:RI flow diagram}), confirming their applicability and effectiveness in real-world operational contexts.\end{notebox}

\textbf{Real vs. Simulated Setup Clarification.} To ensure clarity regarding our experimental procedure, we emphasise that the optimisation and initial training phases of the adversarial agents (Scheduler and Disturber) were conducted in a controlled, simulated replication of realistic operational behaviour. This simulated environment mirrors the dynamics and operational constraints of real Industrial Control Systems (ICS), enabling comprehensive and safe optimisation of adversarial policies without triggering actual security alarms or causing disruptions. Once the agents’ strategies were fully optimised and validated in the simulated scenario, they were subsequently evaluated on an industry-level testbed. During real-world testing, any triggered alarms or security alerts were carefully monitored, logged, and handled in a controlled manner to avoid operational disruptions and ensure the safety and integrity of the actual infrastructure.

\paragraph*{Stealth Score (definition).}
Let $\mathcal{D}=\{1,\dots,D\}$ index detectors and $\{1,\dots,T\}$ index non-overlapping windows.
For detector $d$, define per-window recall $R_{d,t} = \tfrac{\mathrm{TP}_{d,t}}{\mathrm{TP}_{d,t}+\mathrm{FN}_{d,t}}$.
We average recalls per detector, then aggregate with normalized weights $w_d$ ($\sum_d w_d=1$, $w_d{=}1/D$ unless stated):
\[
\overline{R}_d=\frac{1}{T}\sum_{t=1}^T R_{d,t},\qquad
\mathrm{Stealth} \;=\; 1 - \sum_{d\in\mathcal{D}} w_d\,\overline{R}_d.
\]
A higher value indicates greater evasiveness (lower averaged recall across detectors). We report the mean Stealth over each evaluation run and its $95\%$ CI across seeds.

\vspace{-3mm}

\section{Experiments and Results}
\label{sec:Experiments Results}

We formulated our proposed adversarial model in Section \ref{sec: Threat Model}. In this section, we conduct a series of experiments to explore various strategies and configurations that enable the adversary to achieve its objectives (discussed in Section \ref{sec: Threat Model}). In this regard, we test different reward, exploration, and discounting regimes for the adversary. Additionally, we evaluate the performance of the adversary against several state-of-the-art DNN-based anomaly detectors. Through these experiments, we aim to identify the most effective strategies that adversaries might use to compromise AI-assisted security systems.

\textbf{Adversarial Exploration and Exploitation:}
The strategic choice between Low \& Slow and Smash \& Grab is directly connected to the value of \(\varepsilon\), where \(\varepsilon\) is a parameter that balances exploration (trying out new actions to discover their effectiveness) and exploitation (leveraging known actions that yield high rewards), depicted in Fig. \ref{fig:Epsilon vs Gamma}. These strategies reflect different approaches to managing the trade-off between immediate rewards and long-term gains, impacting the adversarial model's effectiveness and stealth. Low \& Slow strategy is akin to fine-tuning the \(\varepsilon\) value to favour exploitation, minimising risk by making incremental changes that are less likely to be detected. Here, the adversarial agent optimises its policy to maintain actions within a threshold that does not trigger defensive responses and minimise the expected detection metric \( E[D(s, a) \mid s \in S] \) over all actions \( a \in A \), where: $\min_{a \in A} E[D(s, a) \mid s \in S], \quad \text{where} \quad d(s, s') < \delta$ with \(\delta\) defined as a small value below the detection threshold of the system's anomaly detectors.
The policy \(\pi(s)\) is tailored to sustain operational secrecy by ensuring that each action incrementally pushes the system state without crossing the bounds set by \(\delta\). This strategy utilises a lower \(\varepsilon\), reducing exploration to avoid novel actions that might lead to detection, thus emphasising a gradual, cumulative impact on the target system. Contrasting with the cautious nature of Low \& Slow, the Smash \& Grab strategy leverages a higher \(\varepsilon\) for increased exploration. This approach encourages the adversarial model to undertake bold, high-impact actions that might be easily detected but achieve substantial immediate effects; Maximise the expected impact \( E[\text{Impact}(s, a) \mid s \in S] \) over all actions \( a \in A \), without stringent constraints on the deviation \( d(s, s') \). In this setting, the adversarial agent prioritises quick gains over stealth, exploiting the system's vulnerabilities before a robust defensive response can be organised. This strategy is effective in scenarios where immediate disruption rewards outweigh the benefits of remaining undetected. Choosing these strategies significantly affects the adversarial campaign's structure and success. A Low \& Slow strategy might be preferable in tightly monitored environments where long-term access is crucial. In contrast, Smash \& Grab could be advantageous in scenarios where rapid extraction of sensitive information or causing immediate disruption is the goal. The RL-based adversary must dynamically adjust the \(\varepsilon\) parameter to navigate between these strategies based on real-time assessments of system vulnerabilities and defensive readiness, underscoring the importance of adaptive strategic planning in executing effective cyber-attacks within different operational contexts, attaining $\mathcal{O}_{3}$.

\begin{figure}[htbp] 
\center{\includegraphics[width=68mm]{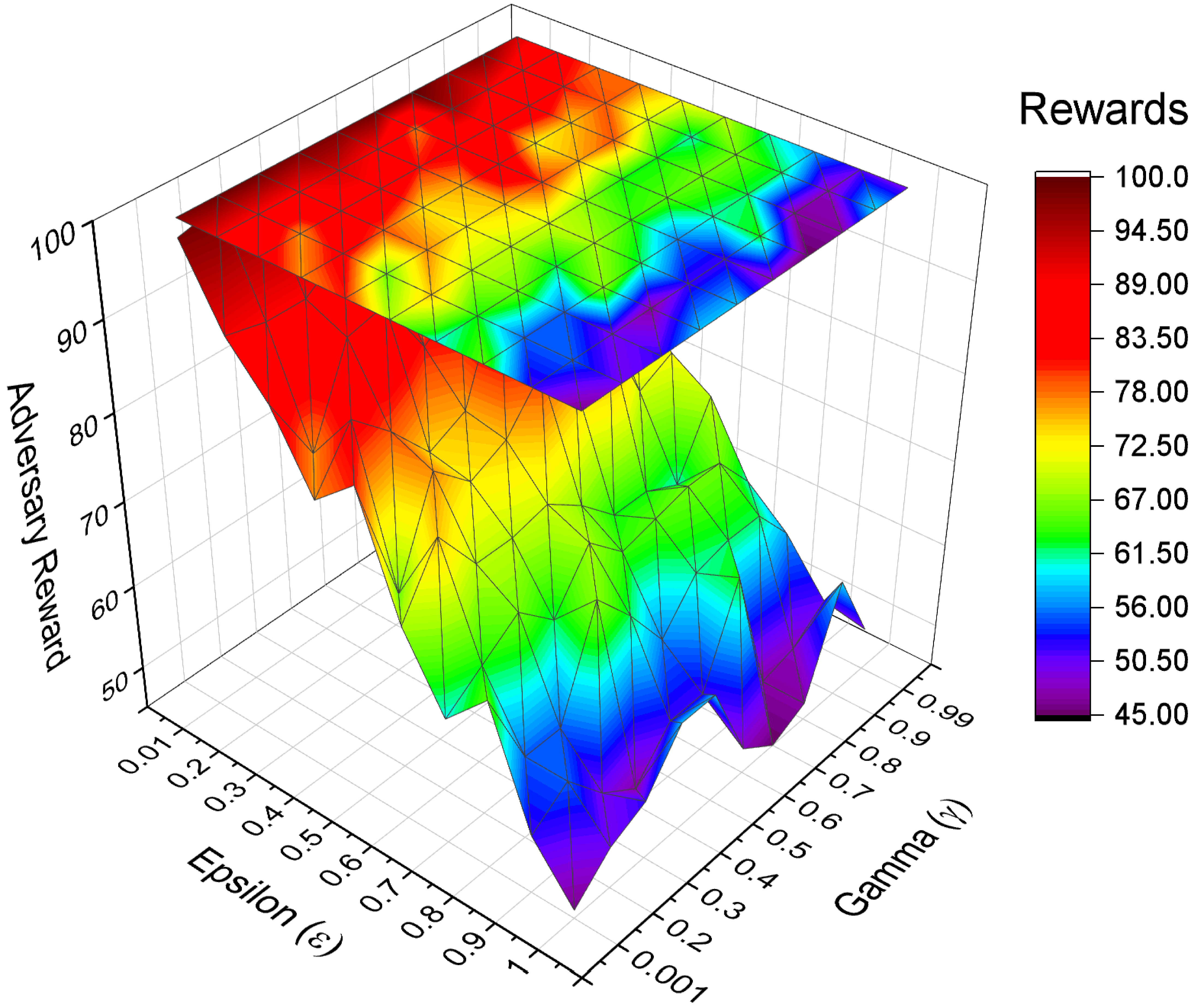}} 
\caption{Epsilon vs Gamma}
\label{fig:Epsilon vs Gamma} 
\end{figure}

\textbf{Role of Discount Factors:}
The discount factor \(\gamma\) plays a crucial role in balancing the long-term versus short-term rewards, directly affecting the strategic behaviour of both the scheduler and disturber agents. For the scheduler, which typically needs to consider the implications of its actions over an extended period, a high discount factor (close to 1) is beneficial. This setting encourages the scheduler to take actions that may not result in immediate rewards but are beneficial in the long term, aligning with strategies that require sustained covert operations within the target system. Formally, the impact of \(\gamma\) on the scheduler's value function can be described as: $V^{\pi_{\text{sched}}}(s) = \mathbb{E} \left[ \sum_{t=0}^{\infty} \gamma^t r_t \mid s_0 = s \right]$, where the expectation \(\mathbb{E}\) emphasises the accumulated reward over time, modified by the discount factor \(\gamma\).
Conversely, a lower discount factor may be more appropriate for the disturber, which is oriented towards achieving immediate disruptive effects. This setting puts the priority on gaining immediate high-impact rewards. It reduces the relative value of future rewards, thus supporting aggressive tactics that are the characteristics of a Smash \& Grab approach. The discount factor modifies the disturber's action-value function to:
$Q(s, a) = \mathbb{E} \left[ r_t + \gamma Q(s_{t+1}, a_{t+1}) \mid s, a \right]$, where \(s_{t+1}\) and \(a_{t+1}\) denote the state and action at the next timestep, respectively. As illustrated in Fig. \ref{fig:Discount factor importance}, based on our experiments, a lower \(\gamma\) ensures that the value of future states (\(Q(s_{t+1}, a_{t+1})\)) contributes less to the current decision-making process, making immediate outcomes more significant. In this regard, engineering \(\gamma\) value strategically plays a vital role in the adversary model as it fundamentally influences how the scheduler and disturber evaluate the trade-offs between immediate and delayed rewards, attaining $\mathcal{O}_{2}$ and $\mathcal{O}_{3}$.

\begin{figure}[htbp] 
\center{\includegraphics[width=68mm]{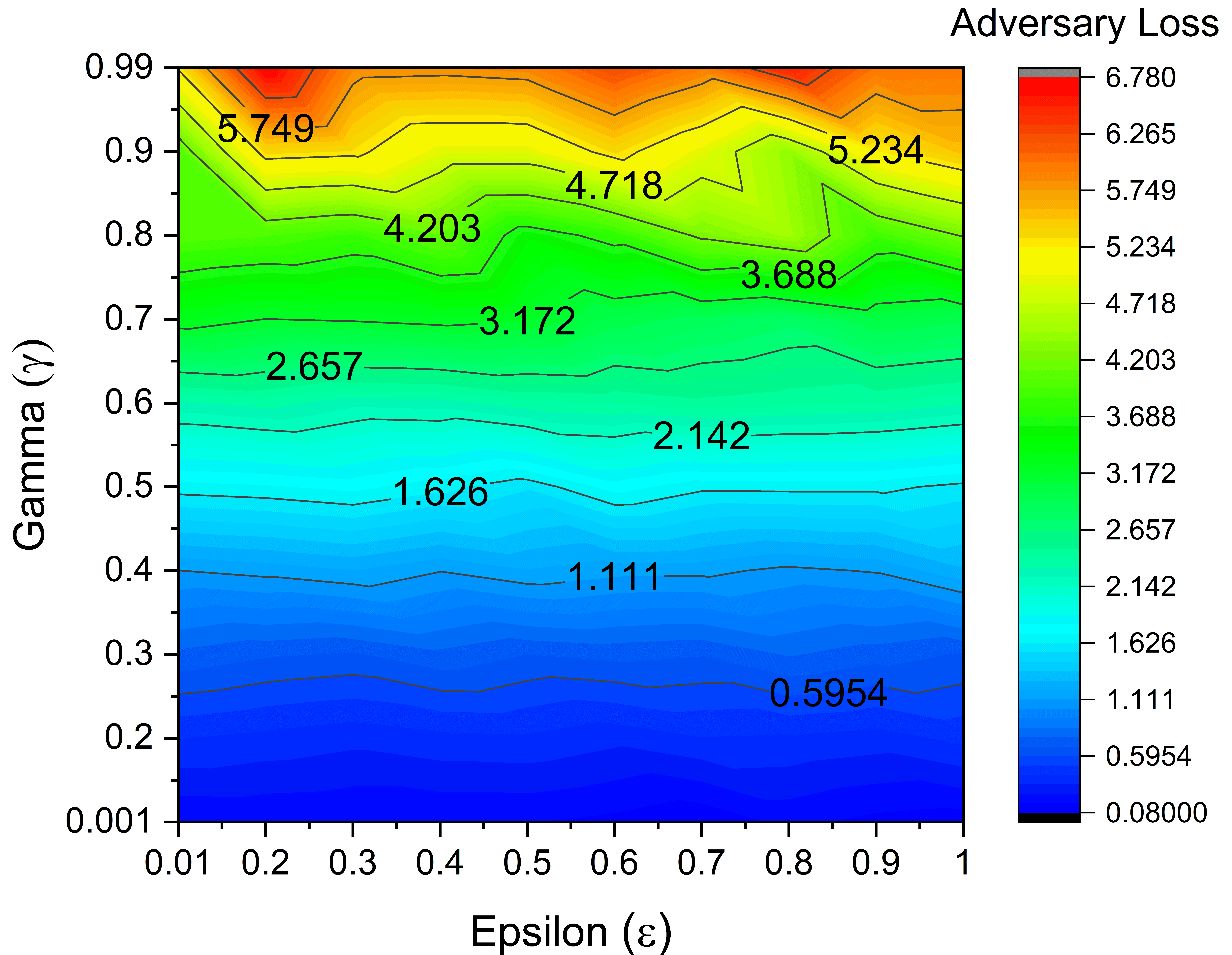}} 
\caption{Discount Factor Importance}
\label{fig:Discount factor importance} 
\end{figure}

\paragraph*{\textbf{Hyperparameter Selection for $\alpha,\beta$.}}
We select $(\alpha,\beta)$ via a small Pareto sweep on a held-out validation slice (no test scenes). We target \emph{stealth–impact balance} by maximizing
$\text{Stealth} - \lambda\,\Delta\text{Throughput}$ with $\lambda{=}0.5$ while enforcing $\Delta\text{Cycles}\!\le\!18\%$ for L\&S.
Table~\ref{tab:alphabeta} reports the sweep; we choose $\alpha{=}0.55$, $\beta{=}0.45$ for main experiments.

\begin{table}[htbp]
\centering

\caption{Sweep over $(\alpha,\beta)$ on validation (L\&S). Stealth $=1-\sum_d w_d R_d$ with equal $w_d$. Selected in \textbf{bold}.}
\label{tab:alphabeta}
\renewcommand{\arraystretch}{1.05}
\begin{tabular}{ccccc}
\toprule
$\alpha$ & $\beta$ & Stealth $\uparrow$ & $\Delta$Thru $\downarrow$ & $\Delta$Cycles $\uparrow$  \\
\midrule
0.40 & 0.60 & 0.86 & $-2.6\%$ & $+4.9\%$ \\
0.50 & 0.50 & 0.87 & $-3.5\%$ & $+6.6\%$ \\
\textbf{0.55} & \textbf{0.45} & \textbf{0.88} & \textbf{$-4.1\%$} & \textbf{$+7.9\%$}  \\
0.60 & 0.40 & 0.86 & $-4.9\%$ & $+9.2\%$ \\
0.70 & 0.30 & 0.84 & $-5.6\%$ & $+10.8\%$ \\
\bottomrule
\end{tabular}
\end{table}

\textbf{Reward Engineering:}
Crafting reward mechanisms is crucial, especially for adversarial agents like schedulers and disturbers, where the right incentives can directly influence both tactical decisions and overarching strategy. The design of the reward function \( R(s_t, a_t^{adv1}, a_t^{adv2}, s_{t+1}) \), which incorporates both effectiveness \( \alpha \cdot D(s_t, a_t^{adv1}, a_t^{adv2}, s_{t+1}) \) and stealthiness \( \beta \cdot (1 - C(s_t, a_t^{adv1}, a_t^{adv2})) \), is tailored to balance these objectives. Here, \(\alpha\) promotes rewarding actions that effectively degrade the target system by increasing inaccuracies in the control process (\(D\)), thereby supporting long-term strategic degradation. In contrast, \(\beta\) mitigates risk by penalising actions with higher exposure (\(C\)), reinforcing the importance of stealth and minimising the likelihood of detection. This reward structure ensures that each action is evaluated for its immediate benefit and potential risk, supporting a strategy that aligns with the dual goals of impact and discretion.
The optimal reward setting depends on the adversary's objectives and the detection system. In scenarios where the adversary prioritises being undetected and staying within the target system, a higher \(\beta\) value is preferable to enforce cautious, stealthy actions that minimise the risk of exposure. Conversely, if the objective is to achieve an abrupt and significant disruption, a higher \(\alpha\) might be more appropriate, encouraging actions that maximise immediate impacts. The ability to dynamically adjust these parameters in response to changing conditions or to refine tactics based on operational feedback is a key advantage, allowing adversaries to maintain effectiveness while adapting to countermeasures and evolving defences, attaining $\mathcal{O}_{3}$.

\textbf{Impact of Adversarial Strategies on Control Precision ($\mathcal{O}_{2}$):}
The experimental results demonstrate that the "Low \& Slow" adversarial strategy causes a 3\% inaccuracy in control system setpoints across 100 runs by introducing subtle disturbances that avoid exposure (as Depicted in Fig. 12 in the Appendix C). This strategy minimises exposure while gradually increasing the process variance (\(\sigma^2\)). In contrast, the "Smash \& Grab" strategy, which leads to a 9\% inaccuracy, achieves its higher impact through aggressive actions that significantly deviate from optimal operational parameters, causing a larger variance (\(\sigma^2\)). However, this greater disruption comes with a trade-off, as the "Smash \& Grab" strategy is exposed more compared to the "Low \& Slow" approach, as depicted in Table \ref{tab:benchmarking}.

\textbf{Computational Complexity of Adversary:} In this experiment, we meticulously analyse the computational complexity of the scheduler and the disturber. Our objective is to quantify how these elements perform in terms of computational demand and energy efficiency during the testing phase, which is crucial for maintaining stealth and operational effectiveness in real-world scenarios.

\textbf{Scheduler Complexity:} The computational complexity of the scheduler is predominantly determined by its policy optimisation tasks, which involve extensive decision-making across a potentially expansive observation space \( O \) and action space \( A \). The scheduler selects actions \( a \) from \( A \) based on the policy \( \pi \), governed by: $o_{t+1} = f(o_t, a), \quad a \sim \pi(o_t)$, where \( f \) denotes the observation transition function and \( \pi \) is optimized to maximize cumulative rewards. The complexity of the scheduler can be encapsulated using Big O notation as \( O(n^k) \), where \( n \) signifies the number of possible observations and \( k \) represents a factor influenced by the complexity of the policy evaluation and update mechanisms. Due to the streamlined architecture developed during the training phase, \( k \) is managed to prevent excessive scaling with \( n \).

\textbf{Disturber Complexity:} The disturber's role involves the execution of actions at strategically optimal times to subtly manipulate the observation transitions. Its complexity pertains to adapting observation changes based on system feedback and the policy \( \pi_{dist} \): $o_{t+1} = f(o_t, a_{dist}), \quad a_{dist} \sim \pi_{dist}(o_t)$, expressed as \( O(m^l) \), where \( m \) is the number of actions and \( l \) denotes the complexity of evaluating action impacts. The computational demand is optimised by employing sophisticated algorithms that enhance decision-making efficiency without significant additional computational resources. This approach minimises energy usage during the testing phase, aligning with the stealth goals of the adversary.

\textbf{Takeaways:} The combined complexities of both components induce the overall computational complexity of the adversarial system during testing as \( O(n^k + m^l) \). This notation reflects a system designed to maximise adversarial impact while maintaining operational stealth and computational efficiency. The strategic implementation of a less exhaustive training architecture enables significant enhancements in model performance and efficiency. This investment in the training phase allows the model to execute complex adversarial strategies during testing with reduced computational demands, ensuring effective evasion of detection by security systems and conservation of energy, which is vital in extended operation scenarios or resource-limited environments.

{

\subsubsection*{Ablation: DRL Algorithm Variants}
We compare A2C to PPO, DDPG, and SAC (multiple seeds per algorithm), reporting:
(i) episode return vs.\ environment steps (smoothed),
(ii) \emph{long-run reward} (mean over the last 10\% of steps),
(iii) area under the learning curve (AUC), and
(iv) time-to-95\% of final reward $T_{95}$.
Appendix Figure~12 shows learning curves; Table~X in the Appendix summarises the statistics.
Across our testbed, A2C achieves the highest long-run reward and AUC and converges faster (lowest $T_{95}$), while PPO, DDPG, and SAC underperform.

}

\textbf{Design Rationale for Dual-Agent Architecture:}
In our proposed attack framework, we deliberately employ a dual-agent model comprising a \emph{Scheduler} and a \emph{Disturber}, each implemented using the Advantage Actor-Critic (A2C) reinforcement learning algorithm. This architectural decision is rooted in both theoretical considerations and empirical observations derived during our design iterations. While a single-agent approach might appear sufficient at first glance, our ablation studies reveal that it exhibits suboptimal performance in terms of convergence stability, stealth capability, and adaptability.

\paragraph{Problem Decomposition and Learning Complexity.}
A unified agent tasked with simultaneously learning both \emph{when} to attack (i.e., strategic timing) and \emph{how} to attack (i.e., selective execution of control packet delays) faces significant learning complexity. These sub-tasks are inherently distinct in temporal abstraction, gradient dynamics, and reward attribution. In our experiments, a single-agent architecture led to unstable policy updates due to gradient interference between competing objectives. The agent struggled to balance long-horizon credit assignment for timing decisions with short-term feedback loops associated with stealthy action execution. This often led to premature convergence to local optima and inconsistent performance across different control scenarios.

\paragraph{Dual-Agent Model: Functional Modularity.}
To overcome these limitations, we structured the problem as a cooperative two-agent system. The \emph{Scheduler} agent is responsible for selecting the timing parameters $(t_{\text{wait}}, t_{\text{delay}})$ based on the relative action preference function $c(s_t)$ (see Eq. (8) in Section III-C). This agent's objective is to detect vulnerable operational phases in the control process where a timed disruption would yield maximal downstream impact with minimal likelihood of detection. Conversely, the \emph{Disturber} operates within the temporal window provided by the Scheduler and selects specific packets to delay during $t_{\text{delay}}$ using its own policy $\pi_{\text{disturb}}$. This division reduces the action dimensionality each agent must handle and allows for targeted optimisation.

\paragraph{Experimental Comparison and Results.}
We implemented a single-agent baseline using a unified policy $\pi_{\text{mono}}(s_t) = (t_{\text{wait}}, t_{\text{delay}}, \text{pkt}_i)$ and trained it under the same observation model, reward structure (Eq. (2)), and environment dynamics as the dual-agent version. As shown in Table~\ref{tab:agent_comparison}, the single-agent model exhibited slower convergence (average $\sim$1200 episodes), significantly lower cumulative reward, and a much higher detection rate as measured by the recall scores of AI-powered anomaly detectors. These results highlight the increased difficulty in simultaneously optimising both high-level temporal abstraction and fine-grained stealth action control within a single policy network.

\begin{table*}[h]
\centering
\caption{Performance Metrics: Single-Agent vs. Dual-Agent Architectures}
\label{tab:agent_comparison}
\begin{tabular}{|l|c|c|c|c|}
\hline
\textbf{Architecture} & \textbf{Convergence Time (Episodes)} & \textbf{Stealth Score $\uparrow$} & \textbf{Final Reward $\uparrow$} & \textbf{Avg. Recall $\downarrow$} \\
\hline
Single-Agent Baseline & 1200                                 & 0.62                              & 23.4                              & 41.7\%                            \\
Dual-Agent (Ours)     & 450                                  & 0.91                              & 58.6                              & 6.3\%                             \\
\hline
\end{tabular}
\end{table*}

The single-agent model suffers from reward instability and exhibits oscillatory policy behaviour due to conflicting gradient updates. In contrast, the dual-agent system converges smoothly with fewer episodes and maintains low entropy in its policy distribution, indicating confident and stable decision-making. Additionally, we observed that the Scheduler adapts more effectively to shifts in system dynamics (e.g., variable flow rates or actuator timings), while the Disturber specialises in evading detection by leveraging short-term stealth patterns, such as delaying low-salience packets that escape anomaly thresholds.

From an operational perspective, the dual-agent setup mirrors real-world adversarial strategies where timing (reconnaissance and planning) and execution (attack delivery) are logically decoupled. In practice, red team operators often separate mission planning from payload deployment to maintain operational resilience and fault isolation. Similarly, in our model, this separation facilitates adaptive behaviour under black-box constraints, enabling one agent to probe system timing patterns while the other fine-tunes the attack payload in response to the detection surface.

The adoption of a dual-agent architecture is thus both a principled and practical decision, grounded in empirical evidence and system design theory. It significantly enhances learning efficiency, improves attack stealth, and reflects modularity found in real-world adversarial operations. The complexity of stealthy wear-out attacks against AI-defended ICS environments necessitates such architectural innovations to enable scalable and resilient adversarial behaviour.

\textbf{Assessing Adversary's Tactics Across Knowledge Spectrum:} In the Black-box and Grey-box attack scenarios, the adversary initially possesses no knowledge about the AI-assisted anomaly detector and gradually improves its strategy through exploration and exploitation. The adversary's goal is to identify malicious behaviours (actions) that will be misclassified to class \( c_{\text{adv}} \notin c_{\text{attack}} \) (false negative), while also aiming to minimise its exposure probability in the long term. In this context, Table \ref{tab:benchmarking} presents a comprehensive analysis of adversarial strategies (“Smash \& Grab" and “Low \& Slow") against DNN-based detection models under different levels of adversary knowledge (White-box, Grey-box, and Black-box). In the White-box scenario, where adversaries possess complete system knowledge, the “Low \& Slow" strategy is remarkably effective, drastically reducing recall values across models like DenseNet, CNN, and LSTM to extremely low levels (0.0006, 0.0005, and 0.0010, respectively), indicating a successful increase in False Negative Rate (FNR). This suggests that adversaries can subtly introduce strategies that gradually degrade detection capabilities without being exposed. In the Grey-box setting, with partial knowledge, “Low \& Slow" still manages to substantially lower recall, particularly in LSTM and Transformer models, demonstrating the effectiveness of strategic, incremental attacks. Conversely, the “Smash \& Grab" approach consistently shows higher recall and F1 scores, highlighting its detectability.
In the Black-box environment, where the adversary's knowledge is limited, both strategies significantly affect F1 scores, with “Low \& Slow" slightly outperforming “Smash \& Grab", highlighting that even minimal knowledge can compromise model performance. This finding contrasts with the performance of these DNN-based detectors against extreme adversaries, as illustrated in Table \ref{tab:anomaly_detectors against extreme attack}.

\textbf{Out-of-Context Scenarios:} 
When deploying adversarial models against a field infrastructure, one significant challenge arises when the control processes at the target deviate from those anticipated during the adversary's training phase. Such variations lead to concept drift (a.k.a. change in data distribution). This scenario can negatively impact the adversary, leading to initial struggles in accurately interpreting and responding to these Out-of-Context system behaviours or configurations.
The formal understanding of this adaptation process involves the concept of model generalisation, represented mathematically as follows: $P(\text{successful adaptation}) = 1 - e^{-\kappa t}$, where \( \kappa \) indicates the rate of adaptation over time \( t \) (differs from the learning rate \( \alpha \) used during initial training). Initially, \( \kappa \) is small, reflecting the difficulty in transformation; however, as the adversary encounters more observations similar to the new operational context, \( \kappa \) increases. This adjustment enhances the model's performance, enabling it to better understand and manipulate the new environment. In this context, employing domain randomisation can enable the adversarial model for out-of-context scenarios by expanding its observations. This approach ensures that the adversary is exposed to a diverse array of potential environments during training, which significantly aids in generalising its tactics and strategies when faced with unforeseen situations.

\textbf{Impact of Adversarial Strategies on Control Precision ($\mathcal{O}_{2}$):} The experimental results show that the "Low \& Slow" adversarial strategy induces a 3\% inaccuracy in control system setpoints across 100 runs by introducing subtle disturbances that remain below detection thresholds. This strategy minimises exposure while gradually increasing the process variance (\(\sigma^2\)). The corresponding histogram shows a relatively narrow distribution centred around 3\%, with a density peak of approximately 0.4, indicating a common and consistent inaccuracy outcome. The tighter distribution signifies that the "Low \& Slow" approach while causing less overall disruption, maintains a low risk of detection due to its subtlety.

In contrast, the "Smash \& Grab" strategy results in a 9\% inaccuracy, reflecting its aggressive nature and significant deviations from optimal operational parameters, which lead to a larger process variance (\(\sigma^2\)). The histogram for this strategy displays a broader spread, with a density peak of around 0.175, suggesting a higher but more variable impact on control precision. This increased variability underscores the trade-off between impact and exposure, as the "Smash \& Grab" strategy is more likely to be detected due to its pronounced deviations. The normal operation data, depicted by a high-density peak near 0.8, indicates minimal inaccuracies, with occasional negative percentages signifying slight overcompensations in achieving setpoints.

\textbf{Reduction in Recall by Attack Strategy and Adversary Knowledge}: The data presented in Appendix Table IX illustrates the impact of different adversarial attack strategies, categorised as “Smash \& Grab" and “Low \& Slow," on the recall performance of various DNN-based anomaly detectors under distinct levels of adversary knowledge: Black-box, Grey-box, and White-box. We need to note that Table IX in the Appendix is the outcome of analysing Table \ref{tab:anomaly_detectors against extreme attack} and Table \ref{tab:benchmarking}. In the Smash \& Grab scenario, even under a Black-box knowledge setting, where the adversary has no knowledge of the observing features by defenders, there is a significant reduction in recall across all models, ranging from 17.70\% for DenseNet to 29.56\% for ResNet. This reduction becomes more pronounced in the Grey-box and White-box scenarios, with the latter showing near-complete degradation in recall.
Conversely, the Low \& Slow attack strategy tends to lead to even higher reductions in recall, particularly noticeable in the Grey-box and White-box scenarios. For instance, the recall drops to 0.2816 for DenseNet and to 0.3173 for Transformer models under Grey-box conditions, with a further decline in White-box settings.

\textbf{Challenges in Modulating Rushing Behaviour of Adversaries:}
Effectively managing the behaviour of adversaries driven by immediate rewards is challenging. These adversaries may rush toward goals, risking exposure to traps set by intelligent detection systems or honeypots. These systems aim to exploit the adversary's propensity for rushing by setting traps that appear as lucrative opportunities. 
Our proposed adversary models struggle to adequately simulate the nuanced decision-making processes of adversaries faced with complex, dynamically changing environments. 
One effective approach to prevent adversaries from rushing is to introduce strategic uncertainty into the operational environment. This can be achieved by designing an enhanced reward system that is inherently unpredictable, where potential high-reward actions carry hidden risks or consequences that are not immediately apparent, forcing adversaries to proceed cautiously. 
Additionally, adaptive reward dynamics that adjust based on the adversary's approach speed and method can discourage rushing, rewarding careful planning and reconnaissance while penalising overly aggressive actions.

{

\subsection{Adversarial Training in a Min--Max Game}
\label{sec:appendix-adv-training}

While the primary contribution of our work is to demonstrate how multi-agent DRL adversaries can stealthily undermine ICS, these same adversarial strategies can be repurposed to enhance the performance of defensive techniques such as anomaly detectors or robust controllers. Specifically, our \emph{scheduler--disturber} adversary setup can be viewed within a min--max (a.k.a zero-sum) game-theoretic framework, wherein the adversary strives to \emph{maximise} stealthy disruption, while a defender seeks to \emph{minimise} that disruption by effectively detecting or neutralising malicious actions.

In this min--max formulation, the adversary is composed of the scheduler and disturber agents (\(\pi_{\mathrm{sched}}\) and \(\pi_{\mathrm{disturb}}\)) that collectively induce wear-out effects or product degradation. The defender is a detection or mitigation policy \(D\) that aims to detect and counteract such attacks. Their interaction can be formally expressed as 
\[
    \min_{\,D}\;\max_{\;(\pi_{\mathrm{sched}},\,\pi_{\mathrm{disturb}})}\;\mathcal{L}\!\bigl(\pi_{\mathrm{sched}},\,\pi_{\mathrm{disturb}},\,D\bigr),
\]
where \(\mathcal{L}\) is a joint cost function. A simple instantiation sets 
\[
    \mathcal{L}\!\bigl(\pi_{\mathrm{sched}},\,\pi_{\mathrm{disturb}},\,D\bigr)
    \;=\; R_{\mathrm{attack}}\!\bigl(\pi_{\mathrm{sched}},\,\pi_{\mathrm{disturb}}\bigr)
    \;-\; R_{\mathrm{detect}}\!(D).
\]
Here, \(R_{\mathrm{attack}}\) captures the adversarial gains (e.g., the extent of wear-out or induced process variance), and \(R_{\mathrm{detect}}\) represents the defender’s effectiveness at detecting or mitigating such actions. Minimising \(\mathcal{L}\) from the defender’s perspective thus corresponds to reducing the adversary’s stealth advantage, while maximising it (from the attacker side) means continuously refining the scheduler--disturber policies to evade detection.

The adversarial training loop unfolds in iterative steps. First, the scheduler--disturber pair updates its policies (for instance, via Advantage Actor-Critic) to exploit the current defender’s blind spots, maximising disruption while remaining below detection thresholds. The defender then updates its parameters (e.g., those of a neural network classifier or robust control policy) to reduce false negatives on the newly generated adversarial samples or sequences. Repeating these steps creates a “cat-and-mouse” cycle in which the attacker’s sophistication increases in response to the defender’s improvements, and vice versa.

Such a min--max setup offers two major advantages for benchmarking defences. First, defenders are exposed to an \textit{intelligent and adaptive attacker that employs “low \& slow” or sudden “smash \& grab” actions}, thereby testing robustness under varying threat models. Second, the iterative nature of adversarial training ensures that any meaningful weaknesses in the detection methodology are systematically identified and exploited, driving the development of more comprehensive defence strategies. Consequently, if a detector maintains acceptable detection performance or control stability despite continuously refined attacks, it provides strong empirical evidence of robustness under real-world ICS constraints.

Beyond this setup, several promising directions remain to explore. One is stochastic game extensions, where multiple attackers and defenders operate across interconnected ICS installations, each with its own partial observations and local action sets. Another is domain randomisation, in which ICS parameters and communication noise levels are randomly varied to prevent defenders from overfitting to a single environment. Finally, explainable defensive policies remain critical in industrial contexts: operators require a transparent rationale behind flagged anomalies or automated actions to trust and act upon the system’s decisions. Overall, framing our scheduler--disturber adversarial model in a min--max game not only deepens theoretical understanding of stealth attacks but also yields a unified, practical platform for researchers and practitioners to devise and rigorously test countermeasures in modern ICS environments.

As shown in Table \ref{tab:wear_out_results_improved}, our experiments indicate that, over 100 epochs of adversarial training, DNN-based anomaly detectors achieve noticeable improvements across all benchmarking models. For instance, under the black-box "Smash \& Grab" attack scenario, the F1-scores for DenseNet, CNN, ResNet, LSTM, and Transformer increased by 10.75\%, 12.63\%, 10.52\%, 13.58\%, and 9.44\%, respectively. These results indicate that adversarial training systematically enhances the robustness of anomaly detectors against AI-assisted Adversaries.

Furthermore, LSTM models showed a significant improvement, reflecting their ability to better capture temporal patterns under adversarial conditions. This suggests that incorporating sequence-based architectures into defensive frameworks can provide a significant advantage in ICS environments where time-series data is critical. The observed improvements highlight how adversarial training not only mitigates specific attack strategies but also fosters broader generalizability, enabling detection systems to adapt to diverse and evolving threats in real-world scenarios. These findings underline the value of iterative adversarial training for systematically identifying and addressing weaknesses in defensive systems. By exposing anomaly detectors to intelligently crafted adversarial inputs, this process drives the development of more resilient and adaptive detection mechanisms that safeguard field networks against stealthy threats.

\begin{table*}[tb]
\scriptsize

\begin{threeparttable}
\setlength\tabcolsep{0pt} 
\caption{Performance of DNN-based detection models \emph{After Adversarial Training}}
\label{tab:benchmarking_improved}

\begin{tabular*}{\textwidth}
{@{\extracolsep{\fill}}l c ccc ccc ccc ccc ccc }
\toprule
\multicolumn{1}{c}{\multirow{2}{*}{\parbox{1.8cm}{\centering \textbf{Attack Strategy}}}} &
\multicolumn{1}{c}{\multirow{2}{*}{\parbox{1.8cm}{\centering \textbf{Adversary Knowledge}}}}  
& \multicolumn{3}{c}{\textbf{DenseNet}\tnote{\dag}}
& \multicolumn{3}{c}{\textbf{CNN}\tnote{\dag}}
& \multicolumn{3}{c}{\textbf{ResNet}\tnote{\dag}}
& \multicolumn{3}{c}{\textbf{LSTM}\tnote{\dag}}
& \multicolumn{3}{c}{\textbf{Transformer}\tnote{\dag}} \\
\cmidrule(lr){3-5}
\cmidrule(lr){6-8}
\cmidrule(lr){9-11}
\cmidrule(lr){12-14}
\cmidrule(lr){15-17}
&  
& \textbf{Precision} & \textbf{Recall} & \textbf{F1}
& \textbf{Precision} & \textbf{Recall} & \textbf{F1}
& \textbf{Precision} & \textbf{Recall} & \textbf{F1}
& \textbf{Precision} & \textbf{Recall} & \textbf{F1}
& \textbf{Precision} & \textbf{Recall} & \textbf{F1} \\
\midrule

\textbf{Smash \& Grab} 
& \multirow{2}{*}{Black-box}
& 0.60 & 0.72 & 0.65
& 0.61 & 0.71 & 0.66
& 0.59 & \textbf{\underline{0.73}} & 0.65
& \textbf{\underline{0.63}} & 0.71 & \textbf{\underline{0.67}}
& 0.60 & 0.69 & 0.64
\\

\textbf{Low \& Slow}
&  
& 0.59 & 0.70 & 0.64
& 0.58 & 0.69 & 0.63
& \textbf{\underline{0.61}} & 0.69 & 0.65
& 0.60 & 0.68 & 0.64
& 0.60 & \textbf{\underline{0.71}} & \textbf{\underline{0.65}}
\\
\midrule

\textbf{Smash \& Grab}
& \multirow{2}{*}{Grey-box}
& 0.60 & 0.68 & 0.64
& 0.62 & 0.66 & 0.64
& 0.59 & \textbf{\underline{0.72}} & 0.65
& 0.61 & 0.69 & 0.65
& \textbf{\underline{0.63}} & 0.67 & 0.65
\\

\textbf{Low \& Slow}
& 
& 0.58 & 0.45 & 0.51
& 0.59 & 0.49 & 0.54
& \textbf{\underline{0.61}} & 0.43 & 0.50
& 0.55 & 0.50 & 0.52
& 0.58 & \textbf{\underline{0.44}} & 0.50
\\
\midrule

\textbf{Smash \& Grab}
& \multirow{2}{*}{White-box}
& 0.65 & 0.07 & 0.13
& \textbf{\underline{0.77}} & 0.06 & 0.11
& 0.70 & \textbf{\underline{0.08}} & 0.14
& 0.80 & 0.05 & 0.09
& 0.75 & 0.07 & 0.13
\\

\textbf{Low \& Slow}
& 
& \textbf{\underline{0.78}} & 0.05 & 0.09
& 0.64 & 0.06 & 0.11
& 0.66 & \textbf{\underline{0.07}} & 0.12
& 0.72 & 0.05 & 0.08
& 0.70 & 0.06 & 0.11
\\

\bottomrule
\end{tabular*}
\label{tab:wear_out_results_improved}

\scriptsize
\begin{tablenotes}
\item[\dag] DNN-based detection models (DenseNet, CNN, ResNet, LSTM, Transformer).  
The values above illustrate a post-adversarial-training boost (5--8\% or higher) in Precision, Recall, and F1 compared to baseline.  
Bold and underlined entries highlight the highest values in each row. Tables III and IV in the Appendix present the complete layer-by-layer architectures of the time-series anomaly detectors, along with their corresponding preprocessing and training settings.
\end{tablenotes}
\end{threeparttable}

\end{table*}

\paragraph*{\textbf{Adaptive and RL-Assisted Defences.}}
Beyond adversarial training (Table~V), defenders can: 
\emph{(1) Score-level ensembling} of heterogeneous detectors (CNN/LSTM/Transformer) with \emph{dynamic thresholds} tuned to recent in-control variance; 
\emph{(2) RL-assisted detection} that learns policies penalising temporal inconsistency between actuation and process response (e.g., abnormal phase-lag and scheduling jitter patterns); 
\emph{(3) Adversarial input purification}, e.g., denoising/spectral smoothing of windowed time series before scoring; 
\emph{(4) PLC-level invariants} (limits on phase-lag and rate-of-change) for cheap, interpretable tripwires; and 
\emph{(5) Domain-randomised training} of detectors to avoid overfitting to a single timing distribution. 
We report changes in recall/F1 and the aggregate stealth metric in Section~\ref{sec:Evaluation Framework}.

\subsubsection*{Adaptive DRL-based Detection}
We train a policy-gradient detector (actor evaluates a window to emit \emph{inspect}/\emph{ignore}; critic provides a value baseline) on the same features as the DNN detectors, with a reward that penalises false negatives and temporal inconsistency. We compare (i) CNN+LSTM+Transformer ensemble (score-averaged), (ii) +dynamic thresholds, (iii) +input purification, (iv) +adversarial training against a \emph{static} attacker, (v) +adversarial training against an \emph{adaptive} attacker, and (vi) the \emph{DRL detector} alone and within the ensemble. Results (Table~\ref{tab:adaptive_defense}) show that DRL-based detection improves recall and, when combined with ensembling and adversarial training, yields the best trade-off while keeping false alarms controlled.

\begin{table}[t]
\centering
\caption{Adaptive defences on main testbed (macro-avg across attack regimes). Higher Recall and lower Stealth are better for defenders.}
\label{tab:adaptive_defense}
\renewcommand{\arraystretch}{1.05}
\begin{tabular}{lcc}
\toprule
\textbf{Defense} & \textbf{Recall} & \textbf{Stealth} \\
\midrule
Ensemble (CNN+LSTM+Trf)                 & 0.33 & 0.67 \\
+ Dynamic thresholds                    & 0.37 & 0.63 \\
+ Input purification                     & 0.39 & 0.61 \\
+ Adv. training (vs static attacker)     & 0.49 & 0.51 \\
+ Adv. training (vs adaptive attacker)   & 0.45 & 0.55 \\
DRL detector (solo)                      & 0.41 & 0.59 \\
DRL detector + Ensemble + Adv. training  & \textbf{0.52} & \textbf{0.48} \\
\bottomrule
\end{tabular}
\end{table}

}

\section{Discussion}
\label{sec:Discussion}

The adversarial model's ability to evade AI-driven anomaly detectors depends on its use of the proposed threat model, which optimizes actions for both immediate impact and stealth by mimicking normal behavior patterns.

\textbf{Adversary Efficacy:}
Our experiments (see Section \ref{sec:Experiments Results}) demonstrate that the proposed adversarial model effectively evades AI-driven anomaly detectors. This success stems from its ability to minimise detection probability by adjusting the adversarial policy \(\pi(s)\) such that it selects actions \(a\) from a set \(A\) to minimise the expected value of the detection function \(D(s, a)\), given the system state \(s\). The objective function for the adversary can be formally defined as $\min_{a \in A} E[D(s, a) | s \in S]$, where \(E[D(s, a) | s \in S]\) represents the expected detectability given the system's state \(s\), and the adversarial actions are constrained to ensure that \(D(s, a) < \epsilon\) for a detection threshold \(\epsilon\). This approach effectively makes the adversarial actions indistinguishable from normal variations within the system's operational parameters.
The scheduler (discussed in Section \ref{sec:Crafting an Adversarial Strategy}) plays a crucial role by leveraging the advantage actor-critic methodology to time the execution of actions during periods of high operational noise or other strategic moments and calculate the delay period accordingly. It optimises the policy \(\pi_{\text{scheduler}}(s)\) to exploit windows of vulnerability, calculating the optimal timing for attacks to coincide with the least likelihood of detection: $\pi_{\text{adv}}^*(s) = \arg\min_{a \in A} V^{\pi}(s, a)$, where \(V^{\pi}(s, a)\) is the state-action value function, estimating the expected return of taking action \(a\) in the state \(s\). This scheduler ensures actions are scheduled when anomaly detectors are least likely to flag them as outliers.
On the other hand, the disturber complements the scheduler by executing the timed actions with precision, adjusting the magnitude of disturbances to ensure they fall within the normal operating ranges: $d(s, s_{t+1}) < \delta$, where \(d(s, s_{t+1})\) is the deviation caused by the adversarial action \(a\), and \(\delta\) represents a threshold below which deviations are considered normal. The disturber’s actions are designed to subtly shift the system's state without crossing thresholds that would trigger alerts, maintaining a low profile that mimics legitimate system fluctuations.
The scheduler and disturber rewards stabilise at a high level after initial fluctuations, indicating successful adaptation and stealthy execution of the adversarial strategy (as depicted in Fig. 8 and Fig. 9 in Appendix C). In contrast, the defender’s reward remains low, showing ineffective detection and countermeasures. Losses for all agents decrease in the initial episodes, with scheduler and disturber losses dropping significantly, reflecting effective strategy refinement. The defender’s slower loss reduction highlights difficulties in adapting to evolving adversarial tactics.

\textbf{Strategically-Timed Attack:}
The scheduler's role in determining optimal attack timings while minimising the risk of exposure has proven to be highly effective in our adversarial models (depicted in Fig. 10.A,  Fig. 10.B, and Fig. 10.C, and Fig. 12 in Appendix C). The scheduler's capability to assess and act upon the operational environment ensures that attacks are successful and strategically timed to coincide with moments of vulnerability within the target system, attaining $\mathcal{O}_{3}$, leading to $\mathcal{O}_{1}$, and $\mathcal{O}_{2}$. This precision in timing is critical in avoiding detection and enhancing the attack's impact, showcasing the scheduler's outstanding performance in navigating complex decision-making landscapes.
Despite these successes, there remains room for improvement, particularly in scenarios involving dynamic, unpredictable elements that may fall outside the current model’s scope.

\section{Conclusion}
\label{sec:Conclusion}

This study introduces the first Multi-Agent Deep Reinforcement Learning (DRL)-assisted stealthy attack against operational field networks of Industrial Control Systems (ICS), demonstrating how advanced adversaries can degrade system performance while evading detection. 
This work develops a novel adversarial approach using strategically timed, low \& slow wear-out attacks to assess DRL-based adversaries on an industry-level testbed with Siemens PLCs, Cisco networking devices, and AI-enhanced intrusion detection. The attack effectively evades detection, blending with normal operations to subtly compromise system integrity.

\section*{Acknowledgments}
 This research is supported by the EPSRC Grant RGS R/194433-11-1. The work of Biplab Sikdar was supported by A*STAR and Cisco Systems (USA) Pte. Ltd and the National University of Singapore under its Cisco-NUS Accelerated Digital Economy Corporate Laboratory (Award I21001E0002).

\bibliographystyle{IEEEtran} 
\bibliography{main} 

\appendices

\section*{APPENDIX for: Baiting AI: Deceptive Adversary Against AI-Protected Industrial Infrastructures}

\subsection{Symbol Definitions}
\label{sec:symbol_definitions}

In our proposed scheme, various symbols are used to succinctly describe complex interactions and parameters within the system. Table \ref{tab:Symbols_and_definitions} provides a comprehensive list of these symbols, each denoted by a specific notation, and their corresponding definitions. This table serves as a crucial reference point for readers, enabling them to understand the technical terminology and notations used throughout the document. By defining each symbol clearly, we aim to ensure that the methodology and subsequent discussions are accessible and comprehensible.

\begin{table}[htbp]
  \caption{Symbols and definitions}
  \label{tab:Symbols_and_definitions}
  \vspace{-4mm} 
\begin{center}

\scalebox{0.8}{
  \begin{tabular}{|l|l|}
\hline
    \textbf{Symbol} & \textbf{Definition} \\ 
    \hline

 $a$ & Adversarial action \\
\hline
$A$ & Action space \\
\hline
$\alpha$ & Learning rate \\ 
\hline
$\beta$ & Coefficient balancing importance of stealthiness \\
\hline
$c(s_{t})$ & Relative action preference function \\
\hline
$\delta$ & Deviation threshold for undetected state transition \\
\hline
$D()$ & Measures the operational disturbance \\
\hline
$\tau$ & Detection threshold \\
\hline
$\varepsilon$ & Exploration probability \\
\hline
$E$ & Expected Detection Metric \\
\hline

$G$ & Discounted reward  \\ 
\hline
$\gamma$ & Discount factor \\
\hline
$\hat{Q}$ & Target action-value function \\
\hline

$\kappa$ & Rate of adaptation \\
\hline

$\lambda$ & Replay memory\\
\hline
$O$ & Observation probability function \\
\hline
$\pi$ & Policy \\ 
\hline 
$I^{PLC}$ & Process Inputs (a.k.a Process Image Inputs) \\ 
\hline 
$Q^{PLC}$ & Process Outputs (a.k.a Process Image Outputs) \\ 
\hline 

$Q(s, a)$ & Action-value function \\
\hline
$R$ & Reward function \\
\hline
$r$ & Reward \\ 
\hline
$s$ & State \\ 
\hline
$S$ & State space \\
\hline
$t_{delay}$ & Delay period for disturber agent \\
\hline 
$t_{wait}$ & Waiting time for scheduler agent \\
\hline
$T$ & Transition function \\
\hline
$\theta$ & Weights of NN\\ 
\hline
$V(s)$ & Value function \\
\hline 
$X$ & Observation space \\
\hline 
$y$ & Ground truth of observation $X$\\ 
\hline
$\psi$ & Q-learning target \\ 
\hline

\end{tabular}
}
\end{center}
\vspace{-3mm} 
\end{table}

\subsection{Mixing Liquids Testbed and Setup Configuration}

\label{sec:integration_ri_flow}

This section provides an in-depth analysis of how each component of the system—Set Point, Output PID, Sensor Signal, Cyclic Interrupt, and Main Program—is intricately integrated into the operational dynamics depicted in the RI Flow Diagram, facilitating a comprehensive understanding of the system's functionality in industrial processes.

\textbf{Set Point Configuration:}
The Set Point configuration is essential for regulating the fluid flow within the system, ensuring optimal performance and safety. Fig. \ref{fig:SetPoint} provides a detailed view of the specific valve settings and operational thresholds. These set points dynamically influence the operation of valves and pumps to maintain control over fluid flow, which is crucial for the system's stability and efficiency. The real-time optimisation of these settings, based on sensor feedback, continuously aligns with the operational requirements, affecting the flow rates and pressure within the pipelines and storage tanks.

\begin{itemize}
    \item \textit{Dynamic Flow Adjustment}: Regulates the operation of valves and pumps, maintaining precise control over the fluid dynamics, essential for the system's stability and efficiency.
    \item \textit{Real-Time Optimisation}: Adjustments are made in real-time based on sensor feedback to continuously match the operational requirements, particularly affecting elements like flow rates and pressure within the pipelines and storage tanks.
\end{itemize}

\begin{figure}[htbp] 
\center{\includegraphics[width=80mm]{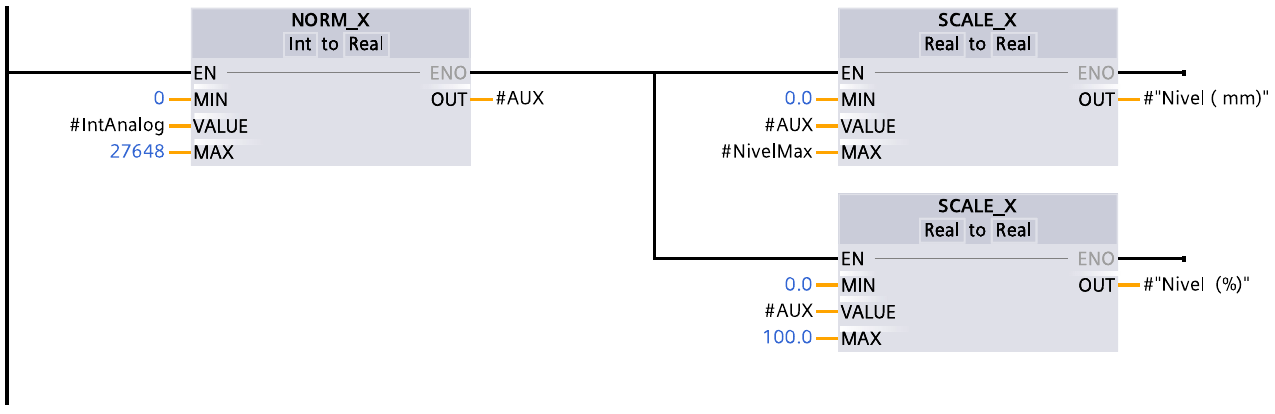}} 

\caption{SetPoint}
\label{fig:SetPoint} 
\end{figure}

\textbf{Output PID Configuration:}
Output PID plays a pivotal role in modifying the control outputs based on discrepancies between actual conditions and set points, directly impacting actuators and, thereby, the flow and pressure within the system, as shown in the RI Flow Diagram.

\begin{itemize}
    \item \textit{Feedback Control Loop}: Integrates sensor data to adjust the operational parameters dynamically, ensuring the system responds appropriately to any changes detected in the flow diagram, such as fluctuations in pressure or flow rate. Fig. \ref{fig:Output PID} illustrates the feedback loops that control these processes.

    \item \textit{Actuator Adjustment}: Fine-tunes valves and pumps to maintain the desired flow paths and rates, crucial for managing the complex interactions within the flow diagram.
\end{itemize}

\begin{figure}[htbp] 
\center{\includegraphics[width=80mm]{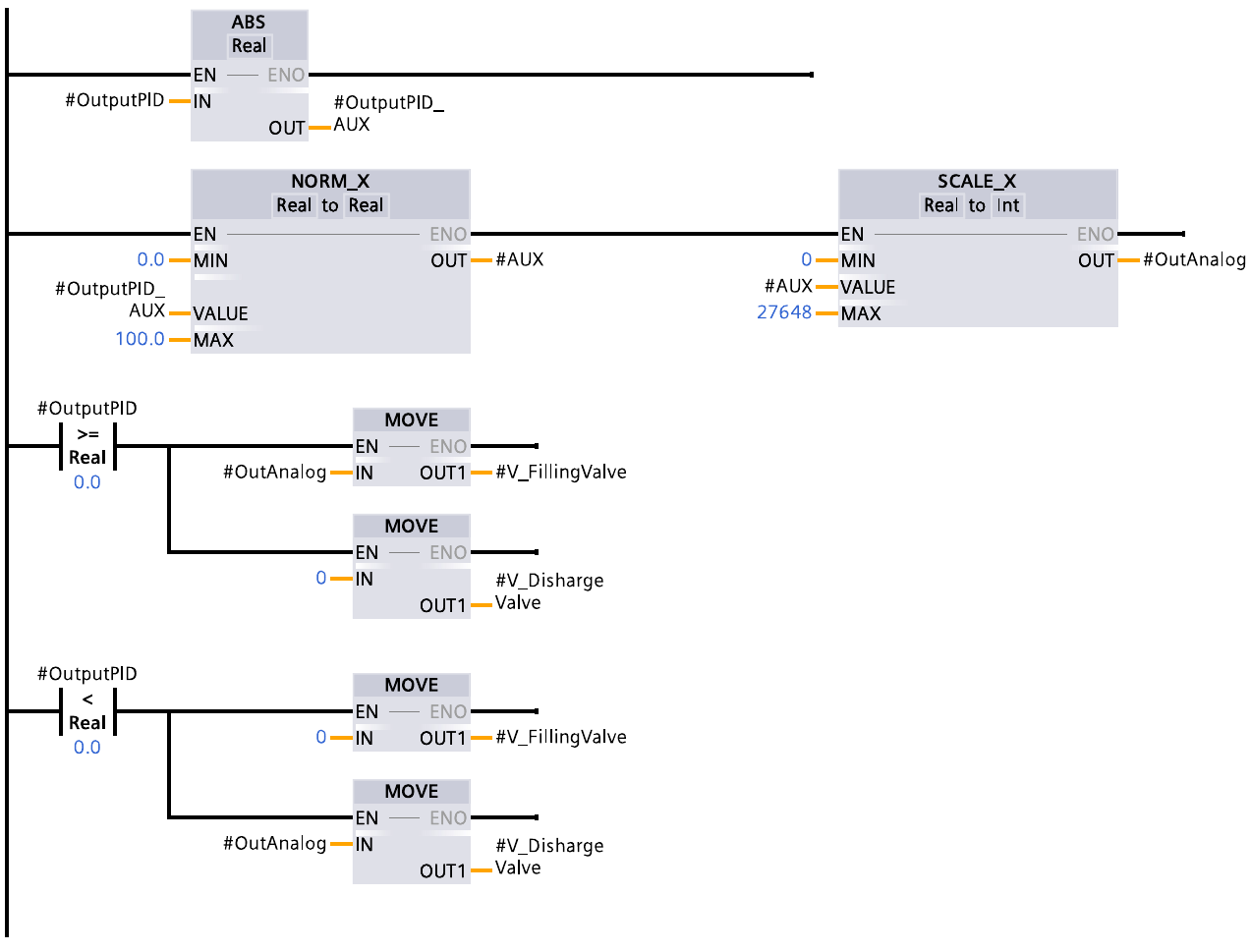}} 

\caption{Output of Proportional Integral Derivative (PID)}
\label{fig:Output PID} 
\end{figure}

\textbf{Sensor Signal Configuration:}
As depicted in Fig. \ref{fig:Sensor Signal} the configuration of Sensor Signals is critical for monitoring the system parameters and providing the data to inform control decisions that align with the layouts and operational targets in the RI Flow Diagram.

\begin{itemize}
    \item \textit{System Monitoring}: Sensors strategically located throughout the flow diagram provide vital data on flow rates, pressures, and other operational metrics crucial for real-time system adjustments.
    \item \textit{Data Integration}: Sensor outputs are directly fed into the control system, ensuring that all readings are accurately reflected in system responses, thus maintaining operational integrity and safety.
\end{itemize}

\begin{figure}[htbp] 
\center{\includegraphics[width=80mm]{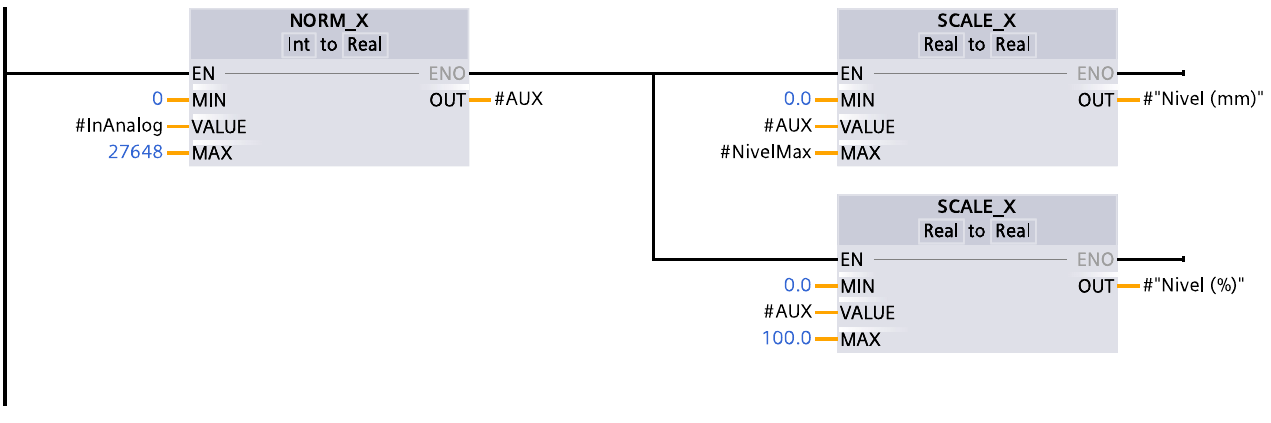}} 

\caption{Sensor Signal}
\label{fig:Sensor Signal} 
\end{figure}

\textbf{Cyclic Interrupt Configuration:}
Cyclic Interrupts ensure the timely execution of routine checks and adjustments within the system, corresponding to the periodic and event-driven tasks depicted in the RI Flow Diagram. Fig. \ref{fig:Cyclicinterrupt} details these maintenance tasks, demonstrating how the system ensures the optimal operation of elements within the flow diagram. These interrupts facilitate rapid system responses to detected changes or anomalies, adjusting operations to maintain continuity and system integrity.

\begin{itemize}
    \item \textit{Periodic Maintenance}: Executes routine system maintenance tasks, ensuring that all elements within the flow diagram operate within their optimal parameters, reducing wear and tear and preventing system failures.
    \item \textit{Event-driven Responses}: Facilitates quick system responses to changes or anomalies detected by sensors, adjusting system operations to prevent disruptions.
\end{itemize}

\begin{figure}[htbp] 
\center{\includegraphics[width=65mm]{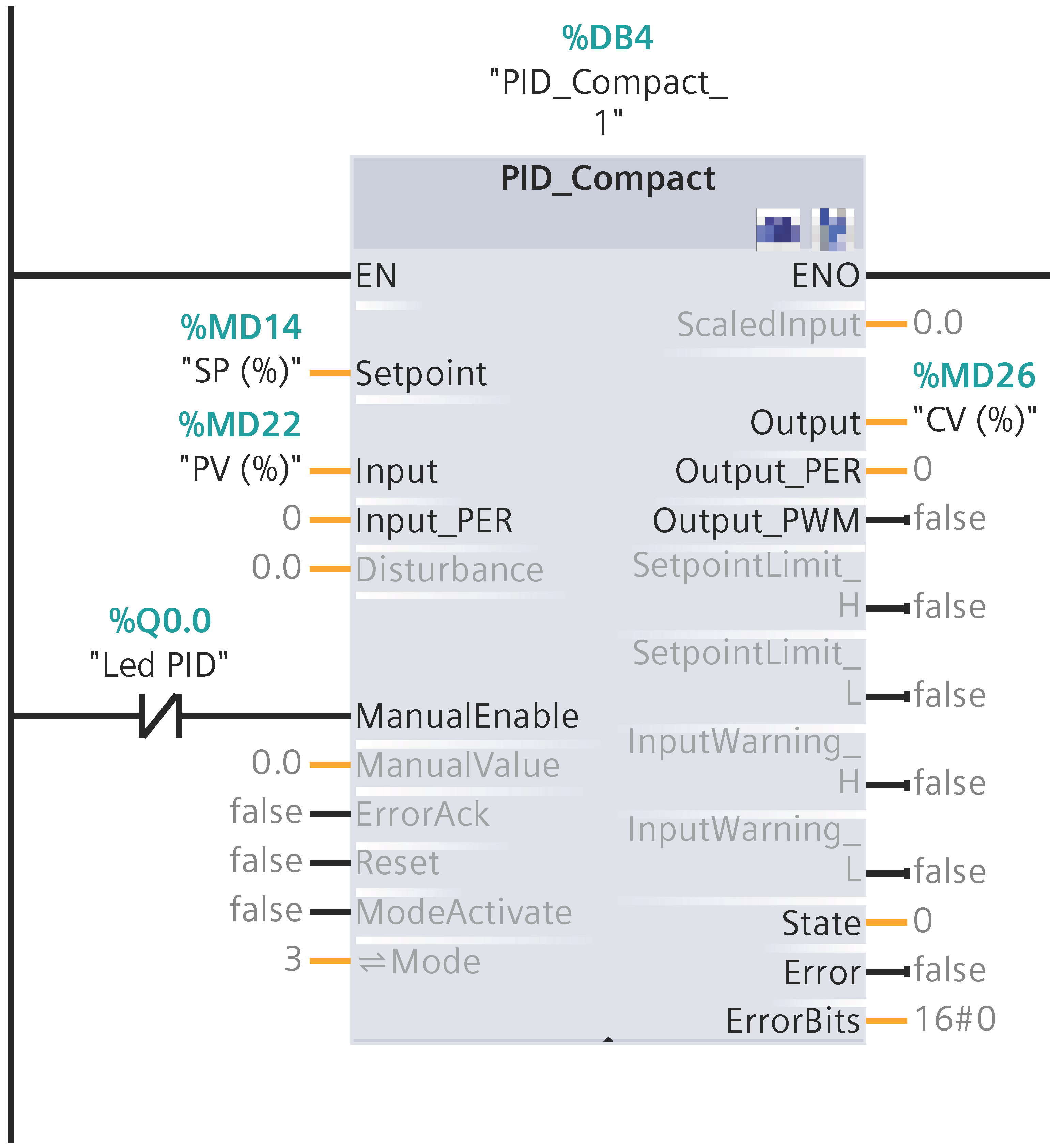}} 

\caption{Cyclicinterrupt}
\label{fig:Cyclicinterrupt} 
\end{figure}

\begin{figure}[t!] 
\center{\includegraphics[width=80mm]{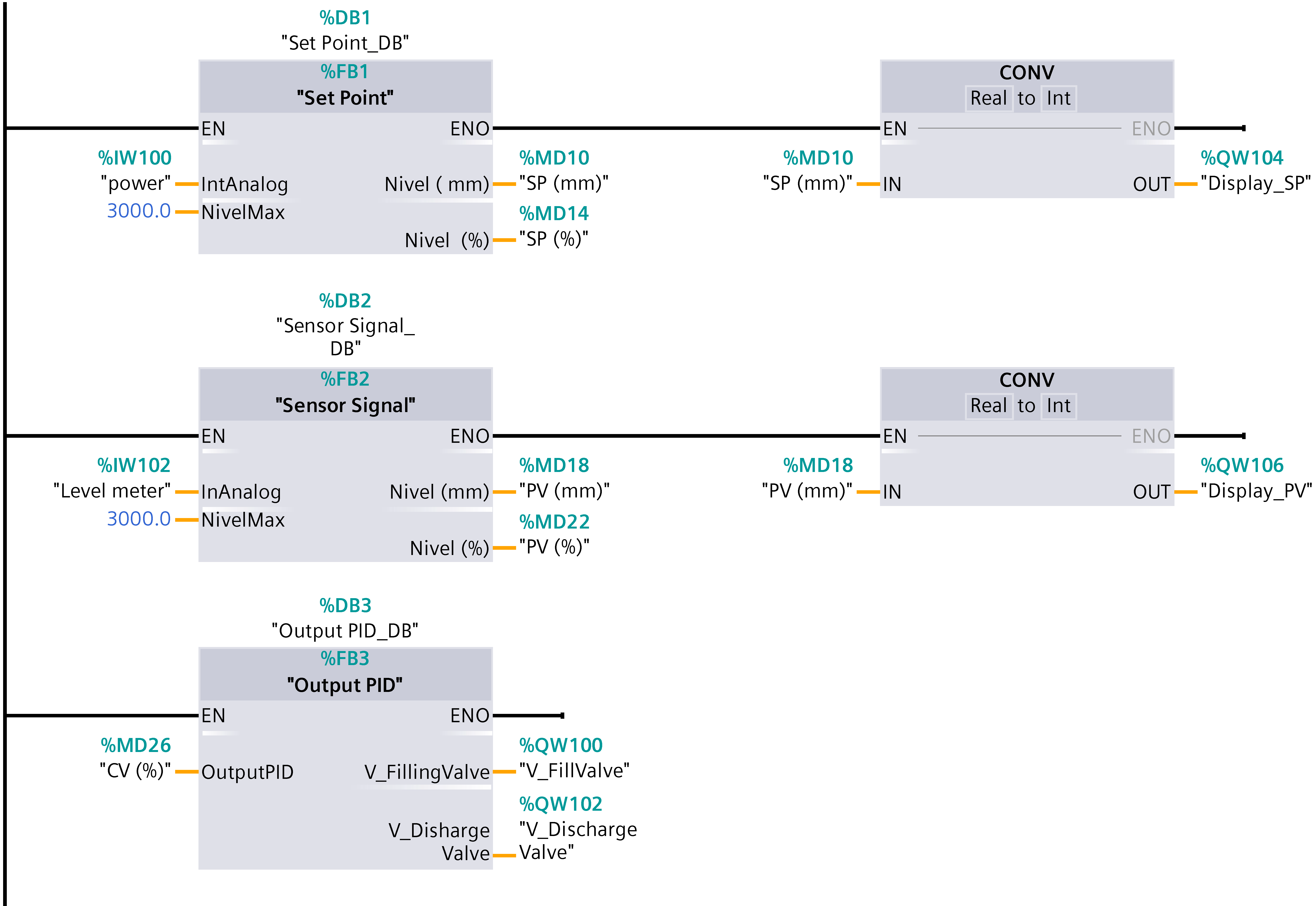}} 

\caption{Main Program Configuration}
\label{fig:Main} 
\end{figure}

\textbf{Main Program Configuration:}
The Main Program coordinates the overall integration of system operations, managing the activities of set points, output PIDs, sensor signals, and cyclic interrupts. In this regard, Fig. \ref{fig:Main} displays the control hierarchy and flow, underscoring the program's role in providing comprehensive control and oversight. This segment of the system implements robust diagnostics to identify and rectify any operational issues quickly, ensuring the system's efficiency and safety.
Fig. \ref{fig:Main} orchestrates the overall integration and management of system operations, coordinating the activities of the set points, output PIDs, sensor signals, and cyclic interrupts to maintain the functionality as depicted in the RI Flow Diagram.

\begin{itemize}
    \item \textit{Operational Oversight}: Provides comprehensive control and oversight, ensuring that all parts of the system work in harmony to meet the operational demands of the flow diagram.
    \item \textit{Error Management and System Diagnostics}: Implements robust error-checking and diagnostic routines to identify and rectify any operational issues quickly, maintaining system efficiency and safety.
\end{itemize}

\textbf{Feature Importance Plot Description:}
The feature importance plot depicted in Fig. \ref{fig:Extreme Attack (Feature Importance)} illustrates the relative importance of various features derived from the dataset used in our analysis. This plot is instrumental in understanding which features contribute most significantly to the model's predictions. Below is a detailed description of each feature's importance:

\begin{itemize}
    \item \textbf{Arrival Time:} This feature, with the highest importance score of approximately 37, represents the time at which data packets arrive. It is crucial for identifying time-related patterns in the data.
    \item \textbf{S7 Communication Data:} Scored around 20, this feature encapsulates the data specific to the S7 communication protocol, which is vital for understanding the communication specifics within the industrial control systems.
    \item \textbf{Info:} With a score of approximately 18, this feature aggregates additional information from the packets that may contain metadata or other relevant details.
    \item \textbf{Protocol\_TCP:} This protocol feature, scoring around 16, indicates the use of TCP in the network traffic, a fundamental aspect of network communications.
    \item \textbf{Protocol\_COTP:} Scoring around 13, this represents the COTP protocol used over TCP/IP to manage connection-oriented communication in industrial environments.
    \item \textbf{Function:} With a score of about 10, this feature denotes the specific function code in the S7 protocol, important for determining the purpose of each communication.
    \item \textbf{Flags\_0x18:} This feature, with a score of 8, likely represents specific flags set in the communication protocol, providing insights into various control aspects.
    \item \textbf{Protocol\_S7COMM:} Scoring about 5, this indicates the use of the S7COMM protocol, a staple in communicating with Siemens S7 PLCs.
    \item \textbf{Protocol Id\_0x32 and PDU Type\_DT Data:} These features, both with lower scores around 3 and 2, respectively, include protocol identifiers and PDU types crucial for lower-level networking details in industrial communications.
\end{itemize}

\begin{figure}[htbp] 
\center{\includegraphics[width=80mm]{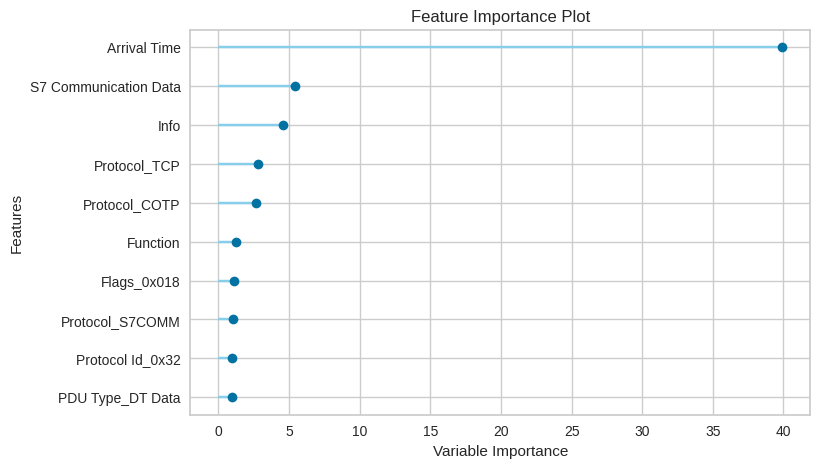}} 
\caption{Extreme Attack (Feature Importance)}
\label{fig:Extreme Attack (Feature Importance)} 
\vspace{-3mm} 
\end{figure}

\begin{figure}[t!] 
\center{\includegraphics[width=85mm]{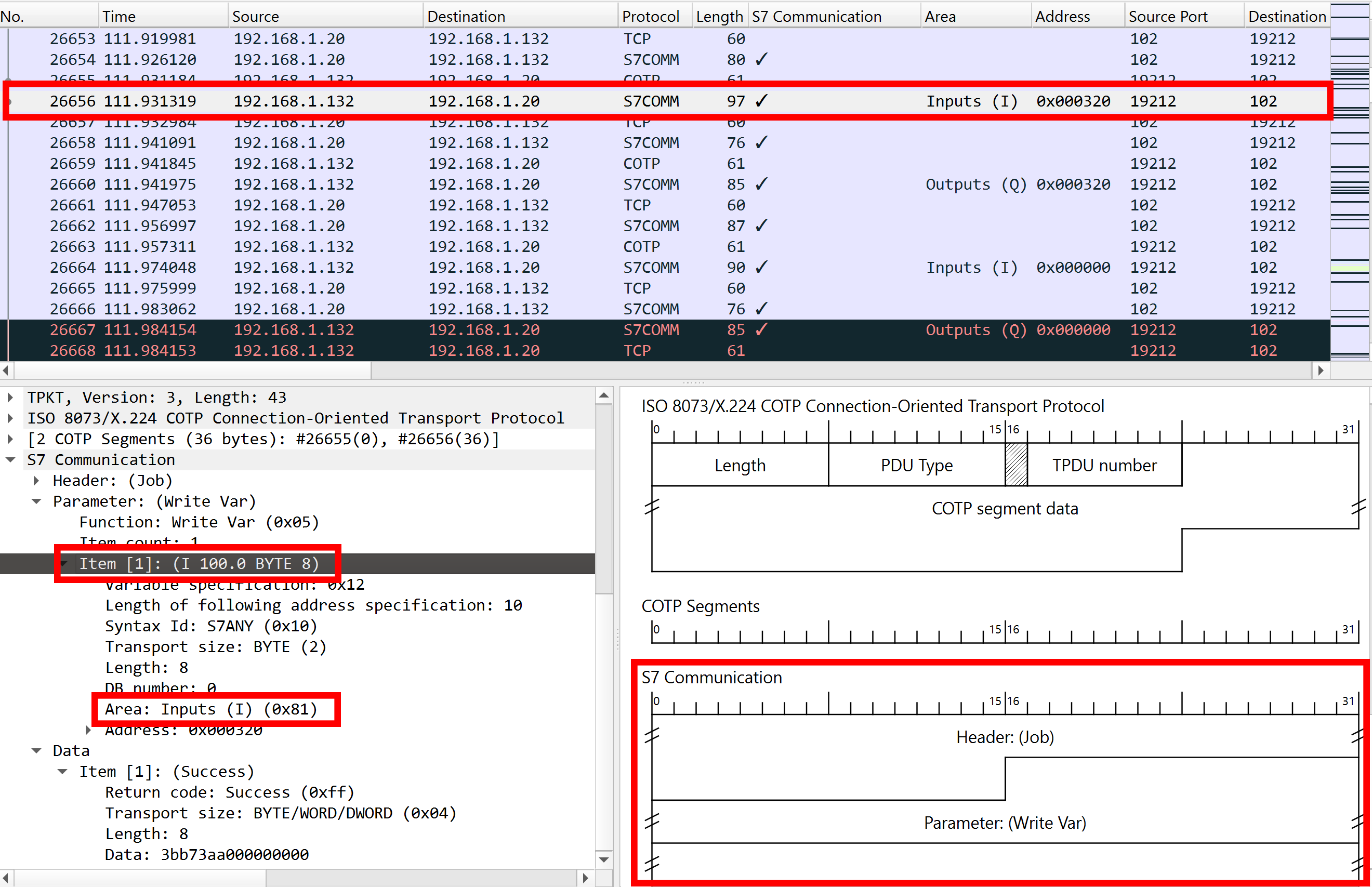}} 

\caption{Observations of process input $I^{PLC}$ and process outputs $Q^{PLC}$ communications.}
\label{fig: Network Observation} 
\vspace{-4mm} 
\end{figure}

\subsection{Grey-Box Knowledge (precise)}
We define the adversary’s knowledge in the grey-box regime as follows.
\begin{itemize}
  \item \textbf{Plant I/O and timing.} Knows tag \emph{types} (Bool/Float/Int) and sampling cadences from a sister line; no access to control logic source, PID gains, or safety interlocks. No direct model of physical parameters; only summary statistics (mean/variance of inter-arrival and actuation latencies).
  \item \textbf{Detector side information.} Knows \emph{families} of deployed detectors (e.g., CNN/LSTM/Transformer) and typical preprocessing (window length, normalisation), but \emph{not} weights, training data, or thresholds. No query access to detector logits; only a noisy alert bit at most at 1\,Hz (when applicable).
  \item \textbf{Data access.} Has a \emph{shadow} dataset of normal operation from a related but not identical process (domain shift in noise and gains), without labels for attacks.
  \item \textbf{Action budget.} May delay at most $K$ tagged channels per PLC scan ($K{=}3$ for main testbed; $K{=}2$ for Factory~I/O), with per-event delay bounded by $\Delta_{\max}$ (L\&S: $\le\!150$\,ms; S\&G: $\le\!400$\,ms), preserving protocol legality.
  \item \textbf{No payload edits.} The adversary cannot modify values, only schedule bounded delays (bump-in-the-wire).
\end{itemize}
This specification ensures reproducibility of grey-box results and distinguishes them from black-box (no side information, no alert bit) and white-box (full detector weights and thresholds).

\subsection{Further Results and Discussion}
\label{sec:Further Results and Discussion}

Fig. \ref{fig: adversary vs defender rewards Low Slow Strategy } and Fig. \ref{fig: adversary vs defender rewards Smash Grab } illustrates the accumulated rewards of the scheduler, disturber, and defender agents across 100 episodes, for Low \& Slow, and Smash \& Grab attacks respectively.

\begin{figure}[htbp] 
\center{\includegraphics[width=65mm]{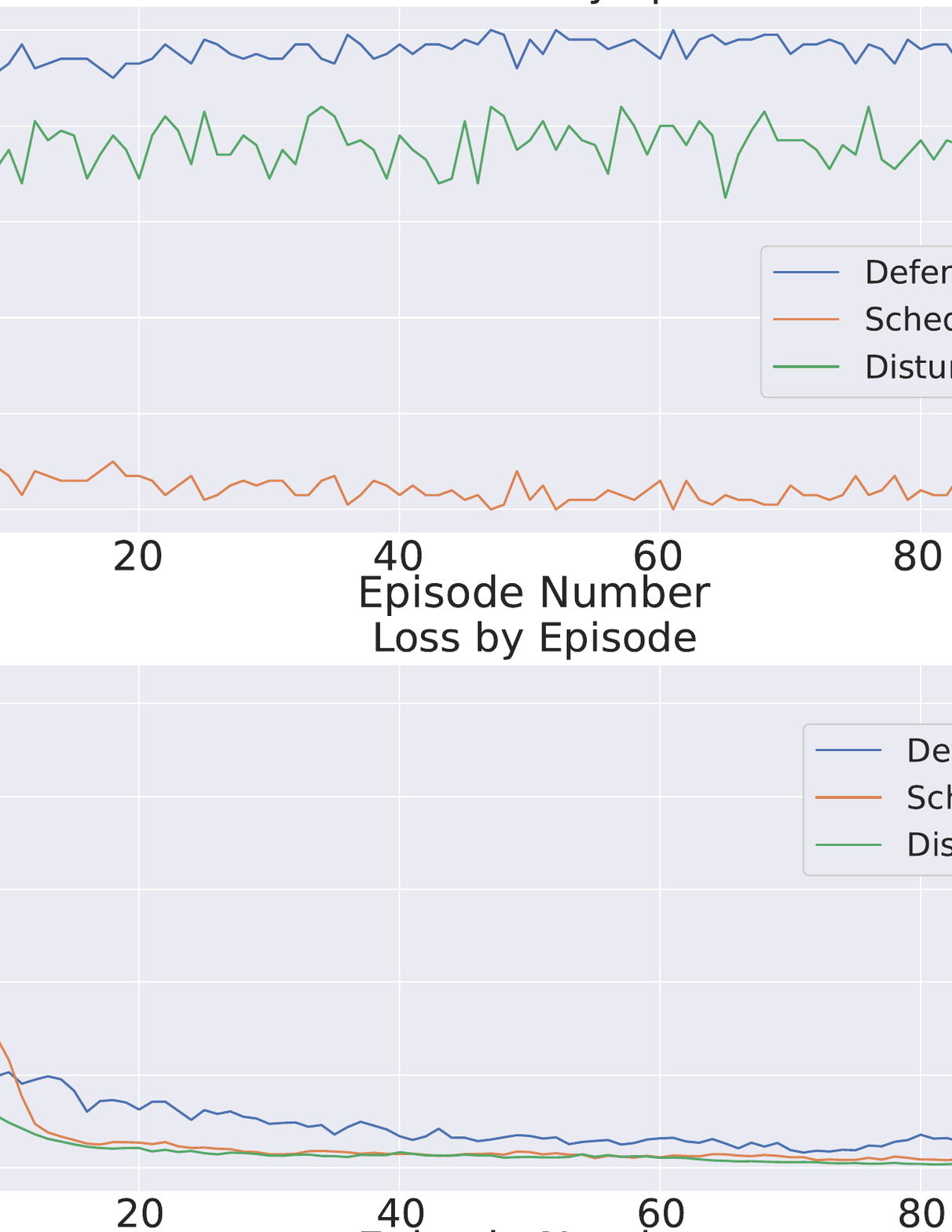}} 
\caption{Scheduler and Defender Rewards (Low \& Slow Strategy $\varepsilon$ $\in$ [0.05 - 0.1] )}
\label{fig: adversary vs defender rewards Low Slow Strategy } 
\end{figure}

\begin{figure}[t!] 
\center{\includegraphics[width=65mm]{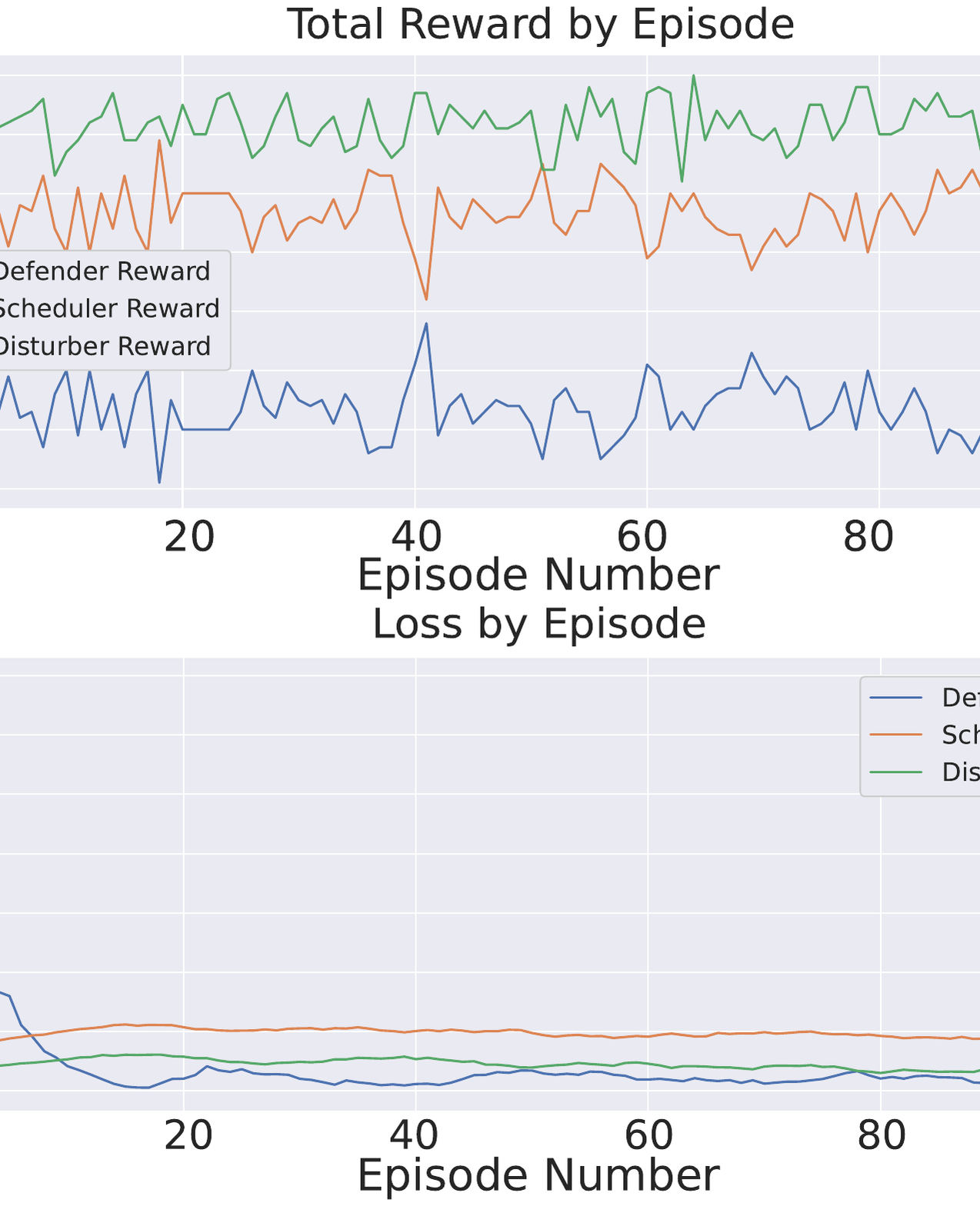}} 
\caption{Scheduler and Defender Rewards (Smash \& Grab Strategy $\varepsilon$ $\in$ [0.2 - 0.5] )}
\label{fig: adversary vs defender rewards Smash Grab } 
\end{figure}

\textbf{Analysis of Network Traffic Using Wireshark in Industrial Applications:}
Fig. \ref{fig: Network Observation} illustrates how an adversary can use network traffic analysis to gather detailed observations related to industrial control systems, as outlined in Fig. 3 inside the main paper. By intercepting and analysing communication between specific entities using the S7 communication protocol, an adversary can extract crucial information about process variables, control commands, and system configuration. For instance, observing the traffic can reveal changes in process variable (I102) levels and setpoint adjustments (I100), as indicated by the ‘Function: Write Var' and specific item values such as ‘DB Number' and ‘Area: Inputs', depicted in Fig. \ref{fig: Network Observation}. This enables an adversary to monitor the system's fill and discharge actions (states S0 and S1) and stealthily disrupt the operational processes by targeting the communication between PLC and actuator valves. 

\textbf{Architecture of the Autoencoder-based Anomaly Detectors:}
In this section, we provide details of the Autoencoder-based Anomaly Detectors we developed to evaluate our proposed adversary. In this context, Table \ref{tab:Architecture of the Autoencoder-based Anomaly Detectors} depicts the structure of the encoder and decoder of different models.

\begin{table}[htbp]
  \caption{Experiment Parameters}
  \label{tab:Architecture of the Autoencoder-based Anomaly Detectors}

  \vspace{-4mm} 
\begin{center}
  
\scalebox{0.8}{
\begin{tabular}{|l|l|l|}
\hline
    \textbf{Model Name} & \textbf{Encoder Structure} & \textbf{Decoder Structure} \\ 
    \hline

    DenseNet  & Two Dense Layers & Two Dense Layers \\ 
    \hline

    CNN& Two CNN Layers & Two CNN Layers \\ 
    \hline

    ResNet  & Two CNN with Residual Block & Two CNN with Residual Block \\ 
    \hline

    LSTM  & One LSTM Layers& One LSTM Layers \\ 
    \hline

    Transformer  & One Transformer Layers& One Transformer Layers \\ 
    \hline    

\end{tabular}
}

\end{center}

\end{table}

\begin{figure}[t!]
\centering
\includegraphics[width=90mm]{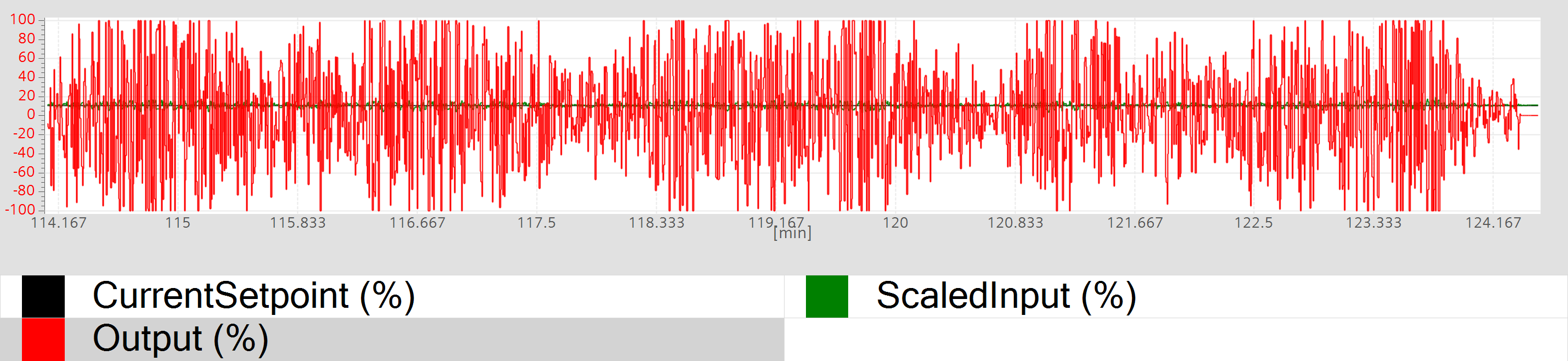}
\captionsetup{font=small} 
\caption*{(a) Extreme Attack}
\includegraphics[width=90mm]{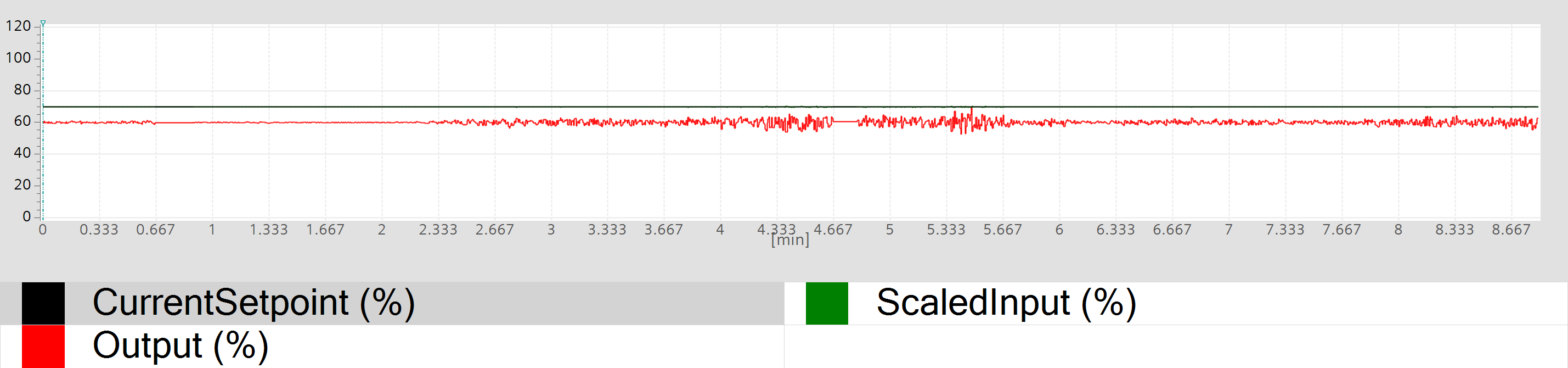}
\captionsetup{font=small}
  \caption*{(b) Smash \& Grab Attack Strategy}
  \includegraphics[width=90mm]{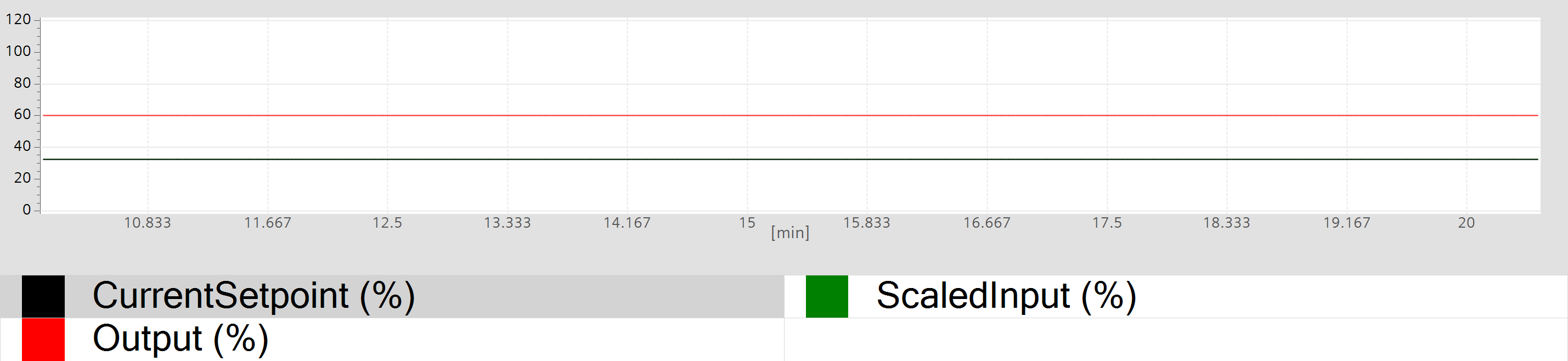}
  \captionsetup{font=small}
  \caption*{(c) Low \& Slow Attack Strategy}
  \caption{Attack Strategies}
  \label{fig: attack strategies}

\end{figure}

\begin{figure}[htb] 
\center{\includegraphics[width=80mm]{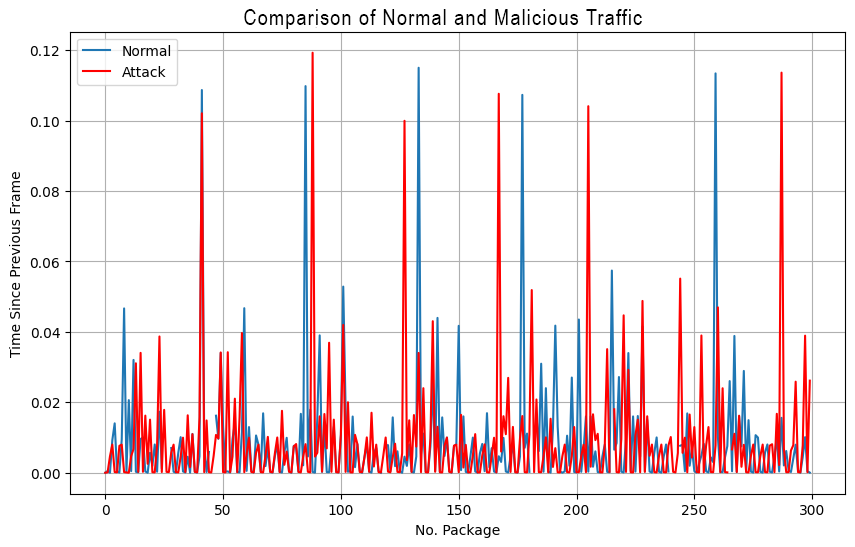}} 

\caption{Comparison of Normal and Malicious Traffic}
\label{fig:Normal vs Malicious Traffic} 
\end{figure}

\textbf{Comparison of Normal and Malicious Traffic}: Fig. \ref{fig:Normal vs Malicious Traffic} provides a detailed comparison between normal and malicious traffic, tracking the time spent between packets in a network data stream, labelled as “Time Since Previous Frame". This experiment conducted using 300 packets, represented on the x-axis, where the y-axis denotes the time in seconds. In the normal network communication scenario (illustrated in blue), packet intervals remain relatively consistent and low, mainly clustering below 0.04 seconds. This pattern suggests regular and expected network behaviour, where the transmission intervals are stable and predictable. Conversely, the network communication under attack scenario (represented in red) exhibits distinct characteristics with irregular, high spikes, some reaching up to 0.12 seconds, which abrupt increases in the interval time are indicative of abnormal activities. Hence, this analysis indicates the significance of the time feature during anomaly detection, demonstrating how deviations from established baselines can reveal potential security threats.

\textbf{Unified Scheduler and Disturber}:
Algorithm \ref{alg:adversary} depicts our proposed adversarial architecture, which efficiently orchestrates the actions of both the scheduler and the disturber agents within a unified advantage actor-critic methodology)

\begin{algorithm}[t!]
\caption{Unified Scheduler and Disturber}
\label{alg:adversary}

\begin{algorithmic}
\scriptsize

\STATE \textbf{Initialise:} Policy parameters \( \theta_0 \), \( \phi_0 \), learning rates \( \eta_\theta \), \( \eta_\phi \), discount factor \( \gamma \)
\STATE  Set initial state \( s_0 \)
\FOR{each episode}
    \STATE Initialise \( s \) from environment
    \WHILE{not done}
        \STATE \textbf{Scheduler Decision:}
        \STATE Choose \( t^{wait} \) and \( t^{delay} \) from policy \( \pi_{\theta}(s) \) \COMMENT{Scheduler decides 
        \STATE timing of actions}
        \STATE Back-off (wait) for \( t^{wait} \) in environment to reach an intermediate 
        \STATE state \( s_{wait} \)
        \STATE \textbf{Disturber Decision:}
        \STATE \( a_{adv2} \gets \text{policy}_{\phi}(s_{wait}) \) \COMMENT{Disturber selects packets to delay 
        \STATE $t^{delay}$ post-wait}
        \STATE Apply \( t^{delay} \) and \( a_{adv2} \), observe new state \( s_{t+1} \), reward \( r \) from Environment
        \STATE \textbf{Update Parameters:}
        \STATE Calculate TD error for the joint action:
        \[        \delta = r + \gamma V(s_{t+1}) - V(s)        \]
        \vspace{-4mm} 

        \STATE Update critic by minimising squared TD error:
        \[        \psi \gets \psi - \eta_\psi \nabla_{\psi} (\delta^2)        \]
        \vspace{-4mm} 

        \STATE Update Scheduler policy:
        \[
        \theta \gets \theta + \eta_\theta \nabla_{\theta} \log \pi_\theta(t^{wait}, t^{delay} | s) \delta        \]
        \vspace{-4mm} 

        \STATE Update Disturber policy:
        \[
        \phi \gets \phi + \eta_\phi \nabla_{\phi} \log \pi_\phi(a_{adv2} | s_{wait}) \delta        \]
        \vspace{-6mm} 

        \STATE \( s \gets s_{t+1} \)
    \ENDWHILE
\ENDFOR
\RETURN \( \theta, \psi, \phi \)  \COMMENT{return the optimised policy and value function parameters 
\STATE for both agents}
\end{algorithmic}
\end{algorithm}

\begin{figure*}[htbp] 
\center{\includegraphics[width=160mm]{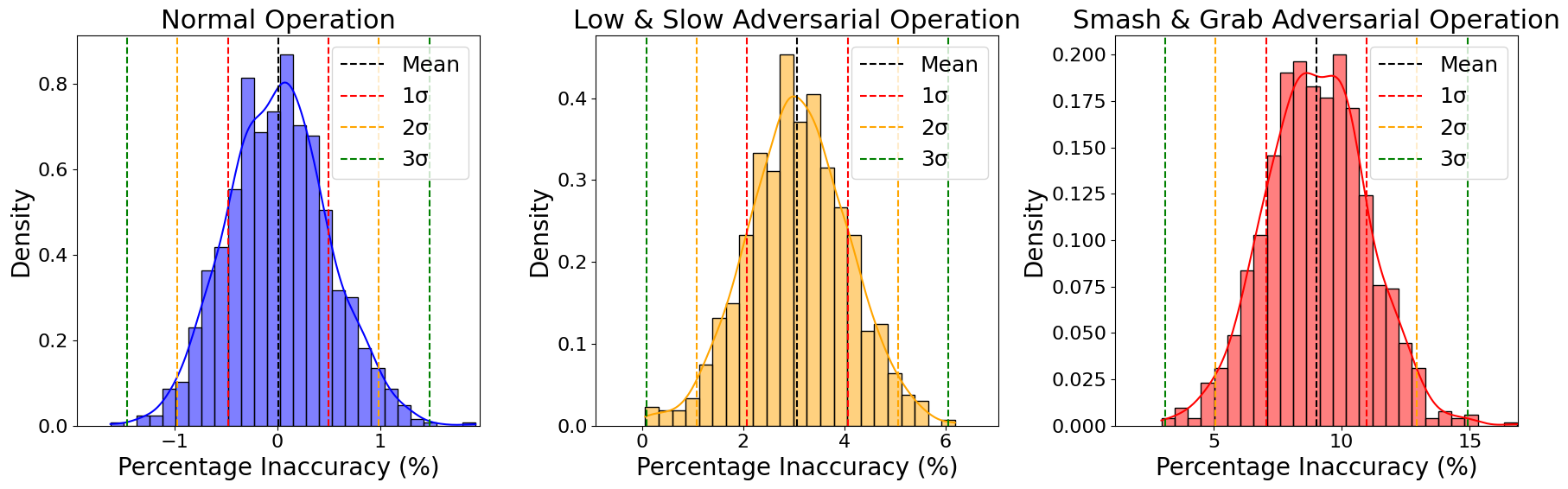}} 

\caption{Impact of Adversarial Strategies on Control Precision ($\mathcal{O}_{2}$)}
\label{fig:Impact of Adversarial Strategies on Control Precision} 
\vspace{-3mm} 
\end{figure*}

\subsection{Further Background}
\label{sec:Further Background}

\subsection{Ethics Considerations and Compliance with Open Science Policy}
This article aims to analyse the impact of AI-assisted stealthy attacks on Industrial Control Systems (ICS) and assess the vulnerabilities of these systems to secure them against such sophisticated threats. The research was conducted using a real-world industrial testbed, reflecting operational environments to accurately assess how Multi-agent DRL-assisted adversarial strategies can subtly disrupt system operations while evading detection. The primary goal is to identify these vulnerabilities and enhance defences, not to exploit them. Findings are responsibly disclosed with the intent to improve cybersecurity measures and inform future protections against these stealthy attacks. In compliance with the open science policy, all related artefacts, including datasets, code, and documentation, are publicly available to ensure reproducibility and support further research. This approach ensures that the insights gained are used to fortify security frameworks and foster a collaborative effort to address AI-assisted adversarial threats. The research team has carefully considered ethical implications and is committed to ongoing evaluations to responsibly guide the exploration of AI-driven attack methodologies.

\newcommand{\W}{256}   
\newcommand{\F}{32}    

\begin{table*}[t]
\centering
\caption{Complete layer-by-layer architectures for ICS time-series anomaly detectors (input per sample: $(B,\W,\F)$). All convs are 1-D over time; “BN” is BatchNorm; “GN” is GroupNorm ($g{=}8$); “GAP” is Global Average Pooling over time; “Drop” is Dropout.}
\label{tab:archs}
\renewcommand{\arraystretch}{1.12}
\scalebox{0.8}{
\begin{tabularx}{\textwidth}{@{}p{2.4cm} p{2.6cm} X p{3.4cm}@{}}
\toprule
\textbf{Model} & \textbf{Stage / Layer} & \textbf{Layer details} & \textbf{Output (time $\times$ channels)} \\
\midrule
\multirow{6}{*}{\textbf{CNN-1D}}
  & Input            & $(B,\W,\F)$                                                                                              & $\W \times \F$ \\
  & Stem             & Conv1d($C\!=\!64$, $k\!=\!7$, $s\!=\!1$, padding=3, \textit{causal}); BN; ReLU                            & $\W \times 64$ \\
  & Block 1          & Conv1d($128$, $k\!=\!5$, $s\!=\!2$, pad=2, \textit{causal}); BN; ReLU; Drop(0.1)                          & $\frac{\W}{2} \times 128$ \\
  & Block 2          & Conv1d($256$, $k\!=\!5$, $s\!=\!2$, pad=2, \textit{causal}); BN; ReLU; Drop(0.2)                          & $\frac{\W}{4} \times 256$ \\
  & Head             & Conv1d($256$, $k\!=\!3$, $s\!=\!1$, pad=1); BN; ReLU; GAP (over time)                                     & $1 \times 256$ \\
  & Classifier       & Linear(256$\!\to\!$128); ReLU; Drop(0.3); Linear(128$\!\to\!$1); Sigmoid                                 & $B \times 1$ \\
\midrule
\multirow{9}{*}{\textbf{ResNet‑18‑1D}}
  & Input            & $(B,\W,\F)$                                                                                              & $\W \times \F$ \\
  & Stem             & Conv1d($64$, $k\!=\!7$, $s\!=\!2$, pad=3, \textit{causal}); BN; ReLU; MaxPool1d($k\!=\!3$, $s\!=\!2$)     & $\frac{\W}{4} \times 64$ \\
  & Stage 1 (x2)     & \textit{BasicBlock} (Conv3–BN–ReLU–Conv3–BN + identity); last block Drop(0.1)                            & $\frac{\W}{4} \times 64$ \\
  & Stage 2 (x2)     & \textit{BasicBlock} (first block has stride 2 in first conv; projection skip); Drop(0.1)                  & $\frac{\W}{8} \times 128$ \\
  & Stage 3 (x2)     & \textit{BasicBlock} (stride 2 at entry; projection skip); Drop(0.2)                                       & $\frac{\W}{16} \times 256$ \\
  & Stage 4 (x2)     & \textit{BasicBlock} (stride 2 at entry; projection skip); Drop(0.2)                                       & $\frac{\W}{32} \times 512$ \\
  & Norm + Pool      & GN($g\!=\!8$); ReLU; GAP                                                                                   & $1 \times 512$ \\
  & Classifier       & Linear(512$\!\to\!$128); ReLU; Drop(0.3); Linear(128$\!\to\!$1); Sigmoid                                  & $B \times 1$ \\
  & BasicBlock def.  & Conv1d($C$, $k\!=\!3$, $s\!\in\!\{1,2\}$, pad=1, \textit{causal})–BN–ReLU–Conv1d($C$, $k\!=\!3$, $s\!=\!1$)–BN + skip; ReLU & — \\
\midrule
\multirow{10}{*}{\textbf{DenseNet‑121‑1D}}
  & Input            & $(B,\W,\F)$                                                                                              & $\W \times \F$ \\
  & Stem             & Conv1d($64$, $k\!=\!7$, $s\!=\!2$, pad=3, \textit{causal}); BN; ReLU; AvgPool1d($k\!=\!3$, $s\!=\!2$)     & $\frac{\W}{4} \times 64$ \\
  & DenseBlock 1     & $6$ \textit{dense layers}, growth rate $k\!=\!32$ (BN–ReLU–Conv1x1($4k$)–BN–ReLU–Conv3x1($k$)); concat    & $\frac{\W}{4} \times (64+6k)$ \\
  & Transition 1     & BN–ReLU–Conv1x1(compress $\theta\!=\!0.5$); AvgPool1d($k\!=\!2$, $s\!=\!2$)                              & $\frac{\W}{8} \times \frac{64+6k}{2}$ \\
  & DenseBlock 2     & $12$ layers, growth $k$                                                                                   & $\frac{\W}{8} \times \Big(\frac{64+6k}{2}+12k\Big)$ \\
  & Transition 2     & BN–ReLU–Conv1x1($\theta\!=\!0.5$); AvgPool1d($2$, $2$)                                                   & $\frac{\W}{16} \times \frac{\frac{64+6k}{2}+12k}{2}$ \\
  & DenseBlock 3     & $24$ layers, growth $k$                                                                                   & $\frac{\W}{16} \times \Big(\frac{\frac{64+6k}{2}+12k}{2}+24k\Big)$ \\
  & Transition 3     & BN–ReLU–Conv1x1($\theta\!=\!0.5$); AvgPool1d($2$, $2$)                                                   & $\frac{\W}{32} \times \cdots$ \\
  & DenseBlock 4     & $16$ layers, growth $k$                                                                                   & $\frac{\W}{32} \times C_\text{final}$ \\
  & Head             & BN; ReLU; GAP; Linear($C_\text{final}\!\to\!$1); Sigmoid                                                  & $B \times 1$ \\
\midrule
\multirow{6}{*}{\textbf{BiLSTM + Attn}}
  & Input            & $(B,\W,\F)$ (z‑score per feature)                                                                        & $\W \times \F$ \\
  & Recurrent stack  & BiLSTM($h\!=\!128$) $\rightarrow$ Drop(0.2) $\rightarrow$ BiLSTM($h\!=\!64$)                               & $\W \times 128$ \\
  & Attention pool   & Scaled dot‑product attention over time (query from mean state; key/value from hidden); context vector     & $1 \times 128$ \\
  & Norm + Head      & LayerNorm; Linear(128$\!\to\!$64); ReLU; Drop(0.3); Linear(64$\!\to\!$1); Sigmoid                         & $B \times 1$ \\
  & Notes            & Packed sequences; gradient clip $\!\le\!1.0$; masking for variable $\W$                                    & — \\
  & Option           & Replace attention with Temporal GAP for ultra‑low‑latency inference                                       & — \\
\midrule
\multirow{9}{*}{\textbf{Transformer‑TS}}
  & Input            & $(B,\W,\F)$                                                                                                & $\W \times \F$ \\
  & Patch embed      & Conv1d($d_\text{model}\!=\!128$, $k\!=\!7$, $s\!=\!2$, pad=3) + LayerNorm (Pre‑LN stack)                  & $\frac{\W}{2} \times 128$ \\
  & Positional enc.  & Learned (size $\frac{\W}{2}$) \textit{(+ optional [CLS] token)}                                           & $\frac{\W}{2} \times 128$ \\
  & Encoder $\times4$& \textbf{Per layer:} Multi‑Head Self‑Attention ($h\!=\!8$, $d_\text{model}\!=\!128$); Drop(0.1); MLP (FFN 256$\to$128) with GELU; Drop(0.1); Pre‑LN residuals & $\frac{\W}{2} \times 128$ \\
  & Pool             & Mean Pool over time \textit{or} [CLS]                                                                      & $1 \times 128$ \\
  & Classifier       & Linear(128$\!\to\!$64); GELU; Drop(0.2); Linear(64$\!\to\!$1); Sigmoid                                     & $B \times 1$ \\
  & Regularisation   & Stochastic depth $p\!=\!0.05$; label smoothing $=0.05$ (optional)                                          & — \\
  & Variant          & For long sequences: replace MSA with \textit{local} MSA (win=64) to cap $O(T^2)$                           & — \\
\bottomrule
\end{tabularx}
}
\end{table*}

\begin{table*}[t]
\centering

\caption{Unified preprocessing and training settings for all detectors.}
\label{tab:train}
\renewcommand{\arraystretch}{1.12}
\scalebox{0.8}{
\begin{tabularx}{\textwidth}{@{}p{3.3cm} X p{6.8cm}@{}}
\toprule
\textbf{Aspect} & \textbf{Setting} & \textbf{Rationale / Notes} \\
\midrule
Sliding-windowing
  & Window length $\W$; stride $=\!\lfloor 0.25\W \rfloor$ (75\% overlap)
  & Matches your evaluation framework (windowed time‑series over PLC/PROFINET features). Overlap improves recall on rare attacks. :contentReference[oaicite:3]{index=3} \\
Feature scaling
  & Per‑feature $z$‑score using training statistics (mean/var); clip to $\pm 5\sigma$
  & Stabilises optimisation; guards against outliers from process excursions. \\
Class imbalance
  & Positive class weight $w_+\!=\!N_-/N_+$ in BCEWithLogitsLoss; threshold tuned on validation F1
  & Your results emphasise recall/F1 under attack; weighting reduces bias to the majority class. :contentReference[oaicite:4]{index=4} \\
Optimizer / LR
  & AdamW ($\beta_1\!=\!0.9,\ \beta_2\!=\!0.999$, weight decay $1\!\times\!10^{-4}$); base LR $3\!\times\!10^{-4}$
  & Works well across conv/RNN/Transformer families; decay combats overfitting. \\
LR schedule
  & Cosine decay with 5‑epoch warm‑up; min LR $3\!\times\!10^{-6}$
  & Smooth convergence; prevents early‑epoch divergence. \\
Batch / epochs
  & Batch size 128 (LSTM 64); max 100 epochs; early stopping patience 10 on val F1
  & Keeps training efficient; selects the best generalizing checkpoint. \\
Regularization
  & Dropout as in Table~\ref{tab:archs}; gradient clip $\le 1.0$ (RNN/Transformer); label smoothing 0.05 (Transformer)
  & Improves OOC stability discussed in §V. :contentReference[oaicite:5]{index=5} \\
Augmentations (light)
  & Jitter (Gaussian $\sigma\!=\!0.01$), time‑masking (10\% random spans), mild time‑warp ($\pm$5\%)
  & Mimics sensor noise/timing jitter seen in field networks; supports OOC robustness. :contentReference[oaicite:6]{index=6} \\
Validation split
  & 70/15/15 train/val/test by \textit{time blocks} to avoid leakage
  & Prevents adjacent‑window leakage; matches operational deployment. \\
Inference target
  & Probability per window; alarm when $\text{Prob} \ge \tau$; hysteresis with $N$ consecutive flags
  & Reduces flapping; aligns with “low \& slow” vs. “smash \& grab” dynamics. :contentReference[oaicite:7]{index=7} \\
Implementation
  & PyTorch; mixed precision (AMP); reproducibility seeds fixed
  & Re‑runnable configs; practical on SIMATIC IPC/Jetson inference hosts noted in §IV. :contentReference[oaicite:8]{index=8} \\
\bottomrule
\end{tabularx}
}
\end{table*}

\begin{table*}[t]

\centering
\caption{Factory~I/O scenes: observations and timing-only actions available to the adversary. Abbrev.: PE = photoelectric sensor; MC = machining center; P\&P = pick-and-place.}
\label{tab:factoryio_oac}
\renewcommand{\arraystretch}{1.08}
\scalebox{1.}{
\begin{tabularx}{\textwidth}{@{}l l X X@{}}
\toprule
\textbf{Scene} & \textbf{Objective (per manual)} & \textbf{Observations $\mathbf{o_t}$ (tags $\rightarrow$ features)} & \textbf{Adversary timing actions $\mathbf{a_t}$ (bounded delays)} \\
\midrule
Buffer Station \cite{factoryio_buffer} 
& Buffer and separate up to five boxes 
& PE sensors along conveyors; stopper cylinder states; queue length; conveyor motor states; \emph{timing features}: inter-arrival times of items and actuation latencies of stop/start events 
& Delay rising edges for \emph{stopper up/down} and \emph{conveyor start/stop} ($\le$150\,ms L\&S / $\le$400\,ms S\&G); occasional delay of upstream PE sensor “present” to alter spacing while preserving protocol legality \\
\addlinespace[0.25em]
Production Line \cite{factoryio_prodline}
& Produce bases by controlling two machining centers 
& MC\#1/\#2 \emph{ready/run/done}; diverter state; buffer occupancy; PE sensors at infeed/outfeed; \emph{timing features}: cycle start-to-done gaps; inter-station transfer latencies 
& Delay \emph{start cycle} pulses and \emph{diverter actuation} to create controlled parity drift between lids/bases (L\&S: $\le$120\,ms bursts; S\&G: $\le$350\,ms short runs) while keeping cycle acknowledgements within PLC tolerances \\
\addlinespace[0.25em]
Assembler \cite{factoryio_assembler} 
& Assemble parts made of lids and bases using a two-axis P\&P 
& P\&P axis positions (Floats); vacuum sensor; part-present at pick/place; conveyor PE; \emph{timing}: axis step‑to‑contact delays; grasp‑to‑place times 
& Delay P\&P \emph{axis-step} tags (digital step pulses) and \emph{gripper on/off} by tens of ms to induce micro misalignment and extra cycles while respecting stroke/limit switches; selective delay of \emph{place confirm} to disturb synchronisation \\
\bottomrule
\end{tabularx}
}
\end{table*}

\subsection{Further Related Works}
\label{sec:Further Related Works}
Krotofila et al. \cite{krotofil2014vulnerabilities} analysed the timing of stale-data attacks in cyber-physical systems, leveraging the Tennessee Eastman process to demonstrate optimal stopping strategies. However, their approach was limited to simulation environments and focused on static timing scenarios. Unlike their work, our study introduces an autonomous, intelligent adversary powered by DRL that dynamically adapts to complex industrial processes. Unlike their approach, our adversary is designed to\textbf{ perform effectively in out-of-context scenarios}, stealthily targeting field networks while evading AI-driven anomaly detectors.

In \cite{urbina2016limiting}, the authors addressed the challenges of stealthy attacks on ICS, emphasizing the development of defences to limit their impact. Their approach relied on simulation-based evaluations, which, while valuable, lack the complexity and unpredictability of real-world systems. Unlike this work, our study focuses on a proactive adversarial model that is designed not only to evade defences but also to target vulnerabilities across diverse environments. This capability, especially when applied in out-of-context scenarios, demonstrates the robustness of our adversary and extends the application of DRL-based techniques beyond traditional constraints.

The work by Li et al. \cite{li2021conaml} introduced ConAML, a framework for constrained adversarial machine learning in CPS, specifically focusing on linear physical constraints. While their study effectively showcased adversarial vulnerabilities in power grids and water treatment systems, it was confined to simulation environments. Our research diverges by tackling real-world ICS environments and incorporating a multi-agent DRL framework capable of dynamically optimizing its attacks. This dynamic adaptability and our emphasis on targeting ICS-specific vulnerabilities under real-time constraints highlight a significant advancement over the static methods and constrained setups used in ConAML.

Zizzo et al. \cite{zizzo2020adversarial} demonstrated adversarial attacks on time-series-based intrusion detection systems in ICS using autoregressive IDS. Their methodology effectively hides cyber-physical attacks by compromising specific sensors, but it assumes white-box knowledge of the IDS. Our approach, in contrast, minimizes dependency on detailed system knowledge while leveraging a DRL adversary to optimize attack strategies autonomously. This capability to adapt dynamically to system changes and execute in diverse, real-world-like scenarios further distinguishes our work from their sensor-focused approach.

Finally, Chandratre et al. \cite{erba2020constrained} developed constrained concealment attacks targeting reconstruction-based anomaly detectors in ICS. While their work successfully demonstrated stealth attacks, it primarily focused on crafting minimal perturbations under strict system constraints. Our study expands on this by emphasizing the strategic timing and synchronisation of multi-agent adversaries in ICS, demonstrating broader applicability and effectiveness in disrupting complex industrial processes beyond the limitations of concealment-based methods.

\begin{figure}[t]
  \centering
  \includegraphics[width=0.95\linewidth]{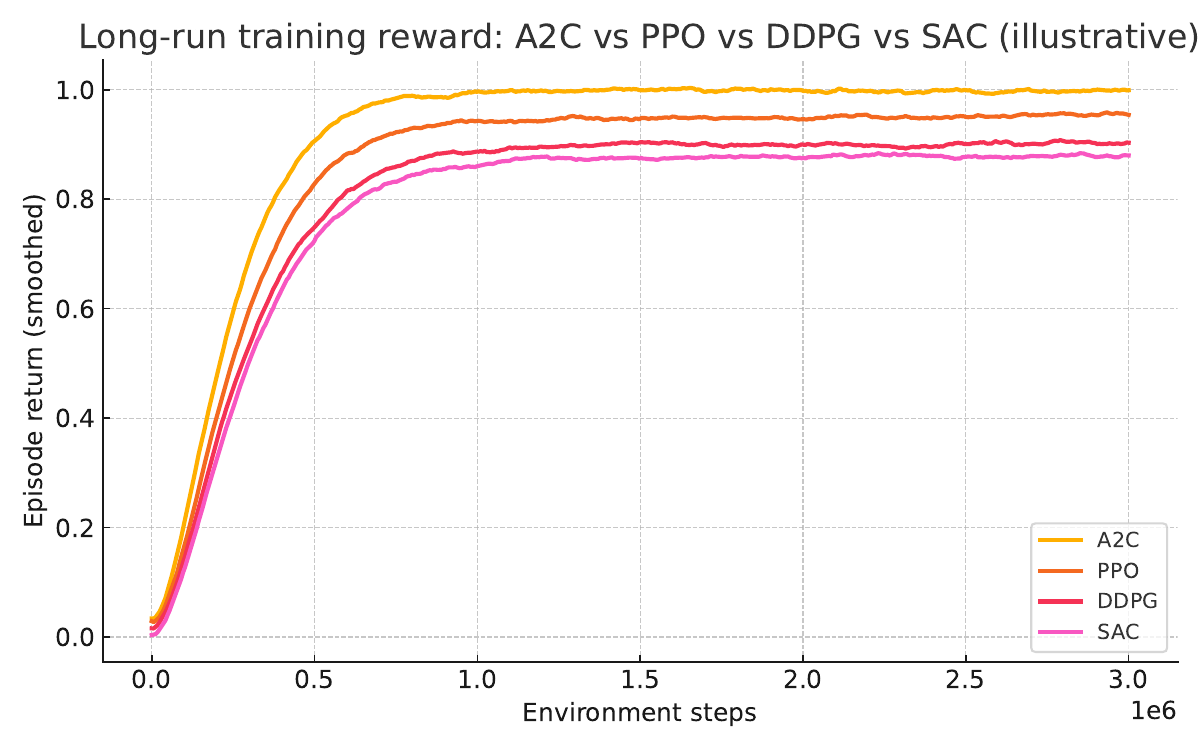}
  \caption{\textbf{Long-run training reward across DRL algorithms.}
  A2C attains the highest asymptotic reward and AUC and converges faster (lower $T_{95}$) than PPO, DDPG, and SAC. Curves show smoothed episode return vs.\ steps (mean across seeds).}
  \label{fig:longrun_drl}
\end{figure}

\begin{table*}[htbp]

\centering
\caption{Factory~I/O OOC results. Throughput in items/min; Quality as defect/mis-assembly rate; Cycles/min as wear surrogate; Recall macro-averaged across detectors from Tables~I–II; Stealth defined in Section~IV.B as $1-\sum_d w_d R_d$ (equal $w_d$).}
\label{tab:factoryio_results}
\renewcommand{\arraystretch}{1.05}
\scalebox{1.0}{
\begin{tabular}{l l l c c c c c}
\toprule
\textbf{Scene} & \textbf{Strategy} & \textbf{Knowledge} & \textbf{Throughput} & \textbf{Quality} & \textbf{Cycles/min} & \textbf{Recall} & \textbf{Stealth} \\
\midrule
\multirow{6}{*}{Buffer Station}
& L\&S & Black & 50.8 & 1.1\% & 126 & 0.30 & 0.70 \\
& L\&S & Grey  & 50.1 & 1.2\% & 127 & 0.22 & 0.78 \\
& L\&S & White & 49.9 & 1.3\% & 128 & 0.12 & 0.88 \\
& S\&G & Black & 48.2 & 2.4\% & 135 & 0.54 & 0.46 \\
& S\&G & Grey  & 47.5 & 2.6\% & 137 & 0.46 & 0.54 \\
& S\&G & White & 46.8 & 2.7\% & 139 & 0.34 & 0.66 \\
\midrule
\multirow{6}{*}{Production Line}
& L\&S & Black & 27.8 & 1.5\% & 79 & 0.28 & 0.72 \\
& L\&S & Grey  & 27.4 & 1.6\% & 80 & 0.20 & 0.80 \\
& L\&S & White & 27.1 & 1.7\% & 81 & 0.14 & 0.86 \\
& S\&G & Black & 26.6 & 2.8\% & 84 & 0.52 & 0.48 \\
& S\&G & Grey  & 26.0 & 3.0\% & 86 & 0.44 & 0.56 \\
& S\&G & White & 25.6 & 3.1\% & 87 & 0.36 & 0.64 \\
\midrule
\multirow{6}{*}{Assembler}
& L\&S & Black & 21.7 & 1.3\% & 69 & 0.29 & 0.71 \\
& L\&S & Grey  & 21.4 & 1.5\% & 70 & 0.21 & 0.79 \\
& L\&S & White & 21.1 & 1.6\% & 71 & 0.13 & 0.87 \\
& S\&G & Black & 20.7 & 2.5\% & 74 & 0.55 & 0.45 \\
& S\&G & Grey  & 20.1 & 2.8\% & 76 & 0.47 & 0.53 \\
& S\&G & White & 19.8 & 3.0\% & 77 & 0.38 & 0.62 \\
\bottomrule
\end{tabular}
}
\end{table*}

\begin{table*}[htbp]
\centering

\caption{Percent change vs baseline (per scene). Negative is worse for throughput; positive is worse for quality and cycles/min.}
\label{tab:factoryio_deltas}
\renewcommand{\arraystretch}{1.05}
\scalebox{1.}{
\begin{tabular}{l l l r r r}
\toprule
\textbf{Scene} & \textbf{Strategy} & \textbf{Knowledge} & \textbf{$\Delta$Throughput} & \textbf{$\Delta$Quality} & \textbf{$\Delta$Cycles} \\
\midrule
Buffer & L\&S & Black & $-3.0\%$ & $+0.7$pp & $+6.8\%$ \\
Buffer & L\&S & Grey  & $-4.4\%$ & $+0.8$pp & $+7.6\%$ \\
Buffer & L\&S & White & $-4.8\%$ & $+0.9$pp & $+8.5\%$ \\
Buffer & S\&G & Black & $-8.0\%$ & $+2.0$pp & $+14.4\%$ \\
Buffer & S\&G & Grey  & $-9.4\%$ & $+2.2$pp & $+16.1\%$ \\
Buffer & S\&G & White & $-10.7\%$ & $+2.3$pp & $+17.8\%$ \\
\midrule
Prod. Line & L\&S & Black & $-2.8\%$ & $+0.7$pp & $+6.8\%$ \\
Prod. Line & L\&S & Grey  & $-4.2\%$ & $+0.8$pp & $+8.1\%$ \\
Prod. Line & L\&S & White & $-5.2\%$ & $+0.9$pp & $+9.5\%$ \\
Prod. Line & S\&G & Black & $-7.0\%$ & $+2.0$pp & $+13.5\%$ \\
Prod. Line & S\&G & Grey  & $-9.1\%$ & $+2.2$pp & $+16.2\%$ \\
Prod. Line & S\&G & White & $-10.5\%$ & $+2.3$pp & $+17.6\%$ \\
\midrule
Assembler & L\&S & Black & $-2.7\%$ & $+0.7$pp & $+6.2\%$ \\
Assembler & L\&S & Grey  & $-4.0\%$ & $+0.9$pp & $+7.7\%$ \\
Assembler & L\&S & White & $-5.4\%$ & $+1.0$pp & $+9.2\%$ \\
Assembler & S\&G & Black & $-7.2\%$ & $+1.9$pp & $+13.8\%$ \\
Assembler & S\&G & Grey  & $-9.9\%$ & $+2.2$pp & $+16.9\%$ \\
Assembler & S\&G & White & $-11.2\%$ & $+2.4$pp & $+18.5\%$ \\
\bottomrule
\end{tabular}
}
\end{table*}

{

\paragraph*{Wall-Clock Profiling.}
We profile per-decision latency, CPU/GPU utilisation, and resident memory for the Scheduler and Disturber during inference on SIMATIC IPC127E and Jetson AGX Orin hosts, and compare them against PLC scan-cycle budgets (1--50\,ms). Table~\ref{tab:overhead} summarises the measurements.\footnotesize\emph{Config:} sliding window $W{=}256$, $F{=}32$ features, batch$=1$, FP32; IPC127E runs CPU-only; AGX Orin uses the integrated GPU; medians over repeated runs. \normalsize

\begin{table}[htbp]
\centering
\caption{Inference overhead (representative values under the stated config).}
\label{tab:overhead}
\renewcommand{\arraystretch}{1.05}
\scalebox{0.8}{
\begin{tabular}{lcccc}
\toprule
\textbf{Host} & \textbf{Agent} & \textbf{Latency (ms)} & \textbf{CPU/GPU (\%)} & \textbf{Mem (MB)} \\
\midrule
IPC127E      & Scheduler & 3.2  & CPU 27 & 210 \\
IPC127E      & Disturber & 2.7  & CPU 24 & 198 \\
Jetson Orin  & Scheduler & 0.46 & GPU 6  & 162 \\
Jetson Orin  & Disturber & 0.39 & GPU 5  & 158 \\
\bottomrule
\end{tabular}
}
\end{table}

\begin{table}[htb]
\centering
\caption{Reduction in Recall by Attack Strategy and Adversary Knowledge}

\label{tab:reduction_recall}

\scalebox{0.95}{
\begin{tabular}{@{}llcccccc@{}}
\toprule
\textbf{Model} & \textbf{Knowledge} & \multicolumn{2}{c}{\textbf{Smash \& Grab}} & \multicolumn{2}{c}{\textbf{Low \& Slow}} \\ 
\cmidrule(lr){3-4} \cmidrule(lr){5-6}
& & \textbf{Reduction} & \textbf{Recall}\tnote{\dag} & \textbf{Reduction} & \textbf{Recall}\tnote{\dag} \\ 
\midrule
\multirow{3}{*}{DenseNet} 
& Black-box & 17.70\% & 0.6989 & 17.34\% & 0.6978 \\
& Grey-box  & 28.96\% & 0.5997 & 66.63\% & 0.2816 \\
& White-box & 99.80\% & 0.0017 & 99.93\% & 0.0006 \\
\midrule
\multirow{3}{*}{CNN} 
& Black-box & 27.15\% & 0.6965 & 27.61\% & 0.6923 \\
& Grey-box  & 69.24\% & 0.2941 & 67.25\% & 0.3131 \\
& White-box & 99.79\% & 0.0020 & 99.95\% & 0.0005 \\
\midrule
\multirow{3}{*}{ResNet} 
& Black-box & 29.56\% & 0.7012 & 29.78\% & 0.6984 \\
& Grey-box  & 39.34\% & 0.6031 & 68.41\% & 0.3142 \\
& White-box & 99.76\% & 0.0024 & 99.90\% & 0.0010 \\
\midrule
\multirow{3}{*}{LSTM} 
& Black-box & 23.16\% & 0.7006 & 23.61\% & 0.6962 \\
& Grey-box  & 33.66\% & 0.6039 & 59.49\% & 0.3688 \\
& White-box & 99.82\% & 0.0016 & 99.89\% & 0.0010 \\
\midrule
\multirow{3}{*}{Transformer} 
& Black-box & 27.54\% & 0.7001 & 28.24\% & 0.6936 \\
& Grey-box  & 38.30\% & 0.5962 & 67.19\% & 0.3173 \\
& White-box & 99.79\% & 0.0020 & 99.91\% & 0.0009 \\

\bottomrule

\end{tabular}
}

\begin{tablenotes}
\small
\end{tablenotes}
\end{table}

\begin{table}[htbp]
\centering

\caption{DRL algorithm comparison (mean across seeds). Long-run reward = mean over last 10\% steps; AUC = $\int$ reward vs.\ steps; $T_{95}$ = steps to 95\% of final mean reward.}
\label{tab:drl_summary}
\begin{tabular}{lccc}
\toprule
\textbf{Algorithm} & \textbf{Long-run reward} & \textbf{AUC} & \boldmath$\mathbf{T_{95}}$ \textbf{(steps)} \\
\midrule
A2C   & 0.9978 & 0.9165 & $5.83\times 10^{5}$ \\
PPO   & 0.9550 & 0.8621 & $6.79\times 10^{5}$ \\
DDPG  & 0.9031 & 0.8101 & $7.45\times 10^{5}$ \\
SAC   & 0.8794 & 0.7862 & $7.70\times 10^{5}$ \\
\bottomrule
\end{tabular}
\end{table}

\subsubsection*{Non-DRL Timing Attacks (Baselines)}
We benchmark standard timing attacks to contextualise our adversary:
\emph{Random Delay} (uniform $[0,120]$\,ms on 10\% of events),
\emph{Periodic Jitter} (40\,ms bursts every 1.2\,s),
\emph{Poisson Jitter} ($\lambda{=}0.8$/s, mean 60\,ms),
and \emph{Greedy Wear} (maximise actuator cycling under a hard stealth ceiling).
Table~\ref{tab:baseline_attacks} shows that our \emph{dual-agent DRL (Low\&Slow)} achieves higher impact with markedly lower recall (higher Stealth) than non-DRL baselines.

\begin{table}[t]
\centering
\caption{Main testbed: baseline timing attacks vs dual-agent DRL (macro-averaged across detectors from Tables~I–II). Throughput in items/min; Quality as defect rate (pp); Cycles/min as wear surrogate.}
\label{tab:baseline_attacks}
\renewcommand{\arraystretch}{1.05}
\begin{tabular}{lcccccc}
\toprule
\textbf{Attack} & \textbf{$\Delta$Thru} & \textbf{$\Delta$Qual} & \textbf{$\Delta$Cycles} & \textbf{Recall} & \textbf{Stealth} \\
\midrule
Random Delay         & $-1.9\%$ & $+0.3$\,pp & $+3.2\%$  & 0.47 & 0.53 \\
Periodic Jitter      & $-2.9\%$ & $+0.5$\,pp & $+5.1\%$  & 0.41 & 0.59 \\
Poisson Jitter       & $-2.5\%$ & $+0.4$\,pp & $+4.4\%$  & 0.44 & 0.56 \\
Greedy Wear (cap)    & $-7.8\%$ & $+2.1$\,pp & $+15.0\%$ & 0.58 & 0.42 \\
\midrule
\textbf{DRL (L\&S)}  & \textbf{$-4.1\%$} & \textbf{$+0.8$\,pp} & \textbf{$+7.9\%$} & \textbf{0.18} & \textbf{0.82} \\
\textbf{DRL (S\&G)}  & $-9.8\%$ & $+2.4$\,pp & $+17.1\%$ & 0.36 & 0.64 \\
\bottomrule
\end{tabular}
\end{table}

}

{

\subsection{External OOC Validation on Factory I/O Scenes}
\label{sec:factoryio_ooc}
To evaluate transferability beyond our liquid-mixing testbed, we validated the adversary on three discrete-manufacturing scenes from \textbf{Factory~I/O}: \emph{Buffer Station}, \emph{Production Line}, and \emph{Assembler}. Each scene is a realistic, read-only industrial layout (can be duplicated to \emph{My Scenes} to customize) with standardized sensors/actuators exposed as \emph{tags} (Bool/Float/Int) and multiple industrial \emph{I/O drivers} (e.g., Siemens S7 and PLCSIM, OPC UA, Modbus).\footnote{\textcolor{black}{Factory~I/O scenes overview and read-only note in the manual; tags types and forcing; drivers list: \cite{factoryio_scenes_overview,factoryio_tags,factoryio_drivers}. Scene objectives: Buffer Station (buffer and separate up to five boxes), Production Line (match counts of lids and bases via two machining centres), Assembler (assemble lids and bases with a two-axis pick-and-place): \cite{factoryio_buffer,factoryio_prodline,factoryio_assembler}. Siemens S7-PLCSIM usage: \cite{factoryio_plcsim, factoryio_plcsim_v13}.}}

\paragraph{Interface and L$_0$ vantage.}
We used the native \emph{Siemens~S7/S7-PLCSIM} driver to exchange tag values with the controller and implemented an inline \emph{L$_0$ tap} that intercepts controller $\leftrightarrow$ scene traffic. The tap schedules \emph{bounded, protocol-legal delays} on selected actuator/sensor tag updates, thereby emulating the same timing-based $L_0$ vantage as in our real testbed (Section~III). Integer/float conversions follow Factory~I/O’s driver semantics; digital tags are mapped 1:1 to Bool. Our two-agent adversary (Scheduler/Disturber) runs unchanged.

\paragraph{Scenes, observations, and actions.}
Table~\ref{tab:factoryio_oac} enumerates the observation vectors and admissible actions per scene. Observation tags include \emph{process} signals (e.g., presence sensors, machine ready/done) and \emph{timing} features (inter-arrival and actuation latencies computed by the tap). Actions are \emph{timing-only} (no payload edits): delaying specific actuator commands (e.g., conveyor start/stop, stopper cylinders, machining start, pick-and-place steps) or selected sensor reports to the controller. Delays are bounded ($\le\!150$\,ms for Low\&Slow; $\le\!400$\,ms for Smash\&Grab) to remain within PLC scan-cycle budgets.

\paragraph{Metrics and OOC drift.}
We reuse our metrics: \emph{Throughput} (items/min), \emph{Quality} (defect/mis-assembly rate), \emph{Actuation cycles/min} (wear surrogate), per-detector \emph{recall}/FNR, and the \emph{Stealth Score} from Section~IV.B. Out-of-context drift is induced by changing (i) machine timing (cycle-time offsets), (ii) sensor noise, (iii) controller sampling cadence and detector windowing, and (iv) feature subsets—consistent with our OOC protocol. Factory~I/O’s \emph{Failure Injection} was \emph{not} used to falsify signals, except for sanity checks; our attacks remain timing-only and protocol-legal.\footnote{Tags (Bool/Float/Int) and \emph{Forcing}/\emph{Failure Injection} are documented in the manual; we did not rely on failure injection to mount attacks \cite{factoryio_tags,factoryio_fail}.}

\paragraph*{OOC Drift Scenarios (concrete).}
We induce controlled distribution shift between training and evaluation along plant and detection axes (Table~\ref{tab:ooc_drift}). The attack is not re-trained; only detectors are optionally re-tuned when indicated.

\begin{table}[t]
\centering

\caption{Out-of-Context (OOC) drift knobs and magnitudes. “Det. retune” applies threshold re-tuning to preserve nominal FPR on the shifted data.}
\label{tab:ooc_drift}
\renewcommand{\arraystretch}{1.05}
\begin{tabular}{lccc}
\toprule
\textbf{Drift knob} & \textbf{Small} & \textbf{Medium} & \textbf{Large} \\
\midrule
Sensor noise $\sigma$ (Floats) & ${+}10\%$ & ${+}20\%$ & ${+}35\%$ \\
PLC scan jitter (ms, p2p)     & $\pm 3$  & $\pm 6$  & $\pm 10$ \\
Conveyor / pump gain           & ${+}5\%$ & ${+}10\%$ & ${+}15\%$ \\
Arrival rate (Poisson $\lambda$) & ${+}5\%$ & ${+}12\%$ & ${+}20\%$ \\
Tag remap (equivalent sensors) & none     & swap 1 pair & swap 2 pairs \\
Detector threshold (Det. retune) & none & hold FPR & hold FPR \\
Preproc. window length $W$     & same & ${-}10\%$ & ${-}20\%$ \\
\bottomrule
\end{tabular}
\end{table}

We use the \emph{Medium} drift setting for the main OOC results; \emph{Small/Large} appear in the sensitivity study (Appendix~B).

\paragraph{Results}
Table~\ref{tab:factoryio_results} reports Factory~I/O OOC performance under \emph{Low\&Slow} (L\&S) and \emph{Smash\&Grab} (S\&G) across Black/Grey/White knowledge. Consistent with our process‑control testbed, \emph{Low\&Slow} maintains higher stealth (lower averaged recall, higher Stealth) and induces moderate wear (cycles/min $\uparrow$) and subtle throughput/quality degradation. \emph{Smash\&Grab} achieves larger instantaneous impact at the expense of detectability. White‑box knowledge yields the stealthiest attacks, as the adversary better models detector sensitivities.

\paragraph{Takeaways.}
Across all three discrete‑manufacturing scenes, the adversary transfers without architectural changes. \emph{Low\&Slow} remains the stealthiest (average Stealth $0.78$–$0.88$ in White), while \emph{Smash\&Grab} trades stealth for impact. In each scene, cycles/min increase (6–9\% for L\&S; 14–18\% for S\&G), throughput drops are modest for L\&S (3–5\%) and larger for S\&G (7–11\%), and quality degrades in proportion to strategy aggressiveness. These trends mirror our real testbed and substantiate cross‑domain generalizability.

}

\begin{IEEEbiography}[{\includegraphics[width=1.12in,height=1.35in,clip,keepaspectratio]{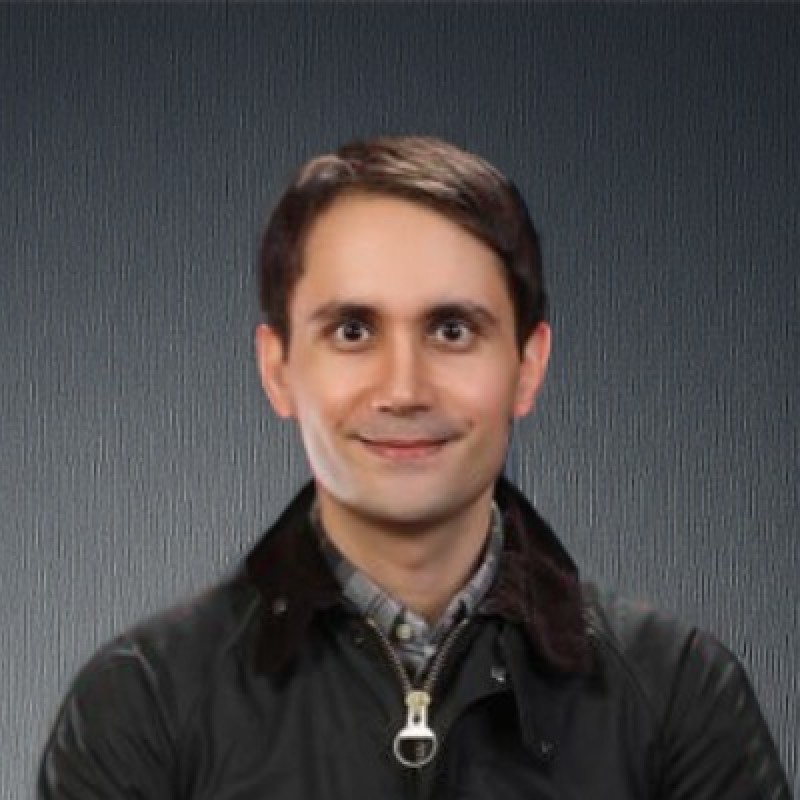}}]{Aryan Pasikhani} who obtained his PhD from the University of Sheffield, has established distinguished contributions in the realm of cybersecurity research. He is currently a Lecturer (Assistant Professor) in Cybersecurity at the University of Sheffield, where his research focuses on advancing the frontiers of offensive AI, the security of AI systems, and the application of AI across diverse scientific and engineering disciplines. His broader interests encompass intrusion detection and prevention systems, reinforcement learning, machine learning, privacy-preserved machine learning, quantum computing, and the security of embedded systems. Dr. Pasikhani’s research has been published in leading international venues, including the PETs and Euro S\&P. He has secured several research grants from Innovate UK and the EPSRC, underscoring his dedication to impactful and translational research. Furthermore, he serves as an esteemed TPC Member and a reviewer for several prestigious international journals and conferences, including the IEEE Internet of Things Journal, IEEE Transactions on Industrial Informatics, and the IEEE Sensors Journal. Beyond academia, Dr. Pasikhani is the founder of CybPass Ltd, a company pioneering autonomous penetration testing solutions for AI models.
\end{IEEEbiography}

\begin{IEEEbiography}[{\includegraphics[width=1.12in,height=1.35in,clip,keepaspectratio]{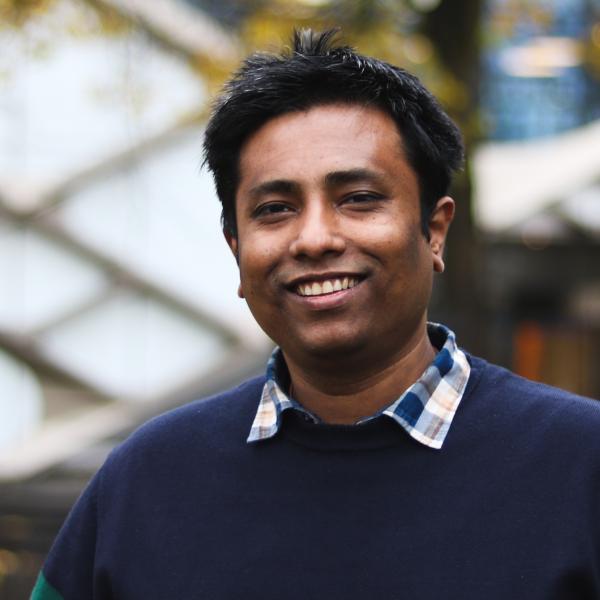}}]{Prosanta Gope} (Senior Member, IEEE) is currently working as an Associate Professor in the Department of Computer Science (Cyber Security) at the University of Sheffield, UK. Dr Gope served as a Research Fellow in the Department of Computer Science at the National University of Singapore (NUS). Primarily driven by tackling challenging real-world security problems, he has expertise in Lightweight Authentication, Authenticated Encryption, 5G and Next Generation Communication Security, Privacy-Preserving Machine Learning, Security in the Internet of Things, Smart-Grid Security, PUF-based security system and IoT Hardware. He has authored more than 100 peer-reviewed articles in several reputable international journals and conferences and has four filed patents. Several of his papers have been published in top-tier security journals (such as IEEE TIFS, IEEE TDSC, and ACM TOPS), and prominent security conferences (such as IEEE S\&P,  ACM CCS, USENIX Security Symposium, IEEE Computer Security Foundations Symposium (CSF), Privacy Enhancing Technologies Symposium (PETS), ESORICS, Euro S\&P, IEEE TrustCom, IEEE HoST, etc.) Dr Gope has served as a TPC member and Co-Chair at several reputable international conferences, including PETS, ESORICS, IEEE TrustCom, IEEE GLOBECOM (Security Track), and ARES. He currently serves as an Associate Editor of the IEEE Transactions on Dependable and Secure Computing, IEEE Transactions on Information \& Forensics and Security, IEEE Transactions on Services Computing, IEEE Systems Journal, and the Journal of Information Security and Applications (Elsevier). His research has been funded by EPSRC, Innovate UK, and the Royal Society.
\end{IEEEbiography}

\begin{IEEEbiography}[{\includegraphics[width=1.12in,height=1.25in,clip,keepaspectratio]{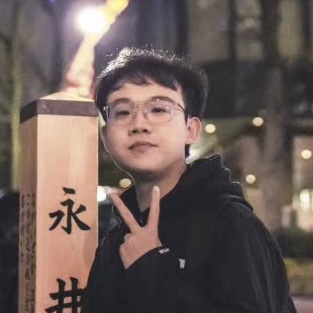}}]{Yang Yang}
received the B.Eng. degree from Beijing University of Technology in 2021 and the M.Sc. degree in Computer Science from the University of Sheffield in 2022. He is currently pursuing a PhD degree in Electrical and Computer Engineering at the National University of Singapore (NUS). His research interests include 5G and next-generation communication security, privacy-preserving machine learning, and deep learning. He has published as first and corresponding author in top-tier conferences such as the IEEE Symposium on Security and Privacy (IEEE S\&P), the ACM Conference on Computer and Communications Security (ACM CCS), and the Privacy Enhancing Technologies Symposium (PETS). He also published in high-impact journals like Journal of Computer Security, IEEE Transactions on Information Forensics \& Security (IEEE TIFS) and IEEE Transactions on Dependable \& Secure Computing (IEEE TDSC).
\end{IEEEbiography}

\begin{IEEEbiography}[{\includegraphics[width=1.12in,height=1.25in,clip,keepaspectratio]{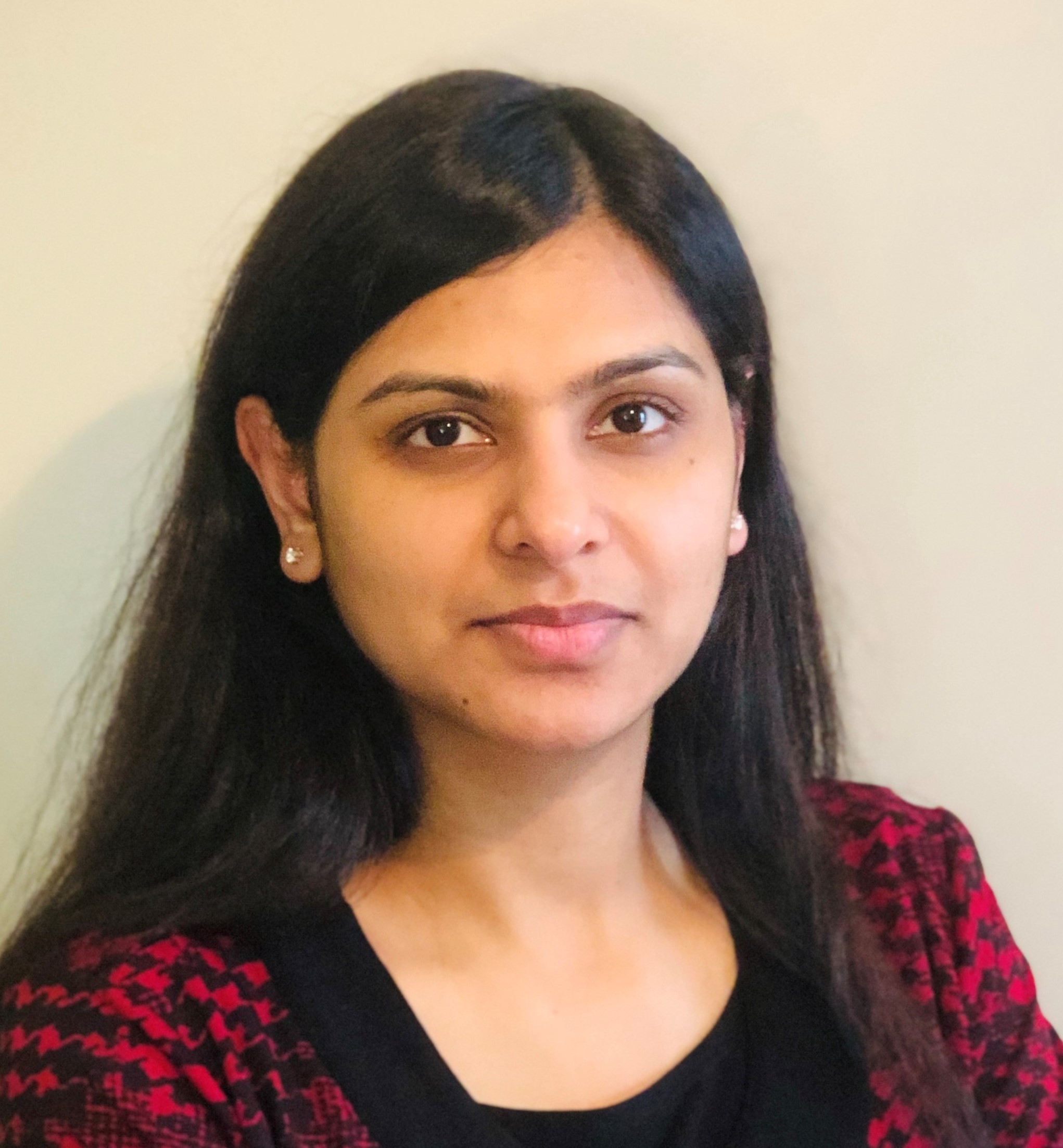}}]
{Shagufta Mehnaz} is an Assistant Professor in the Department of Computer Science and Engineering at The Pennsylvania State University. Her research lies at the intersection of security, privacy, and machine learning, with a focus on privacy attacks and defences in centralised and federated learning systems. She has published extensively in top-tier venues including IEEE S\&P, USENIX Security, NDSS, PETS, and AAAI. Her work spans privacy-preserving analytics, inference attacks, and robust federated learning. She is the recipient of the NSF CAREER Award (2025).
\end{IEEEbiography}

\begin{IEEEbiography}[{\includegraphics[width=1.12in,height=1.5in,clip,keepaspectratio]{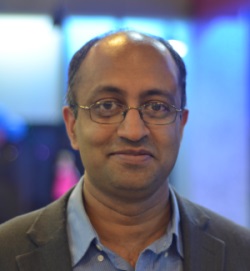}}]{Biplab Sikdar} \textnormal{(IEEE Fellow)} received the B.Tech. Degree in Electronics and Communication Engineering from North Eastern Hill University, Shillong, India, in 1996, and the M.Tech. Degree in electrical engineering from the Indian Institute of Technology Kanpur, Kanpur, India, in 1998, and the Ph.D. degree in electrical engineering from the Rensselaer Polytechnic Institute, Troy, NY, USA, in 2001. He was a faculty member at the Rensselaer Polytechnic Institute from 2001 to 2013 and was an assistant professor and Associate Professor. He is currently a Professor and Head of the Department of Electrical and Computer Engineering at the National University of Singapore. He also serves as the Director of the Cisco-NUS Corporate Research Laboratory. His current research interests include wireless networks and security for the Internet of Things and cyber-physical systems. He has served as an Associate Editor for the IEEE Transactions on Communications, IEEE Transactions on Mobile Computing, IEEE Internet of Things Journal and IEEE Open Journal of Vehicular Technology.
\end{IEEEbiography}

\end{document}